\documentclass[a4paper,usenames,dvipsnames,11pt]{article}
\usepackage{slashed}
\usepackage{mathrsfs,booktabs,multirow,tabularx}
\usepackage{stmaryrd}
\usepackage{xspace}
\usepackage{fancyvrb}
\usepackage{amssymb}    %
\usepackage{verbatim}   
\usepackage{lscape}

\usepackage{listings}
\usepackage{mathtools}
\numberwithin{equation}{section}
 \usepackage{amsmath}
 \usepackage{graphicx}
 \usepackage[merge,numbers,compress]{natbib}
 \usepackage[T1]{fontenc}
 \usepackage{booktabs}
 \usepackage{xcolor} 
 \usepackage{xspace}
 \usepackage{dcolumn}
 \usepackage{placeins}
 \usepackage{amssymb}
 \usepackage{nicefrac}
 \usepackage[colorlinks=true,citecolor=blue,linkcolor=blue]{hyperref}
 \usepackage{caption}
 \usepackage
 [subrefformat=parens,position=top,skip=-15pt,margin=15pt,justification=justified,singlelinecheck=false]
 {subcaption}
 \usepackage{array,multirow}
 \usepackage[makeroom]{cancel}

\makeatletter
\newcommand{\oset}[3][0ex]{%
  \mathrel{\mathop{#3}\limits^{
    \vbox to#1{\kern-3\ex@
    \hbox{\tiny$\scriptstyle#2$}\vss}}}}
\makeatother

\def\beq{\begin{equation}}
\def\eeq{\end{equation}}
\def\beqn{\begin{eqnarray}}
\def\eeqn{\end{eqnarray}}

\def\remove#1#2{#1\hspace{-#2truecm}\backslash}

\def\oset#1#2#3{\left(#1\right)_{#2,#3}}

\newcommand{\bqa}{\begin{eqnarray}}
\newcommand{\eqa}{\end{eqnarray}}
\chardef\MyArticleWithColor=\pdfcolorstackinit page direct{0 g}

\def\cCode#1{\begin{lstlisting}[mathescape,basicstyle=\small
\ttfamily,frame=leftline,aboveskip=4mm,belowskip=4mm,xleftmargin=20pt,framexleftmargin=10pt,
numbers=none,framerule=2pt,abovecaptionskip=0.0mm,belowcaptionskip=3.5mm #1]}

\newcommand\sss{\scriptscriptstyle}

\newcommand\as{\alpha_{\sss S}}

\newcommand\aNLO{{\sc\small MadGraph5\_aMC@NLO}}

\newcommand\HWpp{{\sc\small Herwig++}}

\newcommand\PYe{{\sc\small Pythia8}}
\newcommand\HWs{{\sc\small Herwig6}}
\newcommand\HWsette{{\sc\small Herwig7}}
\newcommand\FJ{{\sc\small FastJet}}

\newcommand\sherpa{{\sc\small Sherpa}}

\newcommand\pt{p_{\sss T}}

\newcommand{\LO}{{\rm LO}}

\newcommand{\NLO}{{\rm NLO}}

\newcommand\FKSpairs{{\cal P}_{\sss\rm FKS}}

\newcommand{\bq}{\bar{q}}

\newcommand{\ord}{{\cal O}}

\def\EWSL{\rm EWSL}
\def\LOQCDEWSL{\rm LO_{QCD+EWSL}}
\def\NLOQCDEWSL{\rm NLO_{QCD+EWSL}}

\def\NLOQCD{{\rm NLO_{QCD}}}
\def\NLOQCDPS{\rm NLO_{QCD}+PS} 
\def\LOQCD{{\rm LO_{QCD}}}
\def\LOQCDPS{\rm LO_{QCD}+PS} 

\def\NLOQCDEW{\NLO_{{\rm QCD+EW}}}
\def\NLOQCDEWPS{\NLOQCDEW+{\rm PS}}
\def\NLOQCDEWSLPS{\NLOQCDEWSL+{\rm PS}}
\def\LOQCDEWSLPS{\LOQCDEWSL+{\rm PS}}
\def\bestprednoPS{\NLOQCD\otimes{\rm EWSL}}
\def\bestpred{\bestprednoPS+{\rm PS}}
\def\PSnoQED{\rm PS_{\cancel{\rm QED}} }

\def\dEWSL{\delta^{\rm EWSL}}
\def\dEWSLH{\dEWSL_{\sss (\clH)}}
\def\dEWSLS{\dEWSL_{\sss (\clS)}}

\def\beq{\begin{equation}}
\def\eeq{\end{equation}}
\def\beqar{\begin{eqnarray}}
\def\eeqar{\end{eqnarray}}
\def\barr#1{\begin{array}{#1}}
\def\earr{\end{array}}
\def\bfi{\begin{figure}}
\def\efi{\end{figure}}
\def\btab{\begin{table}}
\def\etab{\end{table}}
\def\bce{\begin{center}}
\def\ece{\end{center}}

\def\de{\delta}

\def\refeq#1{\mbox{(\ref{#1})}}

\newcommand{\M}{{\cal{M}}}

\def\mathswitchr#1{#1}
\newcommand{\PW}{\mathswitchr W}

\newcommand{\PZ}{\mathswitchr Z}

\newcommand{\PH}{\mathswitchr H}

\newcommand{\Pt}{\mathswitchr t}

\def\mathswitch#1{\relax\ifmmode#1\else$#1$\fi}
\newcommand{\MW}{\mathswitch {M_\PW}}
\newcommand{\MZ}{\mathswitch {M_\PZ}}
\newcommand{\MH}{\mathswitch {M_\PH}}

\newcommand{\Mt}{\mathswitch {m_\Pt}}

\newcommand{\ew}{{\mathrm{ew}}}

\newcommand{\weak}{{\mathrm{weak}}}

\newcommand{\cew}{C^{\ew}}

\newcommand{\GB}{V}

\newcommand{\lrMwithabs}{l(|r_{kl}|,M^2)}

\newcommand{\lsW}{l(s,\MW^2)}

\newcommand{\lWM}{l(\MW^2,M^2)}

\newcommand{\LrM}{L(|r_{kl}|,M^2)}

\newcommand{\Lrs}{L(|r_{kl}|,s)}

\newcommand{\LsW}{L(s,\MW^2)}

\newcommand{\lrs}{\log{\frac{|r_{kl}|}{s}}}
\newcommand{\lrsalpha}{l(|r_{kl}|,s)}

  \newcommand{\TO}{\rightarrow}

\newcommand{\denpoz}{{\sc\small DP}}

\newcommand{\deltaEW}{\delta^{\rm EW}_{\rm LA}}
\newcommand{\deltaQCD}{\delta^{\rm QCD}_{\rm LA}}
\newcommand{\Ltop}{L^t(s)}
\def\Las#1{l^{\as}(#1)}
\newcommand{\dmtQCD}{(\de\Mt)^{\rm QCD}}

\newcommand\clH{{\mathbb H}}
\newcommand\clS{{\mathbb S}}

\newcommand\allproc{{\cal R}}
\newcommand\allprocnpo{\allproc_{n+1}}

\newcommand\PYs{{\sc\small Pythia6}}

\newcommand{\muS}{\mu_S}

 \usepackage{comment}

\begin{document}

\title{\hfill ~\\[-30mm]
\phantom{h} \hfill\mbox{\small TIF-UNIMI-2023-25, LU-TP-23-10}
\\[1cm]
\vspace{13mm}   \textbf{Improving  NLO QCD event generators with high-energy EW corrections}}

\date{}
\author{
Davide Pagani$^{1\,}$\footnote{E-mail:  \texttt{davide.pagani@bo.infn.it}},
Timea Vitos$^{2,3}$\footnote{E-mail:  \texttt{timea.vitos@thep.lu.se}},
Marco Zaro$^{4\,}$\footnote{E-mail:  \texttt{marco.zaro@mi.infn.it}}
\\[9mm]
{\small\it $^1$ INFN, Sezione di Bologna, Via Irnerio 46, 40126 Bologna, Italy} \\[3mm]
{\small\it $^{2}$ Department of Physics, Lund University,} 
\\%
{\small\it Sölvegatan 14A, SE-223 62, Lund, Sweden}\\[3mm]
{\small\it $^{3}$ Department of Physics and Astronomy, Uppsala University,} 
\\%
{\small\it Box  516,  751 20, Uppsala, Sweden  }\\[3mm]
{\small\it $^4$ TIFLab, Universit\`a degli Studi di Milano \& INFN,} 
\\%
{\small\it Sezione di Milano, Via Celoria 16, 20133 Milano, Italy}\\[3mm]
        }
\maketitle

\begin{abstract}
\noindent

In this work we present a new method for the combination of electroweak (EW) corrections at high energies, the so-called EW Sudakov logarithms (EWSL), and next-to-leading-order QCD predictions matched to parton-shower simulations (NLO+PS). Our approach is based on a reweighting procedure of NLO+PS events. In particular, both events with and without an extra hard emission from matrix elements are consistently reweighted via the inclusion of the corresponding EWSL contribution. We describe the technical details and the implementation in the {\aNLO} framework. Via a completely automated procedure, events at this level of accuracy can be obtained for a vast class of hadroproduction processes. As a byproduct we provide results for phenomenologically relevant physical distributions from  top-quark pair and Higgs boson associated production ($t \bar t H$) and from the associated production of three $Z$ gauge bosons ($ZZZ$).

\end{abstract}
\thispagestyle{empty}
\vfill

\begingroup
\hypersetup{linkcolor=black}
\tableofcontents
\endgroup

\newpage

\section{Introduction}

 After the first two runs of the Large Hadron Collider (LHC), our knowledge of the fundamental interactions of elementary particles has tremendously improved. Above all, the Higgs boson has been observed~\cite{Aad:2012tfa,Chatrchyan:2012ufa} and its properties have been found compatible  \cite{Aad:2019mbh} with those predicted by the Standard Model (SM), further corroborating the current theory of fundamental interactions. On the other hand, no clear  sign of beyond-the-SM (BSM) physics has been found at the LHC so far, as it has been the case for previous colliders.  In 2022, ten years after the discovery of the Higgs boson, the Run-3 has started and during this period and the subsequent High-Luminosity (HL) runs \cite{Azzi:2019yne,Cepeda:2019klc,CidVidal:2018eel,Cerri:2018ypt,Citron:2018lsq,Chapon:2020heu}, the total amount of recorded data by the LHC will increase by a factor of 20 w.r.t.~the Run-1 and Run-2 data sets combined. Moreover, several options have been proposed for future colliders, involving collisions at higher energies between protons and/or leptons (including also muons). It is therefore clear that the quest for new physics at colliders is only at its initial stage.

 The success of this quest relies on the availability of precise and accurate SM predictions. In other words, the possibility of calculating QCD and electroweak (EW) higher-order corrections. Regarding fixed-order perturbative expansion, QCD radiative corrections at Next-to-Leading-Order (NLO), Next-to-NLO (NNLO) or even Next-to-NNLO ($\rm N^3LO$) accuracy are nowadays available for several processes. In fact,   both NLO QCD and NLO EW corrections can be calculated for processes with high-multiplicity final states, with limitations given only by computing power. This is possible since such corrections have been implemented in Monte Carlo generators and their calculation has been even automated  \cite{Alwall:2014hca, Kallweit:2014xda, Frixione:2015zaa, Chiesa:2015mya, Biedermann:2017yoi, Chiesa:2017gqx,  Frederix:2018nkq, Pagani:2021iwa}, 
 at different levels in the different frameworks, using different one-loop matrix-element providers \cite{Hirschi:2011pa, Cullen:2011ac, Cascioli:2011va, Actis:2012qn, Actis:2016mpe, Denner:2017wsf, Buccioni:2019sur}.  

Another unavoidable ingredient for the  simulation of events at  (hadron) colliders is the modelling of the multiple emission of (QCD) partons and their hadronisation, {\it i.e.}, Parton Shower (PS) simulations. However, while the matching of NLO QCD corrections and parton shower effects has already been achieved \cite{Frixione:2002ik, Frixione:2007nw, Frixione:2007vw} (also for NNLO \cite{Hamilton:2012rf, Alioli:2013hqa, Hoche:2014uhw, Monni:2019whf} and  recently even $\rm N^3LO$  accuracy \cite{Bertone:2022hig} for specific processes) and automated since a long time, in the case of NLO EW corrections a process-independent  approach still needs to be formulated  and either only approximations or case-by-case exact solutions have appeared in the literature so far \cite{Barze:2012tt, Barze:2013fru, Kallweit:2015dum, Granata:2017iod, Gutschow:2018tuk, Chiesa:2019ulk, Chiesa:2020ttl, Brauer:2020kfv, Lindert:2022qdd}. It is therefore desirable that the lack of a general algorithmic procedure for the exact matching of NLO QCD+EW predictions and PS simulations is solved as soon as possible, especially since it is well known that NLO EW corrections can strongly depend on the kinematics, and the naive estimate of their relative impact (NLO EW$\sim\ord(\alpha)\sim\ord(1\%)$ in absolute value) can be easily violated by one or more order of magnitudes. On the one hand, the origin of this violation can be due to a specific mechanism for a specific process and/or  observable \cite{Hollik:2011ps, Baglio:2013toa,Biedermann:2016yds,Biedermann:2017bss, Frederix:2017wme,Denner:2019tmn,Denner:2020zit}. On the other hand, the origin of this violation (NLO EW$\,\gg1\%$  in absolute value) is typically related to two different kinds of effects, which are universal. First, the final-state-radiation (FSR) of photons from light fermions, which is of QED origin and, {\it e.g.}, distorts the Breit-Wigner distributions of the $Z$-boson decay products. The modelling of FSR is available within modern PS simulators, such as \PYe~\cite{Sjostrand:2007gs, Sjostrand:2014zea, Bierlich:2022pfr},   \HWsette~\cite{Bellm:2015jjp}, and {\sherpa} \cite{Sherpa:2019gpd, Schonherr:2008av, Hoeche:2009xc}. Second, the EW Sudakov logarithms  (EWSL) \cite{Sudakov:1954sw} of the form $\alpha^n \log^k(s/\MW^2)$ with $k\le2n$, which are mainly of weak origin and become relevant at high energies ($s\gg\MW^2$). An algorithmic procedure for their evaluation at one- \cite{Denner:2000jv,Denner:2001gw} and two-loop \cite{Denner:2003wi,Denner:2004iz,Denner:2006jr,Denner:2008yn} accuracy is available since long time, the so-called Denner and Pozzorini ({\denpoz}) algorithm for the Sudakov approximation. It has been automated for the first time \cite{Bothmann:2020sxm} in  the {\sherpa} framework, and, after the revisitation and improvement of particular features~\cite{Pagani:2021vyk}, in the {\aNLO} framework  \cite{Alwall:2014hca, Frederix:2018nkq}.

It is clear therefore that  having automated tools for generating events at NLO accuracy matched to PS simulation (NLO+PS) where not only NLO QCD corrections ($\NLOQCD$+PS) but also the dominant NLO EW ones, FSR and EWSL, are taken into account is very useful for current and future experimental analyses, especially if the addition of the EW contributions does not slow down the generation of the events. An example of a tool of this kind based on Ref.~\cite{Bothmann:2020sxm} and the {\sc Meps@Nlo}   \cite{Hoeche:2012yf} method has  already appeared in the literature \cite{Bothmann:2021led} and it has been applied to a specific process: $ZZ$ and $ZZj$ merged production. This work has shown the relevance of such studies and the advantages of a general-purpose automation for (at least) SM processes in general.

The current work precisely presents  the automation, in the {\aNLO} framework, of combined EWSL and $\NLOQCDPS$ accuracy, including QED FSR,  for the event generation of SM processes. We have implemented it in this framework, since it offers all the capabilities for achieving this goal. Indeed, in {\aNLO} the following three features are already available:
\begin{itemize}
\item The automated calculation of matched ${\NLOQCDPS}$ simulation, via the interface with external PS simulators.
\item The automated evaluation of EWSL at one-loop accuracy.
\item The possibility of reweighting events ({\it e.g.}~changing model parameters) from LO and NLO simulations~\cite{Mattelaer:2016gcx}.
\end{itemize}  
Our strategy therefore is based on the reweighting of ${\NLOQCDPS}$ events taking into account the EWSL contribution. FSR can then eventually be simulated directly via the PS. In doing so, we do not reweight with the EWSL only the LO contribution from the hard process, but also the QCD one-loop virtual contribution as well as the first QCD real emission. For the latter, we take into account that both the kinematics and the external states are different. In this way, especially for high-energies ($s\gg\MW^2$), a good approximation of NLO EW corrections is correctly taken into account both for the Born-like process and the one with an extra  hard jet. Automatically, the NLO QCD prediction for the inclusive production is correctly reweighted via the EWSL, and QCD shower effects are taken into account. Moreover, by adopting the so-called $\rm SDK_{weak}$ scheme \cite{Pagani:2021vyk} for the EWSL, which consists of a complete removal of QED effects of infrared (IR) origin, FSR or in general QED effects can be included in the PS simulation avoiding their double counting. 

Since the evaluation of EWSL involves only tree-level matrix elements and compact analytical formulas (see Ref.~\cite{Pagani:2021vyk}), one of the advantages of reweighting via the Sudakov approximation is the speed of this procedure and especially the numerical stability of the results. However, especially for the real emission contributions, it is crucial that the EWSL are damped in phase-space regions where any of the kinematical invariants involving two external states is smaller than $\MW^2$. This is necessary not only because in such phase-space regions the Sudakov approximation is not valid, but also because the soft and collinear limits relating $n+1$ and $n$ final states in QCD must be preserved for a correct matching of ${\NLOQCD}$ predictions and PS simulations also after the reweighting. The approach we have adopted for solving this problem is completely general and, although we will discuss in the paper its implementation in {\aNLO}, could be in principle extended for other matching schemes. Since the approach is based on the reweighting, {\it i.e.}~a step happening after the event generation, it does not rely on the strategy for the event generation itself.  

In this paper we focus on the technical implementation and its validation, leaving phenomenological studies and comparison with exact NLO EW accuracy predictions to dedicated works. Nevertheless,  we show results for two representative SM processes: the top-quark pair and Higgs associated hadroproduction ($pp \TO t \bar t H$) and the hadroproduction of three $Z$ gauge bosons ($pp \TO ZZZ$). We consider for both processes the final-state particles as stable, while for the latter we also consider the case of $Z$ bosons decaying into $e^+e^-$ pairs, enlightening the relevance of considering both EWSL and QED FSR contributions.  In doing so, we use {\sc \small MadSpin} \cite{Artoisenet:2012st} for performing the decays and therefore, while the reconstruction of the tree-level spin correlations are taken into account, we do not preserve the information of the correlation of the helicity-dependent EWSL with the helicities and angular distributions of the decay products.\footnote{During the publication of our work, this issue was addressed for the first time in Ref.~\cite{Lindert:2023fcu}.}

The paper is organised as follow. In Sec.~\ref{sec:general} we discuss the motivations for this work and the general structure of our implementation in {\aNLO} of an automated framework for performing $\NLOQCDPS$ simulations including also the effects of EWSL. In Sec.~\ref{sec:technical} we give the technical details of the implementation of the reweighting approach in {\aNLO}. In Sec.~\ref{sec:results} we present results for $t\bar tH $ and $ZZZ$ hadroproduction at 13 TeV collisions. In Sec.~\ref{sec:conclusion} we give our conclusion and outlook. Finally, in Appendix \ref{sec:EWSL} we briefly summarise the {\denpoz} algorithm as revisited in Ref.~\cite{Pagani:2021vyk}, focusing on the concepts and the formulas that are relevant for the discussion of Sec.~\ref{sec:technical}.

\section{
Overview of the problem and proposed solution \label{sec:general}}

As already mentioned in the introduction, in this section we discuss the motivations for this work and the general structure of the implementation in {\aNLO} of an automated framework for performing $\NLOQCDPS$ simulations including also the effects of EWSL, {\it i.e.}, NLO EW corrections in the Sudakov approximation. For the very interested readers, the technical details can be found in Sec.~\ref{sec:technical}. In the following, we start by introducing the notation, which will be also used in Sec.~\ref{sec:results} where we present numerical results for selected processes. 
\subsection{Notation \label{sec:notation}}
At fixed order, adopting the notation already used in Refs.~\cite{Frixione:2014qaa, Frixione:2015zaa, Pagani:2016caq, Frederix:2016ost, Czakon:2017wor, Frederix:2017wme, Frederix:2018nkq, Broggio:2019ewu, Frederix:2019ubd,Pagani:2020rsg, Pagani:2020mov, Pagani:2021iwa}, the different contributions from the expansion in powers of $\as$ and $\alpha$ to the differential or inclusive cross section $\Sigma$ of a generic (SM) process up to NLO can be denoted as:
\begin{align}
\Sigma^{}_{\LO}(\as,\alpha) &= \Sigma^{}_{\LO_1} + \cdots  + \Sigma^{}_{\LO_k}\, , \label{eq:blobs_LO_general} \\
 \Sigma^{}_{\NLO}(\as,\alpha) &=  \Sigma^{}_{\NLO_1} + \cdots  + \Sigma^{}_{\NLO_{k+1}}\, , \label{eq:blobs_NLO_general} 
\end{align}
 where $k\ge1$ and is process dependent. 
 At LO, meaning tree-level diagrams only, there can be more than one perturbative order $\as^n \alpha^m$ and each one of the orders is associated to a different $\Sigma^{}_{\LO_i}$, but the sum $n+m$ is constant for each process. If $\Sigma^{}_{\LO_i}\propto \as^n \alpha^m$ then $\Sigma^{}_{\LO_{i+1}}\propto \as^{n-1} \alpha^{m+1}$. Similarly, including one-loop corrections, there is more than one perturbative order and each one is associated to a different $\Sigma^{}_{\NLO_i}$, and if $\Sigma^{}_{\LO_i}\propto \as^n \alpha^m$ then $\Sigma^{}_{\NLO_{i}}\propto \as^{n+1} \alpha^{m}$ and $\Sigma^{}_{\NLO_{i+1}}\propto \as^{n} \alpha^{m+1}$. The full set of LO and NLO orders is the so-called Complete-NLO order.
 
In this paper we are interested, from the fixed-order side, in the quantities
\begin{align}
\LOQCD&\equiv \Sigma^{}_{\LO_1} \, ,  \label{eq:NLOdefsb}\\
\NLOQCD&\equiv \Sigma^{}_{\LO_1} + \Sigma^{}_{\NLO_1} \, ,\\ 
\NLOQCDEW&\equiv \Sigma^{}_{\LO_1} + \Sigma^{}_{\NLO_1} + \Sigma^{}_{\NLO_2} \, , \\
\EWSL&\equiv \ord\left(\log^k(s/\MW^2)\right)~{\rm of}~\Sigma^{}_{\NLO_2},~{\rm with}~k=1,2 \, , \label{eq:NLOdefEWSL}\\
\LOQCDEWSL&\equiv \LOQCD+\EWSL \, , \\
\NLOQCDEWSL&\equiv \NLOQCD+\EWSL \,.
\label{eq:NLOdefse} 
\end{align}
The rest of the Complete-NLO prediction, $ \Sigma^{}_{\LO_i}$ with $i>1$ and  $\Sigma^{}_{\NLO_i}$ with $i>2$ is not considered and is not relevant for the discussion in this paper. Still it is worth to mention that, as it is already very well known (see {\it e.g.}~Ref.~\cite{Frixione:2014qaa}), the NLO EW corrections, {\it i.e.}~ $\Sigma^{}_{\NLO_2}=\NLOQCDEW-\NLOQCD$, involve both $\ord(\alpha)$ corrections on top of $\LOQCD$ and $\ord(\as)$ corrections on top of the $\LO_2$, where we have used the abbreviation ${\rm (N)LO}_i\leftrightarrow \Sigma^{}_{{\rm (N)LO}_i} $. Some technical subtleties related to this point are discussed in Sec.~\ref{sec:dEWSLtech}.

\subsection{State of the art and general considerations}

The reason why at high energies, and when all the invariants are large, EWSL are a very good approximation of NLO EW corrections (the $\NLO_2$ term) is that 
\begin{equation}
\frac{\NLO_2-\EWSL}{\LOQCD}\sim \alpha \left(\ord(1)+\ord\left(\frac{\MW^2}{s}\right)\right)\, ,
\label{eq:EWSLgood} 
\end{equation}
which is a non-negligible contribution only if we are considering a process for which a precision of very few percent is relevant. EWSL are precisely the ingredient that is missing in many present and future experimental analyses at the LHC that are targeting particles with  transverse momenta larger than or equal to a few hundreds of GeV.
Due to statistics and other uncertainties of both experimental and theoretical origin, these analyses are sensitive to effects of $\ord(10\%)$. Thus, in those cases, EWSL cannot be ignored. 
The same analyses cannot ignore also the effects due to NLO QCD corrections and clearly PS simulations. 

While the matching of $\NLOQCDEW$ with PS ($\NLOQCDEWPS$) is in general far from trivial, the case of $\NLOQCDEWSL$ ($\NLOQCDEWSLPS$) is indeed trivial once a $\NLOQCDPS$ matching is already available. In the $\NLOQCDPS$ matching, PS  can simulate only effects of $\ord(\as^n)$ with $n>0$ on top of $\NLOQCD$. The matching consists in avoiding the double counting of the case $n=1$ on top of the $\LOQCD$ component. If now we start with the $\NLOQCDEWSL$ at fixed order, PS effects on top of the EWSL components will induce effects of $\ord(\alpha \as^n)$ with $n>0$ on top of the $\LOQCD$, which cannot be double-counted by construction. 

In the previous argument we have implicitly assumed that PS involves only QCD effects, but in modern PS simulators also QED effects such as FSR, possibly involving also photons splitting into fermions, are taken into account. In those cases, the PS simulations will induce effects of $\ord(\alpha ^m \as^n)$ with $m+n>0$ on top of the $\LOQCD$ component, which can instead lead to a possible double-counting of the EWSL in the case $m=1, n=0$. The solution in this case is using the scheme denoted as $\rm SDK_{\rm weak}$ in Ref.~\cite{Pagani:2021vyk}. This scheme completely neglects effects of pure-QED origin in the evaluation of EWSL and therefore takes into account only the purely weak ones, avoiding the double-counting (more details are given in Appendix \ref{sec:EWSL}). The QED component of the EWSL would anyway vanish in observables inclusive in the photon radiation, but otherwise large QED effects are simulated directly by the shower.\footnote{In Ref.~\cite{Pagani:2021vyk} the $\rm SDK_{\rm weak}$ has been devised in order to take into account the cancellation between the virtual and real QED components. This cancellation has not been formally proven but it is supported by several examples that are reported and discussed in Ref.~\cite{Pagani:2021vyk}. One should notice that here we do not want to take into account this cancellation, instead to completely remove the QED component and leave its simulation to the PS. Therefore, the choice of the $\rm SDK_{\rm weak}$ approach is not motivated by the cancellation between virtual and real QED contributions.} These effects include not only the QED component of the EWSL but also FSR effects such as the already mentioned distortion of the Breit-Wigner distribution from the $Z$ boson decay.  This means that it is possible to generate $\NLOQCDEWSL$ events for the production of heavy objects ($W,Z,H$ bosons or top quarks) and let them decay directly via the shower or programs like {\sc \small MadSpin} \cite{Artoisenet:2012st} including QED FSR effects from the shower in $\NLOQCDEWSLPS$ simulations.\footnote{The approach described here would fail in case of the inclusion of purely weak effects directly in the shower \cite{Christiansen:2014kba, Kleiss:2020rcg, Brooks:2021kji, Masouminia:2021kne}. However, the multiple emission of heavy bosons (denoted later in the text as HBR) are not relevant for 10--100\% precision  even at a 100 TeV proton--proton collider \cite{Mangano:2016jyj}. Similar consideration applies for the evolution of proton PDFs involving weak splittings \cite{Bauer:2018arx,Fornal:2018znf}. We note that the case of a high-energy lepton collider is a completely different scenario \cite{Han:2020uid,Han:2021kes,Ruiz:2021tdt,Garosi:2023bvq}.}  We stress the fact that with the notation ``PS" we understand the presence of both  QCD and QED effects in the shower.\footnote{This is the default in the PS input files generated by {\aNLO}.} When we will refer to the purely QCD effects in the PS we will use the notation $\PSnoQED$.

The reader may wonder what would be the problem if in the previous argument instead of considering EWSL, the exact NLO EW corrections were considered, namely the $\NLOQCDEWPS$ case. First of all, it is important to note that claiming NLO EW accuracy means having under control the {\it exact} $\ord(\alpha)$ effects, together with the advantages of shower simulations, meaning {\it e.g.}~the possibilities of setting hard jet-vetoes or studying the transverse momentum of the total final-state system without obtaining the typical pathological results of fixed-order simulations, where large logarithms are not resummed. For this reason not only the QED shower on top of $\LOQCD$ but also the standard QCD part of PS on top of the $\LO_2$ must be taken into account and matched to the NLO EW corrections. Especially the latter contribution poses non-trivial challenges, since the colour flow is not defined as the $\LO_2$ contribution is typically an interference and not a squared amplitude. Intermediate solutions to these problems have been proposed (see Refs.~\cite{Kallweit:2015dum, Gutschow:2018tuk,Granata:2017iod} for applications to phenomenology, possibly extended to multi-jet merged processes, and 
Ref.~\cite{Frixione:2021yim} for more formal aspects related e.g.~to the definition of colour flows for interferences), and results with an exact matching  at $\NLOQCDEWPS$  have been presented only for cases where the $\LO_i$, with $i>1$ are not present \cite{Barze:2012tt, Barze:2013fru,   Chiesa:2019ulk, Chiesa:2020ttl, Brauer:2020kfv}, recently even at ${\rm N}\NLOQCDEWPS$ accuracy \cite{Lindert:2022qdd}. However, a general method has not appeared so far in the literature.

Even if a general $\NLOQCDEWPS$ method were available, the possibility of obtaining $\NLOQCDEWSLPS$ is still desirable for practical reasons: speed and stability. EWSL can be calculated via compact analytical formulas and tree-level matrix elements only, as explained in Appendix \ref{sec:EWSL}. Therefore, unlike exact NLO EW corrections, they can be evaluated in a much faster and 
numerically stable way, without numerical cancellations between real and virtual contributions. 

That said, we want to stress that the EWSL are an approximation of the exact NLO EW corrections and therefore there could be non-negligible effects at high energies that cannot be captured, such as photon-initiated contributions (see, {\it e.g.}, Ref.~\cite{Kallweit:2017khh}). In general, given the possibility of performing at fixed order both the calculations of EWSL and of exact NLO EW corrections, for any phenomenological study, we strongly suggest to compare the two approaches beforehand.
\subsection{Proposed solution: $\bestpred$}
In this work we present the automation not only of the $\NLOQCDEWSLPS$, and also of its simpler version at LO denoted as $\LOQCDEWSLPS$, but also of what we will denote as 
\begin{equation}
\bestpred\, ,
\end{equation}
which will be our best prediction and will be described briefly in the following, leaving the technical details to Sec.~\ref{sec:technical}. 

First of all, predictions at $\bestpred$ accuracy are obtained by showering events at $\bestprednoPS$ accuracy: at variance with the $\NLOQCDEWSL$ case, not only the $\LOQCD$ contribution but also the NLO QCD corrections ($\Sigma^{}_{\NLO_1}$) receive corrections from EW Sudakov logarithms of $\ord\left(\alpha \log^k(s/\MW^2)\right)$ with $k=1,2$.   It is important to note that the NLO QCD corrections originate from both virtual and real contributions. While the former have the same external states and kinematics as the Born contribution, the latter are different. The EWSL have to be therefore evaluated separately for the virtual and real contributions. 

Using a notation that will be exploited in Sec.~\ref{sec:technical}, we denote with  $\dEWSLS$ the relative impact of the EWSL on top of the  $\LOQCD$,
\begin{equation}
\dEWSLS\equiv\frac{\rm EWSL}{\LOQCD}\, \label{eq:dEWSLS}.
\end{equation}
Similarly, considering the same process plus the radiation of a ${\clH}$ard QCD parton, the corresponding quantity is instead denoted with  $\dEWSLH$. As guiding principle, in the $\bestprednoPS$ predictions we want that, similarly to the $\LOQCD$ contribution, also NLO QCD virtual and real contributions in the ${\clS}$oft/collinear regions receive  $\dEWSLS$ corrections. Conversely, the contribution from hard and non-collinear real emissions should be corrected by $\dEWSLH$. At the same time, we need to ensure that in the soft/collinear limits 
\begin{equation}
\dEWSLH\longrightarrow\dEWSLS\, \label{eq:sclimit}.
\end{equation}
Condition $\eqref{eq:sclimit}$ is unavoidable for two different reasons:
\begin{enumerate}
\item{\label{bullet:1}Virtual and real IR poles need to receive the same corrections from EWSL, so that the cancellation of the divergences is preserved.}
\item{In the soft and/or collinear limits at least one of the kinematical invariants involving two external states is by definition smaller than $\MW^2$, invalidating the applicability of the Sudakov approximation and the sensibility of $\dEWSLH$.}
\end{enumerate}
Condition $\eqref{eq:sclimit}$, leaves freedom on how to implement the mapping  between $\dEWSLH$ and $\dEWSLS$ and we will discuss the practical implementation in Sec.~\ref{sec:technical}.

The strategy adopted for correcting the different contributions by either $\dEWSLS$ or $\dEWSLH$ is very similar to the one used in, {\it e.g.}, Ref.~\cite{Maltoni:2014eza} and relies on the general framework introduced in Ref.~\cite{Mattelaer:2016gcx}: reweighting NLO events before showering them. We will give more details in  Sec.~\ref{sec:technical}, but the idea is the following. In the MC@NLO formalism two kinds of events are generated, namely the ${\clS}$ and ${\clH}$ events. The latter class corresponds to the contribution from hard real emission to NLO QCD corrections. It takes into account the contribution of the Monte Carlo (MC) counter term, which is precisely added in order to avoid the double counting from PS effects on top of the $\LOQCD$. On the contrary, the rest of the contributions entering the NLO QCD predictions corresponds to ${\clS}$ events.  Given a process $pp\rightarrow X$, with $X$ having multiplicity $n$,  ${\clS}$ events are of the kind $2\TO n$, while  ${\clH}$ events are of the kind $2\TO n+1$. Denoting the weights of the former as $w_{\clS}$ and the weight of the latter as $w_{\clH}$\footnote{In practice, since MC@NLO events are unweighted up to the sign, one has $|w_{\clH}| = |w_{\clS}|$.}, the events generated at $\NLOQCD$ accuracy with the MC@NLO matching scheme can be promoted to $\bestprednoPS$ accuracy performing the following reweighting before the parton shower:
\begin{eqnarray}
&\clS:~~ w_{\clS}\Longrightarrow & (1+ \dEWSLS) w_{\clS}\,, \label{eq:rewS}\\
&\clH:~~ w_{\clH}\Longrightarrow & (1+ \dEWSLH) w_{\clH}\,. \label{eq:rewH}
\end{eqnarray}
After the reweighting, events can be showered obtaining predictions at $\bestpred$ accuracy.  Again, we will give many more details on the procedure in Sec.~\ref{sec:technical}.

\section{Technical details of the  EW Sudakov reweighting strategy \label{sec:technical}}

In Sec.~\ref{sec:general} we described the general features of the $\bestpred$ approximation and the motivations behind it. 
In this section we provide the technical details of the reweighting procedure, which we have implemented by extending the general-purpose reweighting module of {\aNLO}~\cite{Mattelaer:2016gcx}.
First, we briefly recall the basics of the MC@NLO matching, following a very similar argument of Ref.~\cite{Maltoni:2014eza}. Then we describe how in practice we use the {\denpoz} algorithm for implementing the prescription in Eqs.~\eqref{eq:rewS} and \eqref{eq:rewH} ensuring the condition \eqref{eq:sclimit}. We remind the reader that in Appendix \ref{sec:EWSL} we have summarised the basic structure of the {\denpoz} algorithm and its revisitation in Ref.~\cite{Pagani:2021vyk}, including technical aspects that are also relevant in this section.

\subsection{MC@NLO matching and reweighting}

The structure of a fixed-order NLO calculation of a cross section $\text{d}\sigma$, as performed within {\aNLO},  for a $2\TO n$ production process can be summarised by the following equation
 \begin{equation}
     \text{d}\sigma = \text{d}\phi_n \left( \mathcal B + \mathcal V + \mathcal C^{\rm int} \right) +
     \text{d}\phi_{n+1} \left(\mathcal R - \mathcal C\right)\,.
     \label{eq:nloxsec}
 \end{equation}
The terms $\mathcal{B,V,R}$ are respectively the Born, virtual and real emission contributions. The term $\mathcal C$ is the local
counterterm that renders the integral over the $\text{d}\phi_{n+1}$ phase-space finite, where $\text{d} \phi_{n+1}\equiv \prod_{k=1}^{n+1} \text{d}\bar \phi_k $ and $\text{d} \bar \phi_k$ is the differential of the phase-space integration associated to the particle $k$. The term $\mathcal C^{\rm int}$ is the integrated form of  $\mathcal C$ over $\text{d}\phi_{n+1} /\text{d}\phi_{n} $, such that $ \mathcal C^{\rm int} -\int \mathcal C\, \text{d}\bar\phi_{n+1} =0$. The specific form of the counterterms depends on the subtraction scheme that is used, {\it e.g.}~FKS~\cite{Frixione:1995ms} or CS~\cite{Catani:1996vz}, where the former is the one on which the {\aNLO} implementation is based.
 
In the case of matching of $\NLOQCD$ computations with PS in the MC@NLO formalism, on top of the local counterterm $\mathcal C$ one has to also include the so-called Monte-Carlo counterterm  $\mathcal C_{\rm MC}$~\cite{Frixione:2002ik}, in order to avoid the double counting of PS effects on top of the $ \mathcal B$ contribution. The counterterm $\mathcal C_{\rm MC}$ accounts for
the cross section one obtains from PS simulations by truncating the perturbative expansion at 
${\cal O}(\as^{m+1})$, where $\LO_1$ is of ${\cal O}(\as^m)$.  The MC counterterm depends on the specific PS simulator one interfaces the calculation to\footnote{In {\aNLO} the $\NLOQCDPS$ matching  has 
been fully validated \cite{Torrielli:2010aw,Frixione:2010ra} for \PYe~\cite{Sjostrand:2007gs, Sjostrand:2014zea, Bierlich:2022pfr}, but also 
\HWpp~\cite{Bahr:2008pv,Bellm:2013hwb}, 
\HWs~\cite{Corcella:2000bw,Corcella:2002jc}, \HWsette~\cite{Bellm:2015jjp} and \PYs~\cite{Sjostrand:2006za},
for only strongly-interacting particles in the final state in the case of $\pt$-ordered \PYs.},
 but since the leading IR behaviour
of any PS simulator is the same as the one of $\mathcal R$ (or equivalently $- \mathcal V$ after integrating over $\text{d}\bar\phi_{n+1}$), the analogue of Eq.~\eqref{eq:nloxsec} for $\NLOQCDPS$ simulation is
 \begin{eqnarray}
     \text{d}\sigma ^{(\clS)} & = & \text{d}\phi_{n+1} \left[ \left( \mathcal B + \mathcal V + \mathcal C^{\rm int} \right) \frac{\text{d}\phi_{n}}{\text{d}\phi_{n+1}}
     + \left( \mathcal C_{\rm MC} - \mathcal C \right) \right]\,, \label{eq:sMCS}\\
     \text{d}\sigma ^{(\clH)} & = & \text{d}\phi_{n+1} \left( \mathcal R - \mathcal C_{\rm MC} \right)\,, \label{eq:sMCH}
 \end{eqnarray}
 where $\text{d}\sigma ^{(\clS)}$ and $\text{d}\sigma ^{(\clH)}$ are the cross sections associated to the  $\clS$ and $\clH$ events, respectively.\footnote{The fact that in both classes of events the integration measure $\text{d}\phi_{n+1}$ appears is due to the fact that they are 
 integrated together; in the case of $\clS$ events, the $n+1$-body phase space is simply projected on the underlying $n$-body one.}
 Unlike fixed-order calculations (see Eq.~(\ref{eq:nloxsec})),  MC counterterms are such that the $\text{d}\sigma ^{(\clS)}$ and $\text{d}\sigma ^{(\clH)}$ subtracted cross sections are separately finite and therefore Born-like ($\clS$) and real-emission ($\clH$) events can be unweighted.

The reweighting prescription of Eqs.~\eqref{eq:rewS} and \eqref{eq:rewH} corresponds to 
\begin{eqnarray}
&\clS:~~ \text{d}\sigma ^{(\clS)}\Longrightarrow & (1+ \dEWSLS) \text{d}\sigma ^{(\clS)}\,, \label{eq:rewStech}\\
&\clH:~~ \text{d}\sigma ^{(\clH)}\Longrightarrow & (1+ \dEWSLH) \text{d}\sigma ^{(\clH)}\,. \label{eq:rewHtech}
\end{eqnarray}

As can been easily seen in Eq.~\eqref{eq:sMCH}, the exact cancellation between the term $\mathcal C $ and $\mathcal C_{\rm int}$ is preserved. The cancellation of the $\mathcal C_{\rm MC}$ dependence between Eqs.~\eqref{eq:sMCS} and \eqref{eq:sMCH} is instead more subtle and relies on the condition \eqref{eq:sclimit}. Before giving details on the implementation of $\dEWSLH$ and $\dEWSLS$,  and especially the functional form of the mapping between them  for ensuring this condition, we discuss the implications of   Eq.~\eqref{eq:rewStech} and \eqref{eq:rewHtech}.

First of all, since $\ord(\dEWSLS)=\ord(\dEWSLH)=\alpha$, all features of pure-QCD origin are exactly preserved. Considering EW interactions, the term EWSL in Eq.~\eqref{eq:NLOdefEWSL} is given by $\mathcal{B} \dEWSLS$ combining \eqref{eq:sMCS} and \eqref{eq:rewStech}. In fact, what we have denoted as $\NLOQCDEWSLPS$ in Eq.~\eqref{eq:NLOdefse} corresponds to generating events by setting $\dEWSLH=0$ and multiplying only the term $\mathcal{B}$ in \eqref{eq:rewStech} by $\dEWSLS$. Similarly, the $\LOQCDEWSLPS$ approximation is obtained by keeping only the term $\mathcal{B} \dEWSLS$. Both approximations can be achieved in a much easier way without reweighting, but rather accounting directly for the effects of EWSL in the event generation. This is precisely how we perform such simulations.

The $\bestpred$ accuracy consists instead of showering events generated at $\bestprednoPS$ accuracy, which in turn consists of the reweighting procedure of Eqs.~\eqref{eq:rewStech} and \eqref{eq:rewHtech}  applied on events generated via the MC@NLO approach (Eqs.~\eqref{eq:sMCS} and \eqref{eq:sMCH}). From the arguments of the previous paragraph it is clear that the $\NLOQCDPS$,  $\LOQCDEWSLPS$ and $\NLOQCDEWSLPS$ accuracies are still valid within the $\bestpred$ one, which therefore is superior. More specifically,
\begin{eqnarray}
\frac{\bestpred}{\LOQCDPS}&=&\frac{\NLOQCDEWSLPS}{\LOQCDPS}+\ord(\as \alpha )\,, \label{eq:comparisonwithequations}\\
\frac{\bestpred}{\LOQCDPS}&=&\frac{\NLOQCDPS}{\LOQCDPS}+\ord(\alpha )\,,\\
\frac{\bestpred}{\LOQCDPS}&=&\frac{\LOQCDEWSLPS}{\LOQCDPS}+\ord(\as )\,,
\end{eqnarray}
where in the previous three equations we have specified only the leading term in the combined $\as$ and $\alpha$ expansion. Thus,
what is left for discussion is the consistency of our approach for higher orders and in particular the combination of NLO QCD corrections, EWSL and PS effects: the terms of $\ord(\as \alpha )$ (and higher) indicated in Eq.~\eqref{eq:comparisonwithequations}.

Implicitly in the $\mathcal C_{\rm MC} $ terms there is a dependence on the shower scale $\muS$. Roughly speaking, emissions of partons at an energy smaller than $\muS$ are dealt by the PS simulator, while the first emission at energy larger than $\muS$ is given by the matrix element $\mathcal{R}$.\footnote{In fact, neither a sharp cut for the energy of the first emission at $\muS$ is present, nor the relevant variable is exactly the energy. However in first approximation this is a correct picture of the underlying mechanism and the details are not relevant for the present discussion.} An $\NLOQCDPS$ simulation preserves  $\NLOQCD$ accuracy matching it to  Leading-Log (LL) accuracy in QCD for soft and collinear emissions. Still, a $\muS$ dependence, beyond the aforementioned accuracy, is left and it can be exploited for estimating higher-order effects. 

Since the $\bestpred$ accuracy is an $\it ad~hoc$ approximation for accounting for the dominant EW corrections together with PS effects, the question ``At what order is the $\muS$-dependence emerging?'' is rather academic. We want to elaborate anyway on that in the following, since it will help to understand the consistency of our approach and the relevance of the $\ref{bullet:1}^{\rm st}$ motivation for the prescription \eqref{eq:sclimit}. Concerning the pure-QCD contributions, it is exactly the same situation of $\NLOQCDPS$: it appears beyond NLO QCD accuracy. Taking into account EW corrections, the $\muS$-dependence can emerge only at one order of $\as$ beyond the EWSL 
in Eq.~\eqref{eq:NLOdefEWSL}.

In addition to this, if $\muS\sim\MW\ll \sqrt{s}$, in the relevant region of the matching where the transition between $\clS$ and $\clH$ events take place, condition \eqref{eq:sclimit} actually takes the form $\dEWSLS=\dEWSLH$. As it will be explained in Sec~\ref{sec:dEWSLtech}, the condition \eqref{eq:sclimit} is ensured if any of the invariants is smaller or equal to $\MW^2$. Therefore no dependence on $\muS$ related to EWSL is present at all.

If instead $\muS\sim\sqrt{s}>\MW$, a dependence on $\muS$ can be present at  one order $\as$ beyond the EWSL in Eq.~\eqref{eq:NLOdefEWSL}. It is important to note that this dependence is often due to the unbalance between the $\dEWSLS$ for a given process of the form \eqref{eq:process} and $\dEWSLH$ of the same form with an additional gluon (either in the initial or final state), which does {\it not} interact electroweakly.
Thus,  these effects are in fact expected to be even smaller than their naive estimate: $\ord(\as) \times \ord(\rm EWSL)$.
\begin{figure}[!t]
\begin{center}
\includegraphics[width=0.62\linewidth]{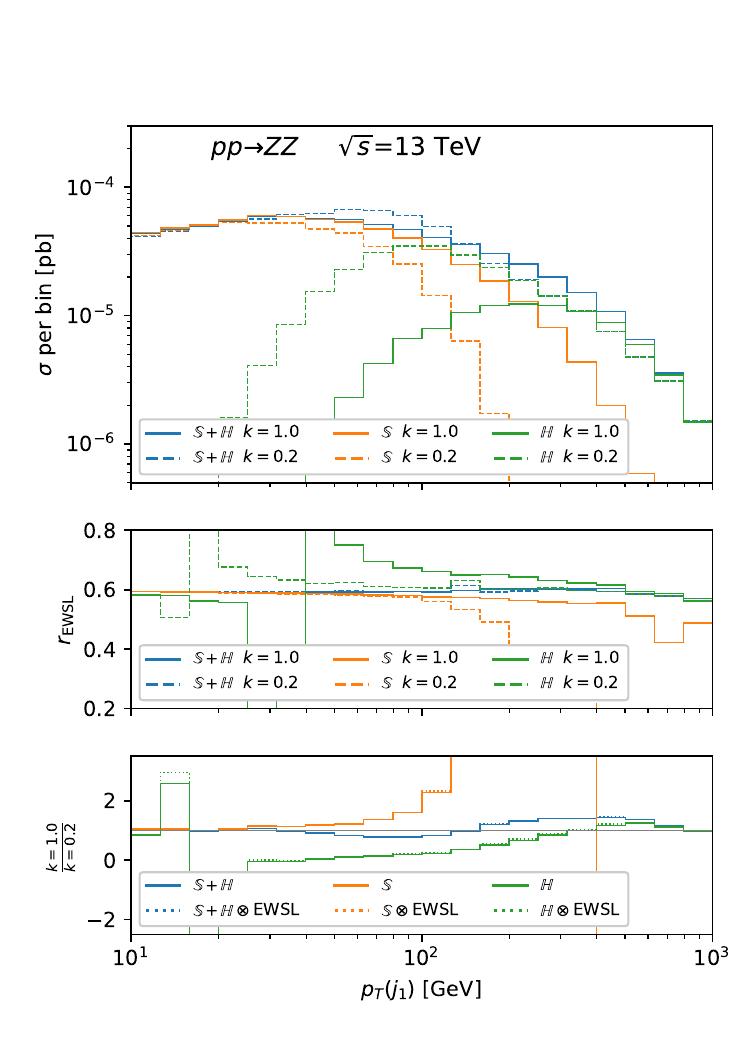}
\end{center}
\caption{Technical test that the relative impact of EWSL on the $\clS+\clH$ samples depends very mildly on the value of $\mu_S$. The representative case of  $ZZ$ hadroproduction with the cuts in \eqref{eq:cutsfig1} is shown.  \label{fig:mustest}}
\end{figure}

We show a concrete example of what we have discussed in this section. In Fig.~\ref{fig:mustest} we consider the case of $ZZ$ hadroproduction where we have set the cuts 
\begin{equation}
p_T(Z)> 600~ {\rm GeV}\,, \qquad m(ZZ)> 1200~ {\rm GeV}\,, \label{eq:cutsfig1}
\end{equation} 
in order to probe the region $\muS\sim\sqrt{s}>\MW$. We show results for two different values of the shower scale $\mu_S$, in particular $\mu_S=k \times H_T/2$  with $k=1$ and $k=0.2$. The case $k=1$ is the standard value\footnote{For more details on the shower scale settings see Refs.~\cite{Alwall:2014hca, Bagnaschi:2018dnh}.}, while $k=0.2$ is an {\it ad hoc} value which has been chosen just for our purpose.

 In the main panel of  Fig.~\ref{fig:mustest} we show the $\NLOQCDPS$ predictions for the two different values of $k$, denoted in the plot as  $\clS+\clH$ (blue) and separately the results for the $\clS$ (orange) and  $\clH$ (green) events alone. The case $k=1$ corresponds to the solid lines, the case $k=0.2$ to the dashed ones. It is important to notice that different values of $k$ return very different contributions from the $\clS$  events and the $\clH$ ones and a non-negligible dependence on $k$, as expected, is left also in the total prediction $\clS+\clH$. In the first inset we show the ratio
\begin{equation}
r_{\rm EWSL}\equiv\frac{\bestpred}{\NLOQCDPS}\, ,
\end{equation}
for the three different sets of events and two different shower scales. It is manifest how the impact of the EWSL is very different for the $\clS$, or $\clH$, events alone when $k=1$ or $k=0.2$, while in the case of the full set of events $\clS+\clH$ there is almost no dependence on the shower-scale choice. This supports our previous argument regarding the fact that although a dependence on $\muS$ of  $\ord(\as) \times \ord(\rm EWSL)$ can be present, it is actually  expected to be even smaller.

In the last inset we show the ratio between the predictions for $k=1$ and $k=0.2$ for the three different sets of events. The case of the $\bestpred$ predictions correspond to the dotted lines while the $\NLOQCDPS$ case to the solid ones. First, we can see that for each set of events (also for the set $\clS+\clH$) there is a visible difference between the case  $k=1$ and $k=0.2$. Second, one can  notice how the ratio is unaffected by the presence of the EWSL contribution.

\subsection{Implementation of $\dEWSLS$ and $\dEWSLH$}
\label{sec:dEWSLtech}

In the following we describe in detail how the $\dEWSLS$ and $\dEWSLH$ functions are implemented, starting with the case of $\dEWSLS$.

In this work, we employ the following approach:
\begin{equation}
\dEWSLS=\deltaEW\Big|_{\rm SDK_{\weak}}(e_\clS)\, \label{eq:dEWSLStech},
\end{equation}
where $\deltaEW$ is the quantity defined in Eq.~\eqref{eq:deltaEW} 
 and we have specified that it is evaluated for an $\clS$ event, denoted as $e_\clS$, which is associated to a process of the form 
\beq \label{eq:processS}
e_\clS:~~ \varphi_{i_1}(p_1)\dots \varphi_{i_{\bar n}}(p_{\bar n})\rightarrow 0\,.
\eeq
In Eq.~\eqref{eq:processS}  we have used the same notation of Eq.~\eqref{eq:process} and understood that for a $2\TO n$ process $\bar n=n+2$.

Equation \eqref{eq:dEWSLStech} is actually refining the definition that was given in Eq.~\eqref{eq:dEWSLS}. The EWSL are calculated in the  $\rm SDK_{\weak}$ scheme, as described in Appendix \ref{sec:EWSL}. This scheme was conceived in Ref.~\cite{Pagani:2021vyk} in order to reproduce as close as possible NLO EW corrections. Here, the final goal is the same, but the $\rm SDK_{\weak}$ is actually employed in order to not double-count QED effects from $\rm PS$ simulations. Equation \eqref{eq:dEWSLStech} implies also that we assume  $\deltaQCD=0$. This assumption has clearly no effect for all the processes for which the $\LO_2$ is zero or anyway smaller than $\alpha/\as\sim 0.1$, {\it i.e.}, the naive expectation for $\ord(\LO_2/\LO_1)$. However it is also a reasonable assumption for a much larger class of processes.
Indeed, even if $\ord(\LO_2/\LO_1)\sim \alpha/\as$, according to Eq.~\eqref{eq:dQCDfinal} and the related discussion, at least one  of the following conditions must be satisfied for $\deltaQCD$ to be in practice relevant:
\begin{itemize} 
\item  $\Sigma^{}_{\LO_2}$ has a sizeable dependence on $m_t$, {\it e.g.}, due to the Yukawa interaction of the top quark, and therefore there is a dependence on the parameter renormalisation of $m_t$ in QCD, $\dmtQCD$. 
\item  The $\LO_2$ involves matrix elements for partonic processes with external gluons.
\item  $\Sigma^{}_{\LO_2}$ depends on $\as$.
\item  $s\gg\mu_R^2$ and $n-1\ne n_g$. 
\end{itemize}

As examples, any purely EW process such as multi-boson production is free of these issues since $\LO_2$ is not present in those cases. The processes involving top quarks in the final state are also typically exhibiting small contributions from the perturbative order $\LO_2$.\footnote{An important exception is four-top production, but in that case not only $\LO_2$ but also $\LO_3$ should be taken into account for sensible results \cite{Frederix:2017wme}. } On the other hand, we reckon that this approximation may miss non-negligible contributions for (multi-)boson production in association with more than one jet, for instance $Z+3j$ studied in Ref.~\cite{Bothmann:2020sxm}, since $\LO_2$ contribution is not negligible in the tails of the distributions.

We discuss now the case of the $\clH$ events, denoted in the following as $e_\clH$. As we mentioned multiple times we wish to ensure that condition \eqref{eq:sclimit} is valid if at least one of the $r_{kl}$ invariants, defined as $r_{kl} \equiv (p_k + p_l)^2$ (see also Eq.~\eqref{eq:Sudaklim}), is such that $|r_{kl}|<\MW^2$.
Actually, since this is a prescription for matching $\dEWSLS$ and $\dEWSLH$ preserving the EWSL accuracy in both the $n$ and $n+1$ final states, a more general condition 
\begin{equation}
|r_{kl}|<c_{\clH\TO\clS}~\MW^2\, , \label{eq:rklltmw}
\end{equation}
is preferable, where $c_{\clH\TO\clS}$  should be chosen of $\ord(1)$ and can be varied around the default value $c_{\clH\TO\clS}=1$ in order to test the dependence on it.

Analogously to Eq.~\eqref{eq:processS}, an $e_\clH$ event is associated to a process of the form
\beq \label{eq:processH}
e_\clH:~~ \varphi_{i_1}(p_1)\dots \varphi_{i_{\bar n+1}}(p_{\bar n+1})\rightarrow 0\,,
\eeq
where we understood again that for a $2\TO n$ process $\bar n=n+2$.
In the FKS language, $e_\clH$ is one of the $r\in\allprocnpo$ partonic processes with $n+1$ particles in the final state.\footnote{In our notation, $r$ denotes already the invariant, so we used $e_\clH$ in the place of it.} If one considers all possible $j\to kl$
branchings ($g\to gg$, $g\to q\bq$, and $q\to qg$, but also $Q\to Qg$, with 
$Q$ being a quark with non-zero mass) a list of new processes with 
$n$ particles in the final state is obtained by removing any possible $(k,l)$ pair and substituting it with $j$ in its place.
In doing so, one also obtains for a given $e_\clH$
the $(k,l)$ pairs that are associated with a soft and/or a collinear
singularity, the set of FKS pairs $\FKSpairs(e_\clH)$ \cite{Frederix:2009yq}, and at the same time the $\clS$ events $e_\clS^{(k,l)}$\ associated to processes of the form:
\begin{eqnarray} \label{eq:processesS}
\begin{split}
e_\clS^{(k,l)}:~~& \varphi_{i'_1}(p'_1)\dots\varphi_{i'_{\bar n}}(p'_{\bar n}) \rightarrow 0 \equiv  \\ & \varphi_{i_1}(p'_1)\remove{\varphi}{0.22}_{\,i_k} \dots \remove{\varphi}{0.22}_{\,i_l}\dots \varphi_{i_{\bar n+1}}(p'_{\bar n+1})\varphi_{i_j}(p'_j) \rightarrow 0\,,
\end{split}
\end{eqnarray}
where we remove the $\varphi_{\,i_k}$ and $\varphi_{\,i_l}$ particles and we add the particle $\varphi_{\,i_j}$ in the position $\bar n$. In the case of hadronic collisions, for a given production process the partonic process \eqref{eq:processH} that can be associated to an event $e_\clH$ is not unique. Thus, for each event  $e_\clH$, the FKS pairs $\FKSpairs(e_\clH)$ and the processes associated to the $e_\clS^{(k,l)}$ events can be different.

It is important to note that the mapping from the $\bar n +1$ momenta $\{p\}$ of $e_\clH$ to the $\bar n$ momenta $\{p'\}$ of $e_\clS^{(k,l)}$ is not uniquely defined. We have used for this purpose techniques for the momentum reshuffling analogous to those of Ref.~\cite{Frixione:2019fxg}.\footnote{These techniques have been originally conceived for the removal   
  (diagram-removal) or subtraction (diagram-subtraction) or resonances, see {\it e.g.} Refs.~\cite{Beenakker:1996ch, Frixione:2008yi, Hollik:2012rc, Demartin:2016axk}.} In particular, given an FKS pair $(k,l)$, 
one particle (which we associate to the index $l$) always belongs to the final state, 
while the other one ($k$) can be either initial or final. 
  \begin{itemize}
      \item If $k$ is a final-state particle, the pair $k,l$ is first replaced by its mother particle $j$
          with $p_j=p_k+p_l$. Then, the energy component of $p_j$ is changed in order to fulfil the
          mass-shell condition. Finally, the initial-state momenta are changed so that the new
          set of momenta satisfies momentum conservation.
      \item If $k$ belongs to the initial state, then $l$ is removed from the event. The remaining
          final-state particles are boosted to their total-momentum centre-of-mass frame and, again,
          the initial-state momenta are changed in order to satisfy momentum conservation.
  \end{itemize}
We now specify the quantity $\dEWSLH$. 
Given all the possible $(k,l)$ pairs of external states for an event $e_\clH$ we define
\begin{equation}
r_{\min,\, \rm abs}\equiv \min(|r_{kl}|)=|r_{\hat k \hat l}|\,,
\end{equation}
with $(\hat k, \hat l)$ being the pair returning the smallest value for $|r_{kl}|$. Introducing the quantity
\begin{equation}
C_{\clH\TO\clS}\equiv c_{\clH\TO\clS}~\MW^2,  \label{eq:Chtos}
\end{equation}
we define $\dEWSLH$  as
\begin{equation}
\dEWSLH\equiv \deltaEW\Big|_{\rm SDK_{\weak}}(e_\clH)~\Theta(r_{\min,\, \rm abs}-C_{\clH\TO\clS}) + \deltaEW\Big|_{\rm SDK_{\weak}}(e_\clS^{(\hat k, \hat l)})~\Theta(C_{\clH\TO\clS}-r_{\min,\, \rm abs})\, , \label{eq:dEWSLHtech}
\end{equation}
where
\begin{equation}
e_\clS^{(\hat k, \hat l)}\equiv e_\clH\Big|_{r_{\hat k \hat l} \Longrightarrow  {\rm sign}(r_{\hat k \hat l})\MW^2}~~{\rm if}~~(\hat k, \hat l) \notin \FKSpairs(e_\clH)\,. \label{eq:H2events}
\end{equation}
In a few words, Eq.~\eqref{eq:dEWSLHtech} says that if the smallest invariant is larger in absolute value than the $\MW^2$ scale the Sudakov contribution $\dEWSLH$ is calculated via the $\deltaEW\Big|_{\rm SDK_{\weak}}$ evaluated for the process $e_\clH$ with the $n+1$ kinematic.
 Otherwise, the Sudakov contribution is calculated via the same quantity evaluated instead for the underlying Born configuration, and the associated $n$-body kinematics, that is obtained via the replacement of the FKS pair giving the smallest invariant with its parent particle.
  In the unlikely (but possible) situation that the smallest invariant is given by a pair not corresponding to a QCD branching, Eq.~\eqref{eq:H2events} says that the EWSL are calculated directly for the process $e_\clH$ with the $n+1$ kinematics, but within the {\denpoz} algorithm the quantity $r_{\hat k \hat l}$ is replaced by $\MW^2$ times the sign of $r_{\hat k \hat l}$. Events of this kind with $|r_{\hat k \hat l}|\ll \MW^2$ are very unlikely, since they are not associated to any divergence. However, this replacement ensures that events are not reweighted via artificially large Sudakov contributions in a region where the approximation is not supposed to work.

The last point concerning the replacement is actually more general and we implemented it in {\aNLO} as a safety feature as 
\begin{equation}
r_{ k  l} \Longrightarrow  {\rm sign}(r_{ k  l})\MW^2\qquad \forall \,r_{ k  l}\,. \label{eq:safetyfirst}
\end{equation}
 Indeed the EWSL approximation should be used only when invariants are large, but we want to prevent that artificially large correction may arise from simulations performed for processes with $|r_{kl}|<\MW^2$ already at $\LOQCD$ accuracy. The replacement is performed not only for simulations in the $\bestpred$ approximation, but also for the $\NLOQCDEWSL$ one.

In conclusion, we can summarise the description of the $\bestpred$ predictions as follows:
\begin{enumerate}
\item Events are generated at $\NLOQCD$ accuracy via the MC@NLO matching scheme.
\item Events are reweighted~\cite{Mattelaer:2016gcx} via the prescription in Eqs.~\eqref{eq:rewS} and \eqref{eq:rewH} using the $\rm SDK_{\rm weak}$ scheme and neglecting the second term in the r.h.s.~of Eq.~\eqref{eq:QCDEWcomb}. This leads to $\bestprednoPS$ accuracy in the MC@NLO matching scheme.
\item  Events are showered via a parton shower including QED effects (possibly after heavy particles are decayed using external tools).
\end{enumerate}

\section{Numerical results \label{sec:results}}

In this section we present numerical results for phenomenologically relevant physical distributions from two different production processes at hadron colliders: the  top-quark pair and Higgs boson associated production ($t \bar t H$), and the associated production of three $Z$ gauge bosons ($ZZZ$). Here
we focus on the presentation of the EWSL-based predictions, and we do not perform any comparison with the exact NLO EW corrections. Still, we remind the reader that EWSL are an approximation for the NLO EW corrections. Hence, the quality of such an approximation should always be checked, at the differential level, before relying on it for phenomenological predictions.  \\
For both processes considered here we show inclusive results (without any cut applied) as well as applying the following cuts:
\begin{equation}  
p_{T}(X)>400~{\rm GeV}\,, \qquad \Delta R(X,Y) > 0.5\,, \label{eq:cuts}
\end{equation}
where $X,Y$ is any of the particle in the final state at the Born level, $p_{T}$ is the transverse momentum and $\Delta R \equiv \sqrt{(\Delta \phi)^2+(\Delta \eta)^2}$ with $\Delta \phi$ being the azimuthal angle between $X$ and $Y$ and $\Delta \eta$ is the difference between their pseudorapidities.

The results have been obtained by generating events with \aNLO{} and using \PYe{} as parton shower. Hadronisation is disabled in the parton shower. Input parameters
are defined in the $G_\mu$ scheme for what concerns EW renormalisation: 
\begin{equation}
M_Z = 91.188 \text{ GeV}, \quad M_W = 80.419 \text{ GeV}, \quad G_{\mu} = 1.16639 \times 10^{-5} \text{ GeV}^{-2},
\end{equation}
and the top quark and Higgs boson masses are set to
\begin{equation}
M_H = 125 \text{ GeV}, \quad m_t = 173.3 \text{ GeV} .   \\
\end{equation}
We employed the {\sc\small NNPDF4.0} parton-distribution-functions~\cite{NNPDF:2021njg}, with NNLO evolution and $\as(M_Z)=0.118$. The renormalisation and factorisation scales have been set equal to $H_T/2$, where $H_T$ is the scalar sum or the transverse energies of all the particles in the final state, before showering the event. For what concerns the
shower starting scale, we use the default setting in \aNLO{}, which, for processes with massive or colourless particles in the final state, is a value proportional to $H_T$ \cite{Alwall:2014hca, Bagnaschi:2018dnh}.  Jets are clustered via the anti-$k_T$ algorithm~\cite{Cacciari:2008gp} as implemented in {\sc \small FastJet}~\cite{Cacciari:2011ma},   with $R=0.4$ and
 are required to have $p_{T,\rm{min}}=10~{\rm GeV}$.

We remind the reader that concerning $t \bar t H$ and $ZZZ$ production several SM calculations including higher-order effects have already been performed. The literature is vast, both for the former \cite{Ng:1983jm,Kunszt:1984ri, Beenakker:2001rj,Beenakker:2002nc, Reina:2001sf,Reina:2001bc,Dawson:2002tg,Dawson:2003zu, Alwall:2014hca, Frixione:2014qaa,Yu:2014cka,Frixione:2015zaa, Frederix:2018nkq, Kulesza:2015vda,Broggio:2015lya,Broggio:2016lfj,Kulesza:2017ukk,Broggio:2019ewu, Ju:2019lwp, Kulesza:2020nfh, Denner:2015yca, Denner:2016wet, Catani:2022mfv} and the latter \cite{Lazopoulos:2007ix, Binoth:2008kt,  Alwall:2014hca, Wang:2016fvj,Frederix:2018nkq}.

\subsection{$t \bar tH$}
\label{sec:ttH}
\begin{figure}[!t]
\begin{center}
\includegraphics[width=0.495\linewidth]{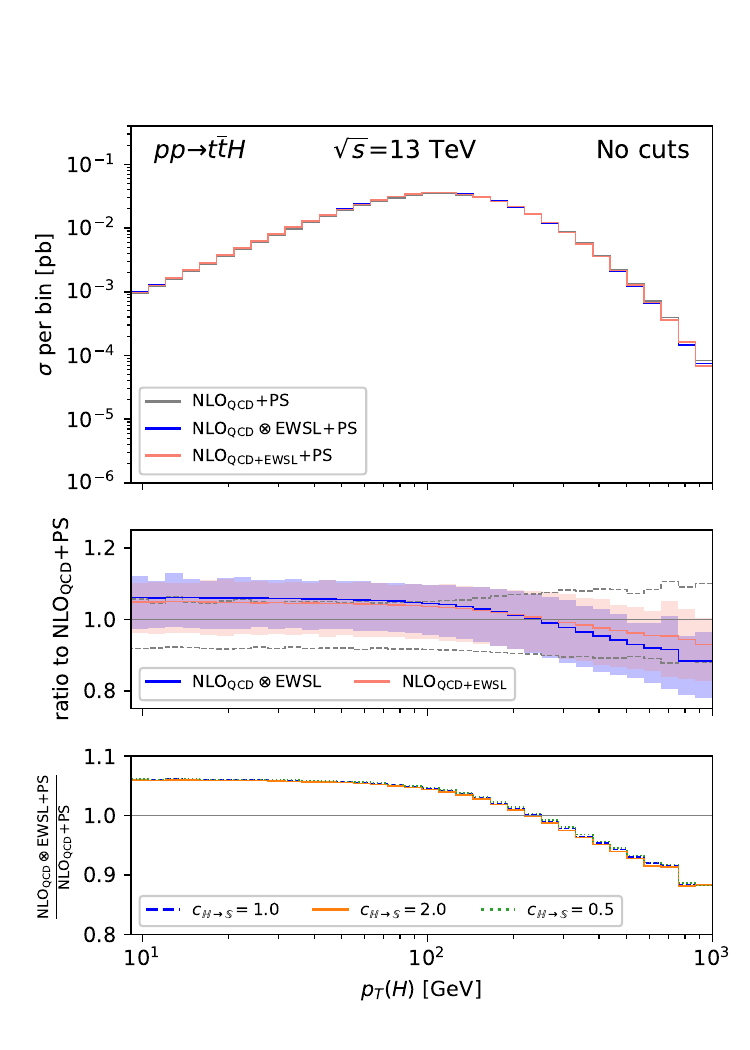}
\includegraphics[width=0.495\linewidth]{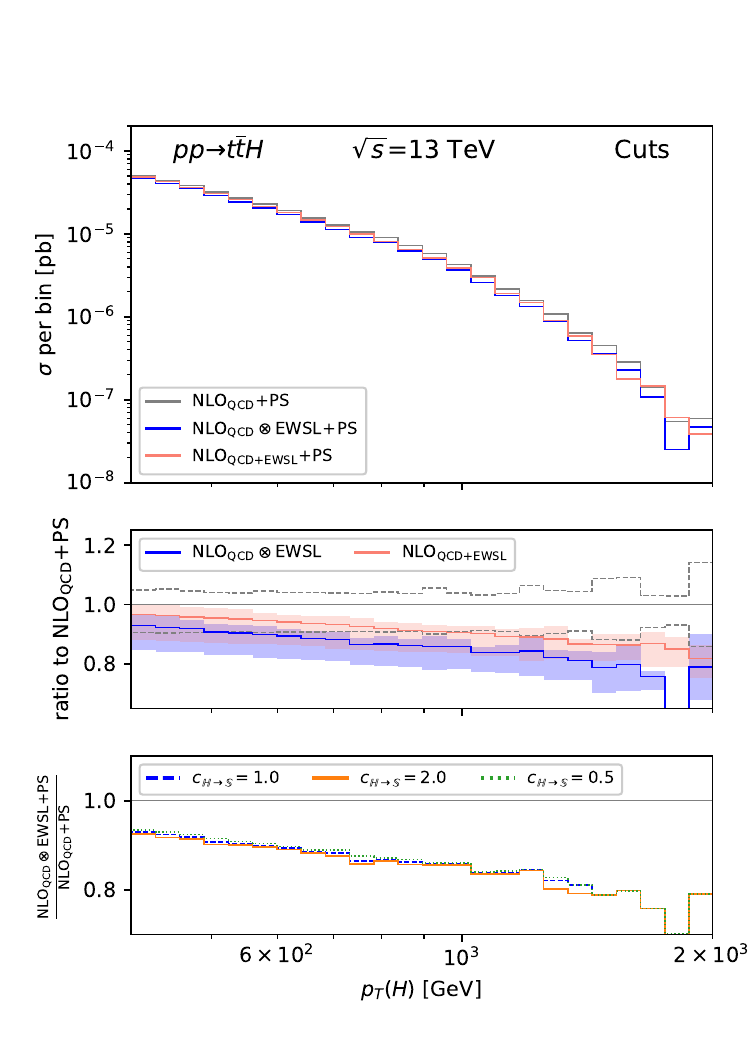}
\end{center}
\caption{Differential distributions for $p_T(H)$ in $t \bar t H$ production at 13 TeV. Left: no cuts applied. Right:   cuts as defined in \eqref{eq:cuts} applied.  
\label{fig:ttH-ptH}}
\end{figure}

\begin{figure}[!t]
\begin{center}
\includegraphics[width=0.495\linewidth]{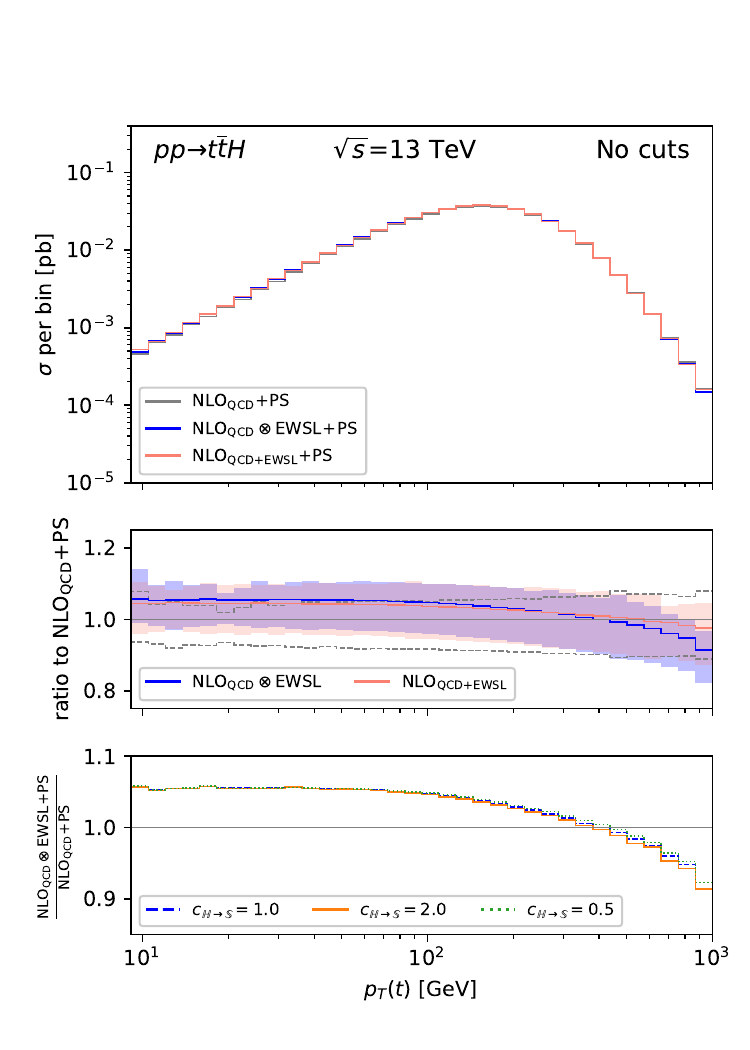}
\includegraphics[width=0.495\linewidth]{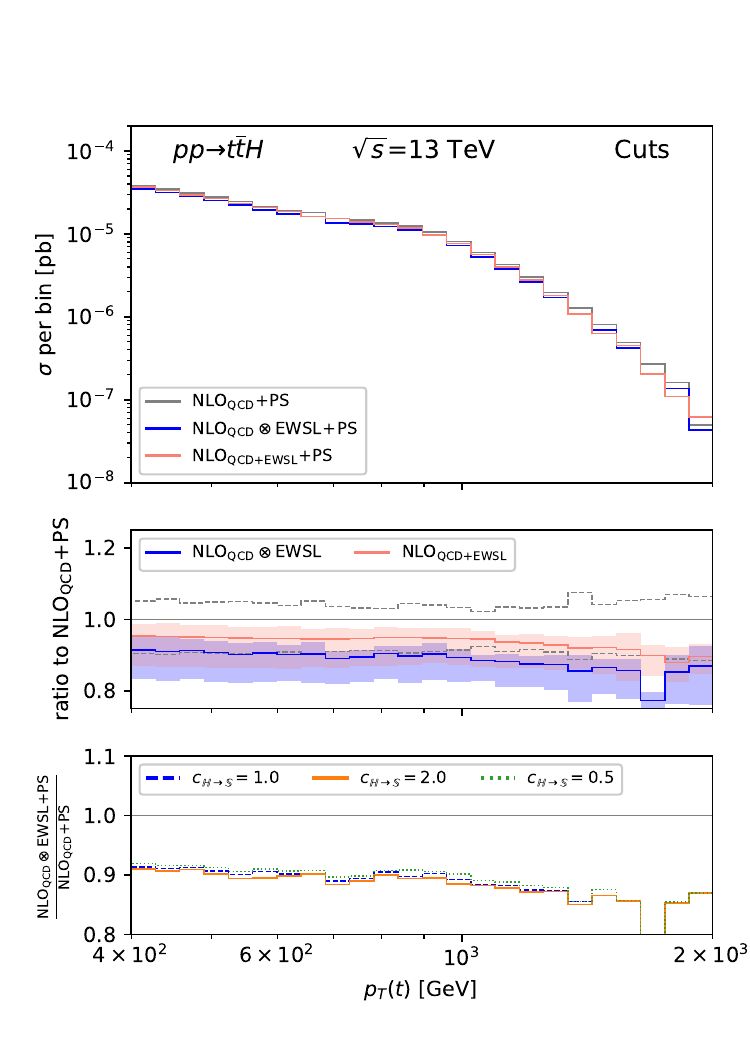}
\end{center}
\caption{Differential distributions for $p_T(t)$ in $t \bar t H$ production at 13 TeV. Left: no cuts applied. Right:   cuts  as defined in \eqref{eq:cuts} applied.   \label{fig:ttH-ptt}}
\end{figure}

\begin{figure}[!t]
\begin{center}
\includegraphics[width=0.495\linewidth]{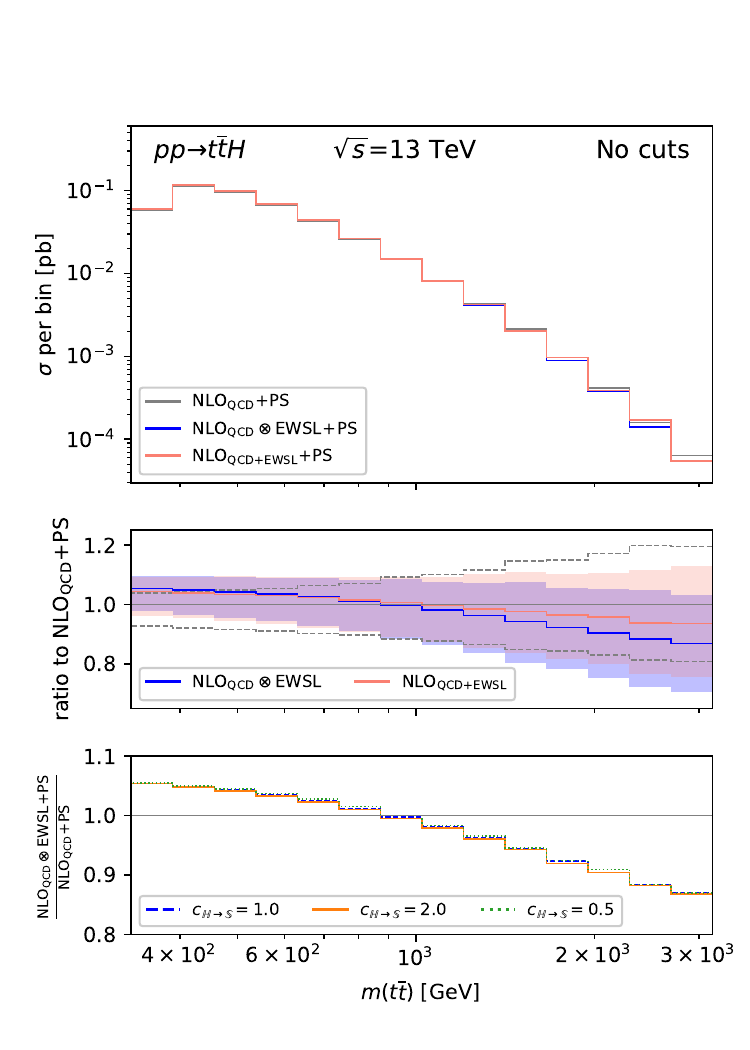}
\includegraphics[width=0.495\linewidth]{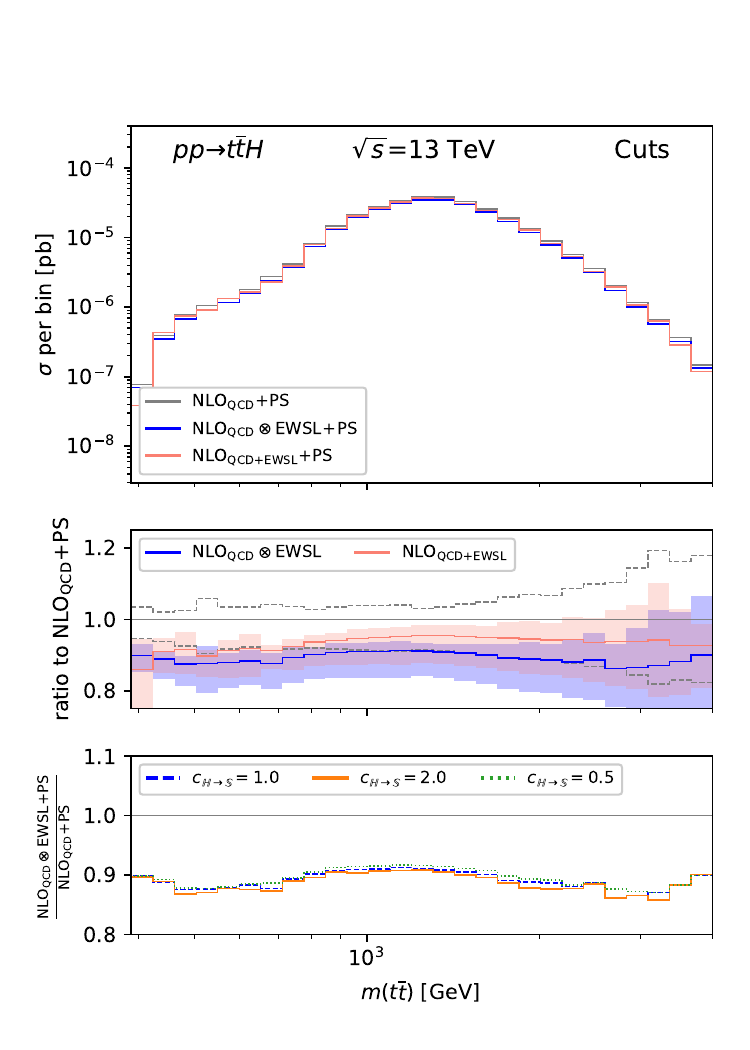}
\end{center}
\caption{Differential distributions for $m(t \bar t)$ in $t \bar t H$ production at 13 TeV. Left: no cuts applied. Right:   cuts  as defined in \eqref{eq:cuts} applied.    \label{fig:ttH-mtt}}
\end{figure}

\begin{figure}[!t]
\begin{center}
\includegraphics[width=0.495\linewidth]{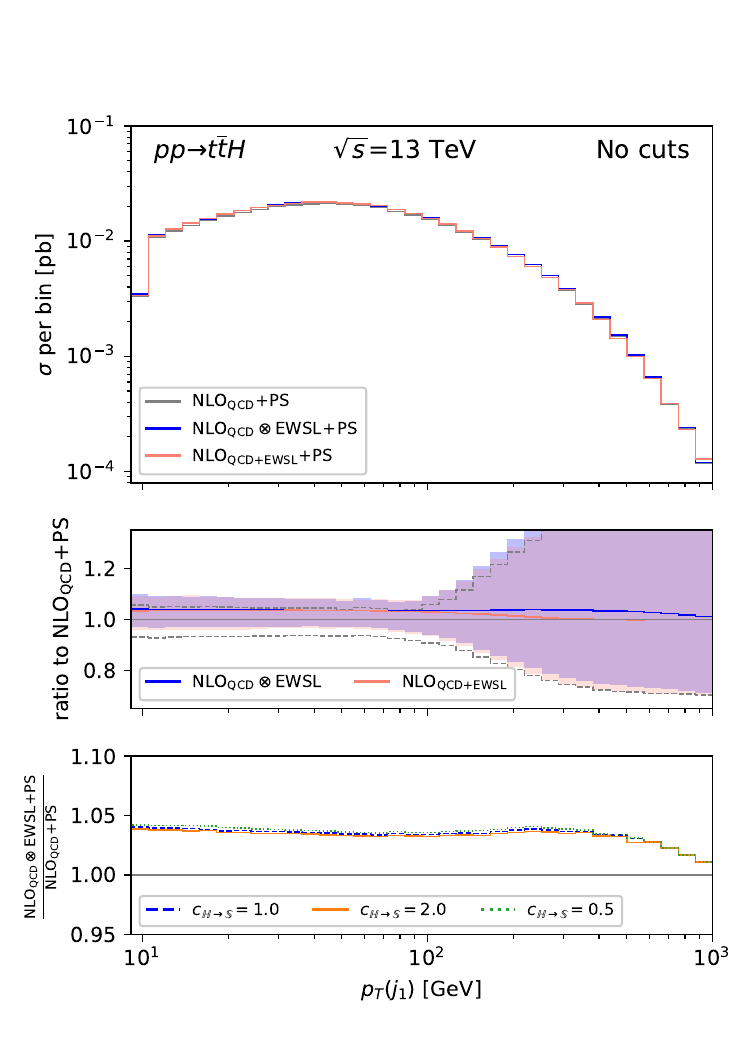}
\includegraphics[width=0.495\linewidth]{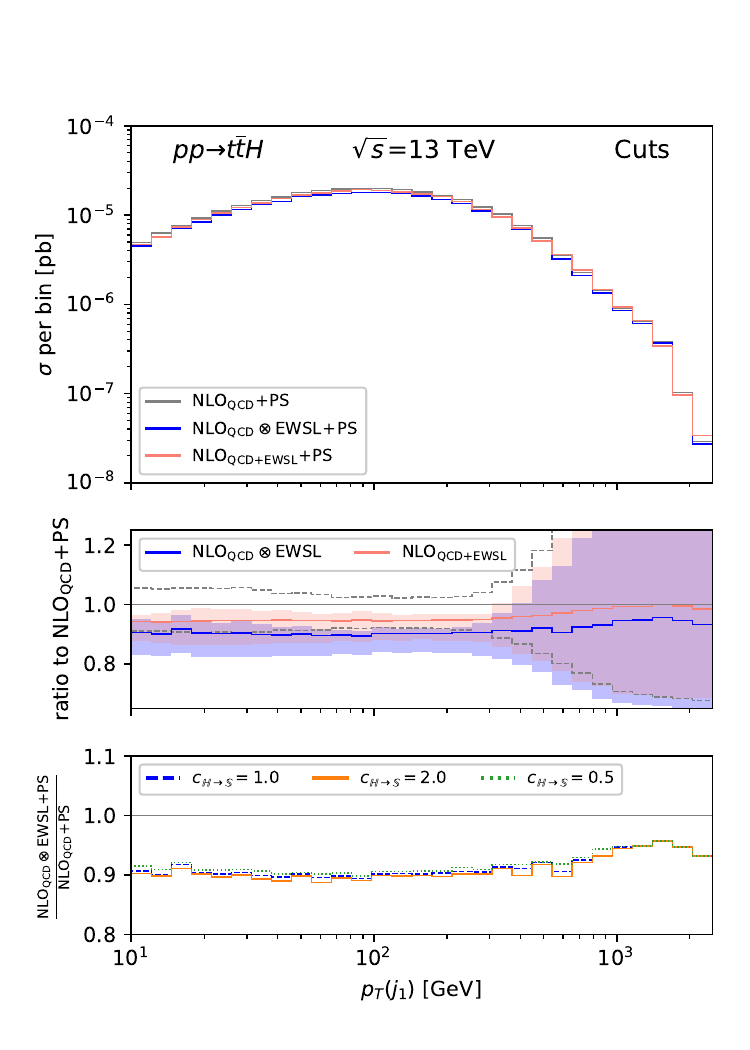}
\end{center}
\caption{Differential distributions for $p_T(j_1)$ in $t \bar t H$ production at 13 TeV. Left: no cuts applied. Right:   cuts  as defined in \eqref{eq:cuts} applied.  \label{fig:ttH-ptj1}}
\end{figure}

We discuss results for the following differential distributions from $t \bar t H$ production in proton--proton collisions at 13 TeV: $p_T(H)$ in Fig.~\ref{fig:ttH-ptH}, $p_T(t)$ in Fig.~\ref{fig:ttH-ptt}, the invariant mass of the top-quark pair $m(t \bar t)$ in Fig.~\ref{fig:ttH-mtt}, and $p_T(j_1)$ in Fig.~\ref{fig:ttH-ptj1}, where $j_1$ is the hardest jet. In each of the Figs.~\ref{fig:ttH-ptH}--\ref{fig:ttH-ptj1} we show inclusive results without cuts in the left plot and those with cuts \eqref{eq:cuts} in the right one.
The layout and the rationale of each plot is the following.
In the main panel we show the central values for predictions with the three different accuracies:
\begin{itemize}
\item $\NLOQCDPS$ (grey),
\item $\bestpred$ (blue),
\item $\NLOQCDEWSLPS$ (red).
\end{itemize}
In the first inset we display the scale uncertainty band for the same three predictions normalised to the central value of $\NLOQCDPS$, where the uncertainty has been evaluated by independently varying the renormalisation and factorisation scale by a factor of two up and down (the usual 9-point scale variation). The band for $\NLOQCDPS$ is obviously centred around one and we show only the upper and lower bounds as dashed grey lines. In the last inset we show the ratio of the $\bestpred$ and $\NLOQCDPS$ predictions for different values of the parameter $c_{\clH\TO\clS}$, introduced in Eq.~\eqref{eq:rklltmw} and entering Eq.~\eqref{eq:dEWSLHtech}. We remind the reader that $c_{\clH\TO\clS}$ parametrises how the condition $\eqref{eq:sclimit}$ is implemented,\footnote{Roughly speaking this means $\dEWSLH=\dEWSLS$ if an invariant connected to a soft/collinear limit is smaller in absolute value than $c_{\clH\TO\clS} \MW^2$.} as can be seen from the aforementioned equations. The default case $c_{\clH\TO\clS}=1$ corresponds to the ratio of the blue and grey lines in the main panel. 

Before commenting the individual distributions we give some general considerations. We first note that the plots without cuts display both distributions for $p_T$'s well below the $\MW$ scale (Figs.~\ref{fig:ttH-ptH}--\ref{fig:ttH-ptj1}) and the $m(t \bar t)$ distribution at the threshold (Fig.~\ref{fig:ttH-mtt}). It is well known that the Sudakov approximation is not expected to hold for these cases and also that additional effects such as Sommerfeld enhancements can be present \cite{Degrassi:2016wml}. However, as already mentioned before, the focus of this paper is not to present phenomenological predictions for $t \bar t H$ (or $ZZZ$ in Sec.~\ref{sec:ZZZ}), but rather to document the features and the technical implementation of the $\bestpred$ approximation in {\aNLO}. We leave discussions and comparisons with NLO EW corrections and/or data  for future detailed studies. We here simply reckon that owing to the reweighting procedure, further classes of EW effects can be naturally incorporated via the redefinitions of the quantities $\dEWSLS$ and $\dEWSLH$ in Eqs.~\eqref{eq:dEWSLStech} and  \eqref{eq:dEWSLHtech}, respectively.

The plots without cuts show interesting features. For small $p_T$ both $\NLOQCDEWSLPS$ and $\bestpred$ corrections to $\NLOQCDPS$ are flat and non-vanishing and the two predictions are almost equal for the entire spectrum. The non-vanishing (still at the level of a very few percent) and flat effect at small $p_T$ is due to the fact that all Born invariants are at least $\sim m_t$ in absolute value and $s\geq(2m_{t}+m_{H})^{2}>M_{W}^{2}$. For this reason, not only  at the threshold are the EWSL non-vanishing: in this range they are positive, and are therefore clearly dominated by the single logarithms. Since (logarithms of the) invariants have very small variations for $p_{T}(t)$ ranging from 0 to $\sim100$ GeV, these corrections are quite flat.  The fact that $\bestpred$ and $\NLOQCDEWSLPS$ are very close is due to the fact that EWSL are relatively small and the QCD  $K$-factor, $\NLOQCDPS/\LOQCDPS$ is very close to one and flat. Indeed, in first approximation, one expects that 
\begin{equation}
\bestpred \simeq \left( \NLOQCDEWSLPS \right) + \EWSL \times\left(\frac{\NLOQCDPS}{\LOQCDPS}-1\right)\,, \label{eq:mult}
\end{equation} 
in other words,  EW and QCD corrections combined in the multiplicative approach.%
\footnote{We explicitly checked the validity of Eq.~\eqref{eq:mult}. 
In particular, in distributions or kinematical regions where QCD corrections are {\it not} dominated by hard emissions, such as {\it e.g.}~in the case of the $p_T(t)$ spectrum, the difference between the r.h.s.~and l.h.s.~of Eq.~\eqref{eq:mult} amounts to typically 1-5\% of the $\NLOQCDPS$ prediction.}

An exception is the case of $m(t \bar t)$ in Fig.~\ref{fig:ttH-mtt}, where we observe the opposite trend: EWSL are not flat and $\NLOQCDEWSLPS$ and $\bestpred$ predictions are different. We verified that indeed the QCD $K$-factor increases at large values of $m(t \bar t)$.
 Similarly to what has been observed and discussed in Refs.~\cite{Czakon:2017wor, Gutschow:2018tuk, Czakon:2019bcq} for the case of $t \bar t$ production in a similar context, one can notice that the scale uncertainty band is smaller in the case of $\bestpred$ predictions than in the case of $\NLOQCDEWSLPS$. This is not a surprise since in the former setup the EWSL multiply NLO QCD corrections, while in the latter one they multiply only the $\LOQCD$ component, which has a LO dependence on the factorisation and renormalisation scales. This improvement is clearly not present in the $p_T(j_1)$ distribution (Fig.~\ref{fig:ttH-ptj1}), since this distribution is not even present at $\LOQCD$ at fixed order. On the other hand, it is interesting to notice that for large values of   $p_T(j_1)$, where $\NLOQCDPS$ is dominated by hard matrix-element contributions and not PS effects, $\NLOQCDEWSLPS$ converges to exactly $\NLOQCDPS$  at variance with $\bestpred$, which includes EWSL corrections also to the first real emission from hard matrix-element.
\begin{figure}[!t]
\begin{center}
\includegraphics[width= 0.495\linewidth]{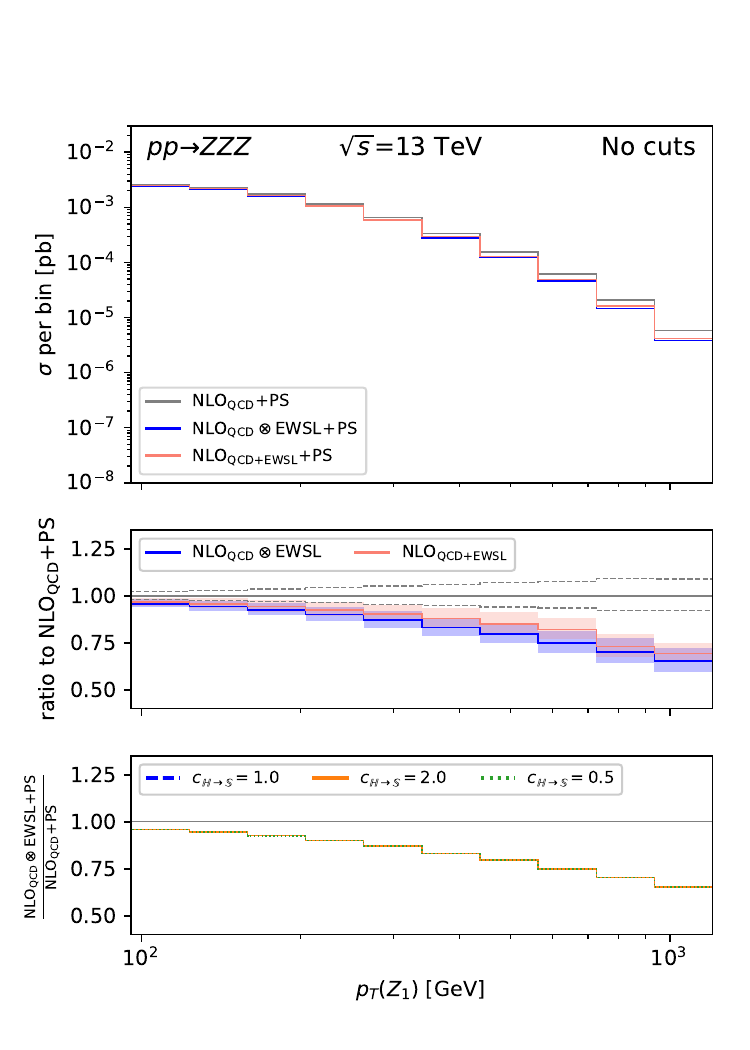}
\includegraphics[width= 0.495\linewidth]{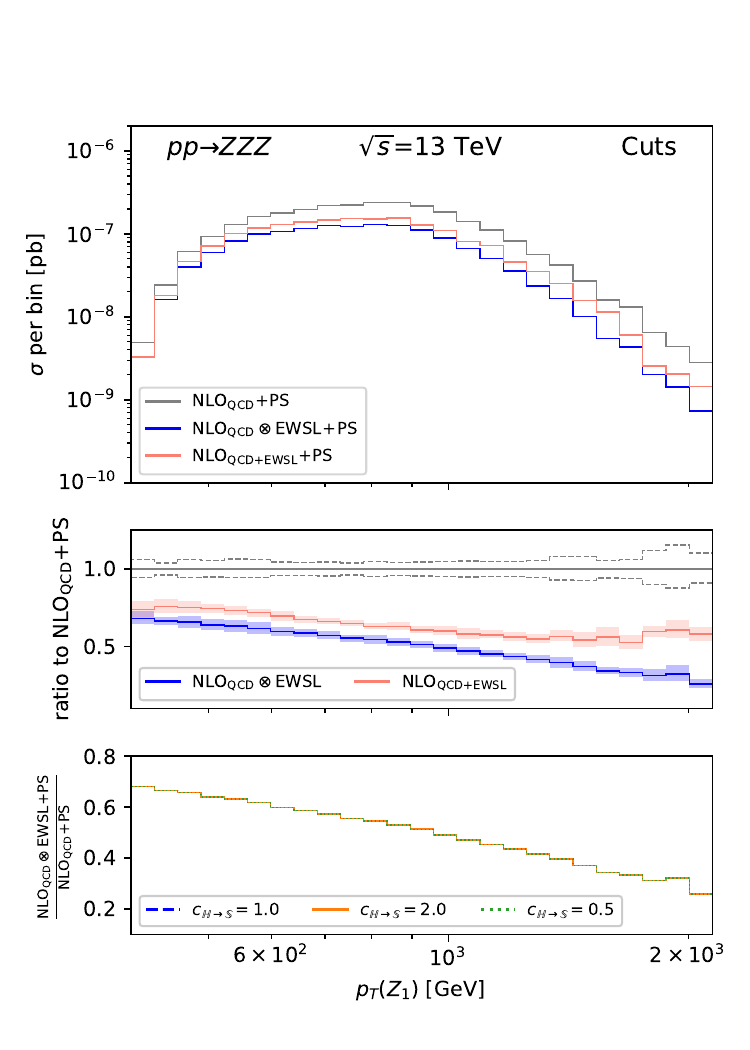}

\end{center}
\caption{Differential distributions for $p_T(Z_1)$ in $ZZZ$ production at 13 TeV. Left: no cuts applied. Right:   cuts as defined in \eqref{eq:cuts} applied.   \label{fig:ZZZ-ptZ1} }
\end{figure}

If the cuts \eqref{eq:cuts} are applied, we can clearly see that the impact of the EWSL increases  and similarly the discrepancy between the $\NLOQCDEWSLPS$ 
and $\bestpred$ predictions increases. In the case of  $p_T(j_1)$ distribution (Fig.~\ref{fig:ttH-ptj1}) we see a flat contribution and a change at very large values for $p_T(j_1)$, where the simulation starts to be dominated by hard matrix-element contributions. One should notice, in particular for this distribution but also for all the remaining ones, that the $\bestpred$ is completely insensitive to the value of $c_{\clH\TO\clS}$ if varied by a factor of two up and down w.r.t.~the reference value $c_{\clH\TO\clS}=1$.

Finally we comment on several checks that we performed and are not directly documented in the plots. The $t\bar t H$ cross section at LO involves contributions not only of order $\as^2 \alpha$ but also of order $\as \alpha^2$ (and $\alpha^3$). Therefore $t\bar t H$ is one of those process that potentially may involve EWSL contributions from the quantity $\deltaQCD$ (see Eqs.~\eqref{eq:orders}--\eqref{eq:dQCDfinal}) that we do not include (see Eq.~\eqref{eq:dEWSLStech}). We have explicitly verified at fixed order that the impact of this term is at most at the permille level and therefore can be safely ignored. We have also verified the effect of not implementing Eq.~\eqref{eq:safetyfirst}, finding no difference for the distributions, although real emission $\clH$  events with an invariant smaller in absolute value than $\MW^2$ and not associated to any QCD splitting have been identified (see Eq.~\eqref{eq:H2events}). Similarly to what has been observed in Ref.~\cite{Pagani:2021vyk}, the inclusion of the logarithms of the form as in Eq.~\eqref{eq:deltasr} has a non-negligible impact.

\subsection{$ZZZ$ }
\label{sec:ZZZ}

\begin{figure}[!t]
\begin{center}

\includegraphics[width= 0.495\linewidth]{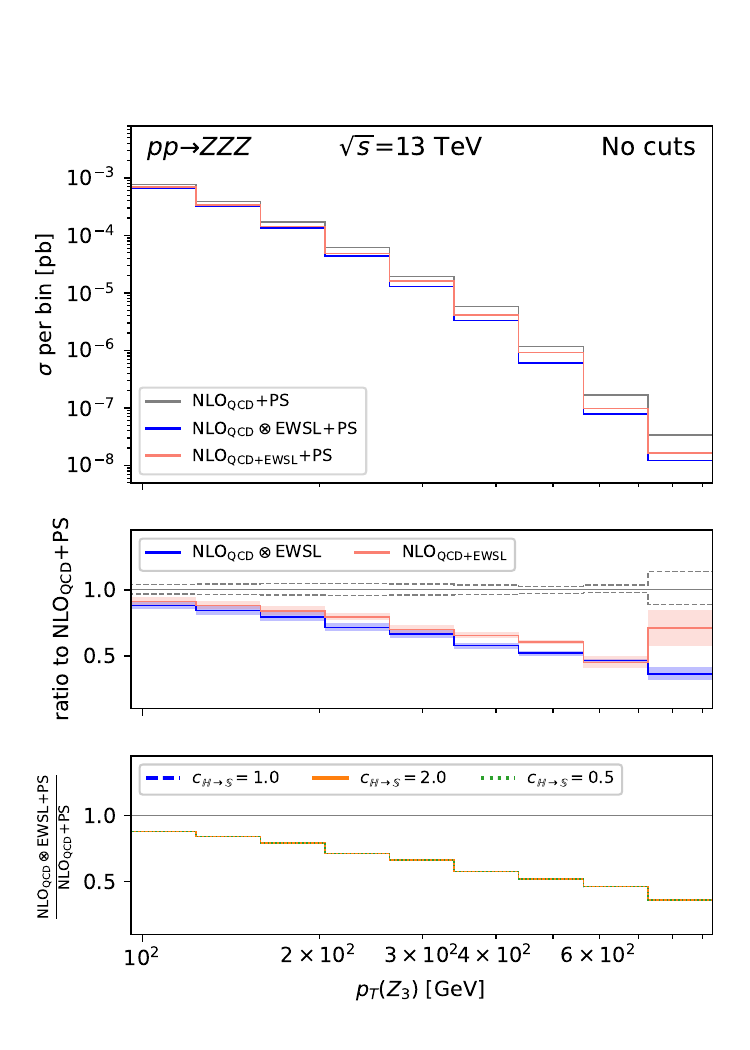}
\includegraphics[width= 0.495\linewidth]{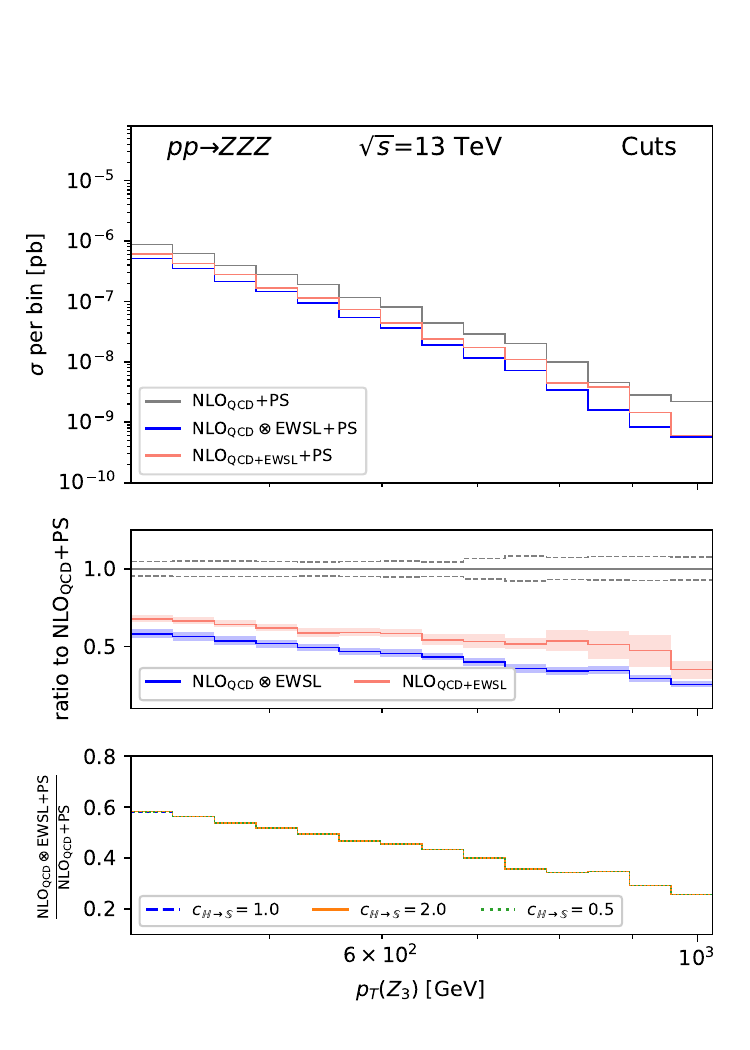}

\end{center}
\caption{Differential distributions for $p_T(Z_3)$ in $ZZZ$ production at 13 TeV. Left: no cuts applied. Right:   cuts as defined in \eqref{eq:cuts} applied.    \label{fig:ZZZ-ptZ3}}
\end{figure}

\begin{figure}[!t]
\begin{center}
\includegraphics[width= 0.495\linewidth]{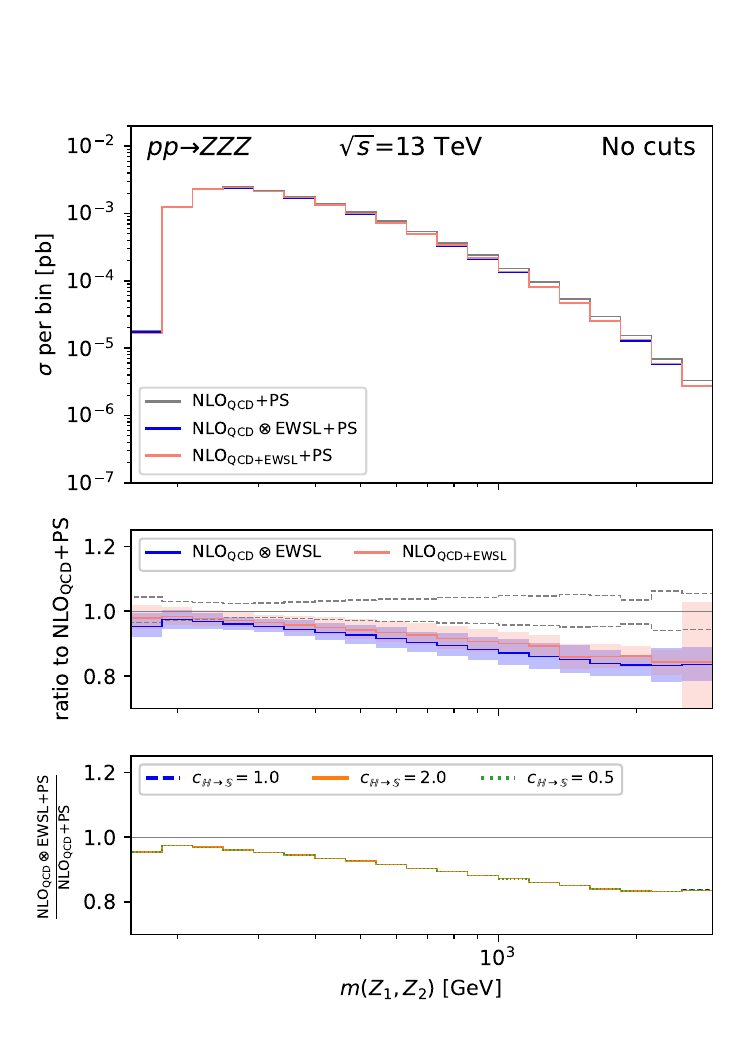}
\includegraphics[width= 0.495\linewidth]{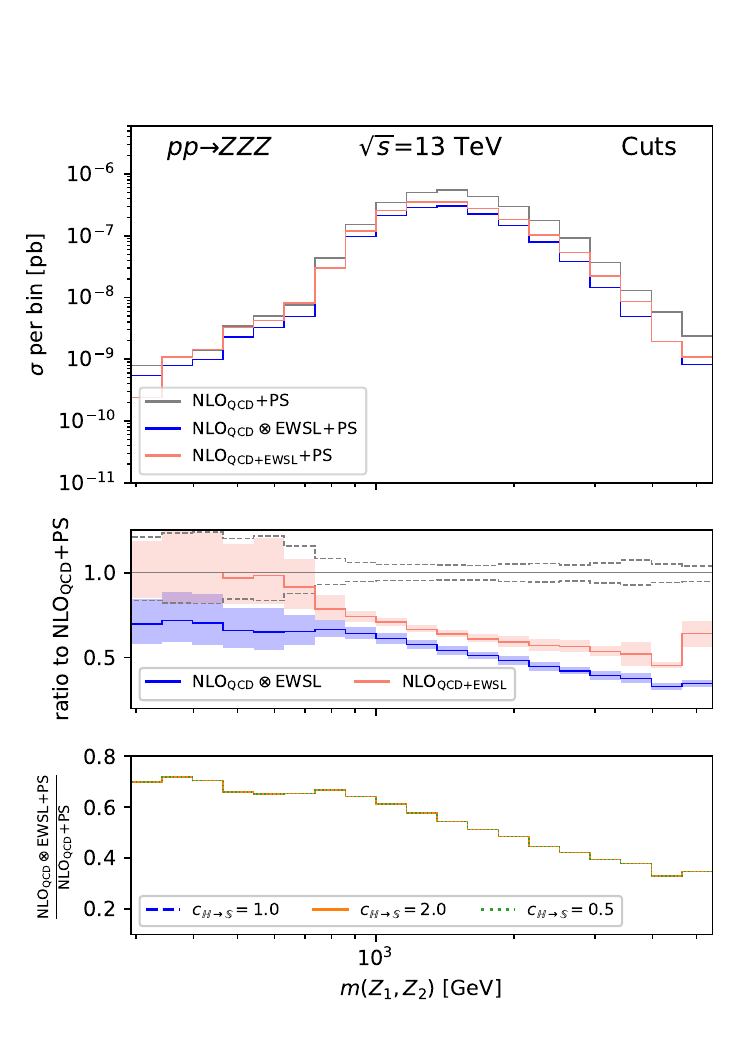}

\end{center}
\caption{Differential distributions for $m(Z_1 Z_2)$ in $ZZZ$ production at 13 TeV. Left: no cuts applied. Right:   cuts as defined in \eqref{eq:cuts} applied.    \label{fig:ZZZ-mz1z2}}
\end{figure}

\begin{figure}[!t]
\begin{center}

\includegraphics[width= 0.495\linewidth]{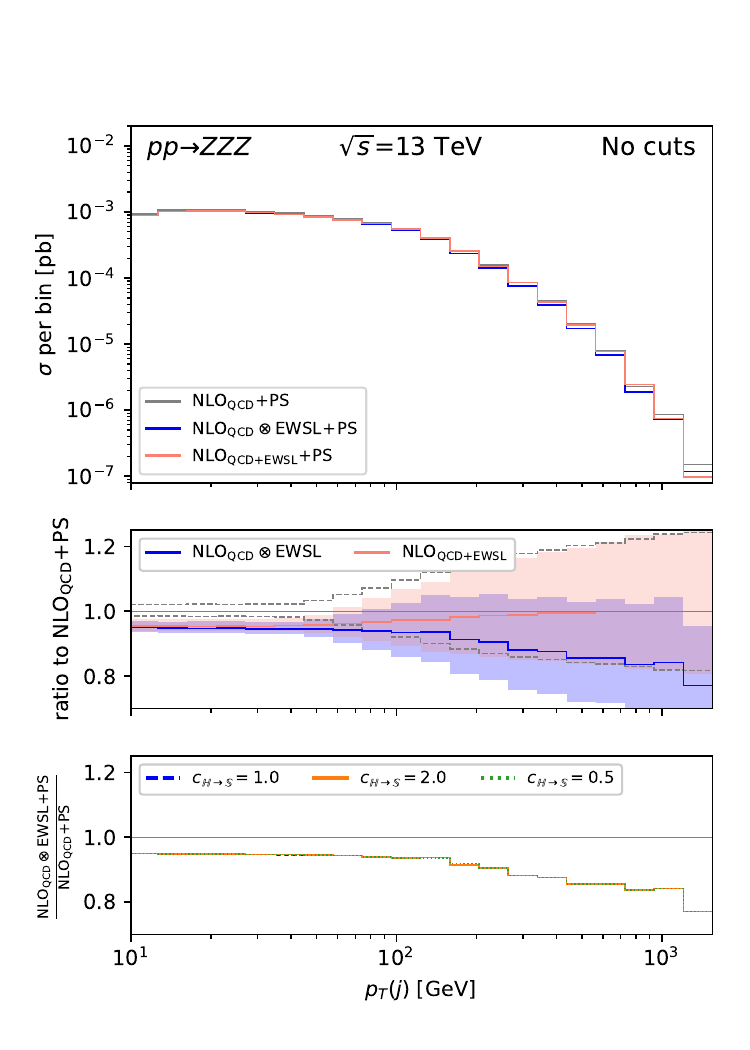}
\includegraphics[width= 0.495\linewidth]{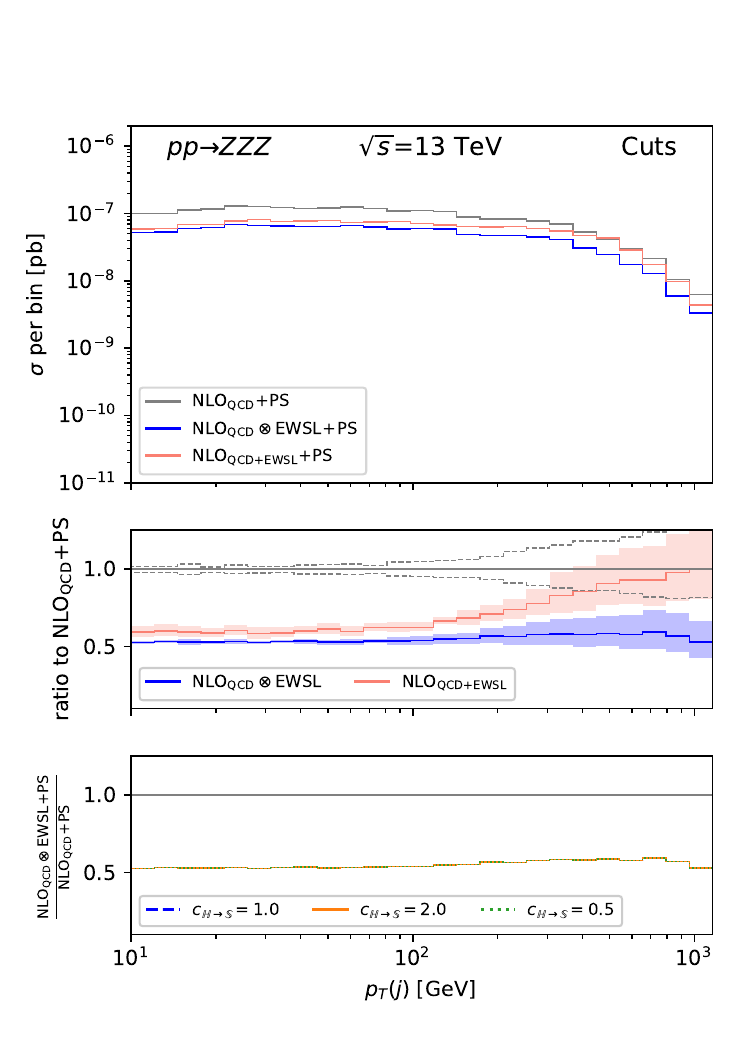}

\end{center}
\caption{Differential distributions for $p_T(j_1)$ in $ZZZ$ production at 13 TeV. Left: no cuts applied. Right:   cuts as defined in \eqref{eq:cuts} applied.    \label{fig:ZZZ-ptj1}}
\end{figure}

In this section, first we discuss results for $ZZZ$ production that are analogous to those of Sec.~\ref{sec:ttH} for $t\overline{t}H$ (Sec.~\ref{sec:ZZZstable}). Then, in Sec.~\ref{sec:ZZZdecayed}, we discuss the case with decays of the $Z$ bosons, scrutinising the impact of the inclusion of QED effects in PS simulations.

\subsubsection{Stable $Z$}
\label{sec:ZZZstable}

We discuss results for the following differential distributions from $ZZZ$ production in proton--proton collisions at 13 TeV: the transverse momentum of the hardest $Z$ boson $p_T(Z_1)$ in Fig.~\ref{fig:ZZZ-ptZ1} and of the softest one $p_T(Z_3)$ in Fig.~\ref{fig:ZZZ-ptZ3}, the invariant mass of the two hardest $Z$ bosons $m(Z_1 Z_2)$ in Fig.~\ref{fig:ZZZ-mz1z2} and of $p_T(j_1)$  in Fig.~\ref{fig:ZZZ-ptj1}. 
The layout of the plots in Figs.~\ref{fig:ZZZ-ptZ1}--\ref{fig:ZZZ-ptj1} is the same of those in Figs.~\ref{fig:ttH-ptH}--\ref{fig:ttH-ptj1}. 

Before discussing the specific distributions we focus on the main differences with the $t \bar t H$ distributions in Sec.~\ref{sec:ttH}. As already observed in the literature,  in the case of multi-boson production EWSL are very large (see {\it e.g.}~Refs.~\cite{Bierweiler:2013dja, Frederix:2018nkq, Bothmann:2021led,Bredt:2022dmm}) and much larger than in the case of $t \bar t H$ production. Indeed, in Figs.~\ref{fig:ZZZ-ptZ1}--\ref{fig:ZZZ-ptj1} we observe a much larger impact of EWSL than in Figs.~\ref{fig:ttH-ptH}--\ref{fig:ttH-ptj1}. Moreover, at variance with $t \bar t H$ production, the LO cross section of $ZZZ$ production does not depend on $\as$. Therefore scale uncertainties are smaller than in the $t \bar t H$ production. Still, NLO QCD corrections can be very large in multi-boson production (see {\it e.g.}~\cite{Frixione:1992pj, Frixione:1993yp,Rubin:2010xp}), therefore also the difference between $\bestpred$ and $\NLOQCDEWPS$ is enhanced, especially when the cuts defined in \eqref{eq:cuts} are applied. We anticipate that, similarly to the case of $t \bar{t} H$ production, we do not observe a dependence on the value of $c_{\clH\TO\clS}$.

In the case of the $p_T(Z_1)$ distribution (Fig.~\ref{fig:ZZZ-ptZ1}) we can clearly see how EWSL are sizeable, especially in the right plot where cuts are present. The same argument applies to the discrepancy between $\bestpred$ and $\NLOQCDEWPS$. We reckon absolute rates are smaller than what could be reasonably measured at LHC, even after the HL program, however, in the tail of the distribution, the effect of EWSL should be not only taken into account but also resummed.\footnote{While the leading EWSL of the form $\alpha^{k} \log^{2k}(s/\MW^{2})$ can be in principle resummed via a simple exponentiation, the next-to-leading case $\alpha^{k} \log^{2k-1}(s/\MW^{2})$  is not straightforward, as can be seen in Refs.~\cite{Denner:2003wi,Denner:2019vbn} and further references therein.} The same considerations are valid and even stronger for the case of   the $p_T(Z_3)$ distribution in Fig.~\ref{fig:ZZZ-ptZ3}.

Turning to the $m(Z_1 Z_2)$ distribution (Fig.~\ref{fig:ZZZ-mz1z2}), we can see that the previous discussion for $p_T$ distributions is valid also here, although with much weaker effects in the case without cuts (plot on the left). When we consider the case with cuts defined in \eqref{eq:cuts} applied, we observe an additional effect for low values of $m(Z_1 Z_2)$.  For $m(Z_1 Z_2) \lesssim$ 700 GeV, the $\NLOQCDPS$ simulation is dominated by the real emission contribution from hard matrix elements. First, this explains why the red line in the first inset of the plot on the right converges to one for small $m(Z_1 Z_2)$ value, similarly to what has been discussed for Fig.~\ref{fig:ttH-ptj1}. Second, since the dominant contribution originates from $ZZZ+1$ jet, it has a much larger dependence on the renormalisation scale. Indeed, the LO cross section for that process is of order $\alpha^3 \as$. For this reason,  the scale-uncertainty bands are larger for $m(Z_1 Z_2) \lesssim$ 700 GeV.

The $p_T(j_1)$ distribution (Fig.~\ref{fig:ZZZ-ptj1}) shows the same features discussed already for the corresponding distributions in $t \bar t H$ production (Fig.~\ref{fig:ttH-ptj1}). However, here the effects are magnified and clearly show the superiority of $\bestpred$ approximation w.r.t.~the $\NLOQCDEWPS$ one.

\subsubsection{$Z\TO e^+e^-$ decays }
\label{sec:ZZZdecayed}

In this section we consider the case in which $Z$ bosons are decayed. Via {\sc \small MadSpin}~\cite{Artoisenet:2012st}, the three $Z$ bosons are decayed into $e^+e^-$ pairs, after including the EWSL in the event as done in the previous section.\footnote{When performing the decay, {\sc \small MadSpin} includes a smearing of the invariant mass of the decay products according
to a Breit-Wigner distribution, for which we employ the width $\Gamma_Z = 2.49877 \text{ GeV}$.}
We also consider the effects induced by EWSL in the $\bestpred$ or $\NLOQCDEWPS$ approximations and the impact of QED effects in PS shower simulations. As already mentioned, we denote with PS when these effects are taken into account and with $\PSnoQED$ when they are ignored.

The process that we consider is therefore $pp\rightarrow e^+e^-e^+e^-e^+e^-$, in the Breit-Wigner approximation emerging from $ZZZ$ production. For 
the sake of simplicity, when the QED shower is enabled, only the photon emissions off charged particles (quarks and leptons) is allowed and the
photon splitting into charged fermions is disabled. Therefore, we always have exactly six electrons/positrons (we will generally call them electrons
in the following) in the final state, plus 
a number of photons, as well as quarks and gluon from the parton shower.\footnote{We recall that in the simulations presented here hadronisation is turned off.} We perform electron-photon recombination as follows:
\begin{enumerate}
    \item Final-state electrons and photons are clustered with \FJ{} into jets using the Cambridge-Aachen algorithm (hence the clustering is purely
        geometric)~\cite{Dokshitzer:1997in,Wobisch:1998wt}, with distance parameter $R=0.1$ and asking for a minimum jet transverse momentum $p_T> 25$ GeV.
    \item Out of the jets returned, we consider as \emph{leptonic jets} those jets for which the sum of the charge of the constituent 
        particles is different from zero.
    \item \label{ref:acceptance6e} The event is kept if exactly six leptonic jets are found, and otherwise it is discarded.
    \item Negatively charged leptonic jets are sorted according to their transverse momentum and they will be dubbed as $e^{-}_i$, $i=1,2,3$, where
        $e^{-}_1$ is the hardest one. 
    \item To each of the $e^{-}_i$, the corresponding positively-charged leptonic jet $e^{+}_i$ is assigned such that the quantity
    \begin{equation}
    \sum_{i=1}^{3}\big[m(e^+_i e^-_i)^2-\MZ^2\big]^2\, ,
    \end{equation}
    is minimised (this means that, in general, $e^+_i$ will not be sorted according to their transverse momentum). 
\end{enumerate}
In Fig.~\ref{fig:ZZZ-ee} we display the following distributions: the transverse momentum of the hardest lepton-jet, $p_T(e^-_1)$ (top-left),  
the invariant mass of the $e^+_1 e^-_1$ pair, $m(e^+_1e^-_1)$, in a range close to $\MZ$ (top-right), and the same distributions for the
softest lepton jets, $p_T(e^-_3)$ (bottom-left) and  $m(e^+_3e^-_3) $ (bottom-right). 

\begin{figure}[htb!]
\begin{center}
\includegraphics[width= 0.39\linewidth]{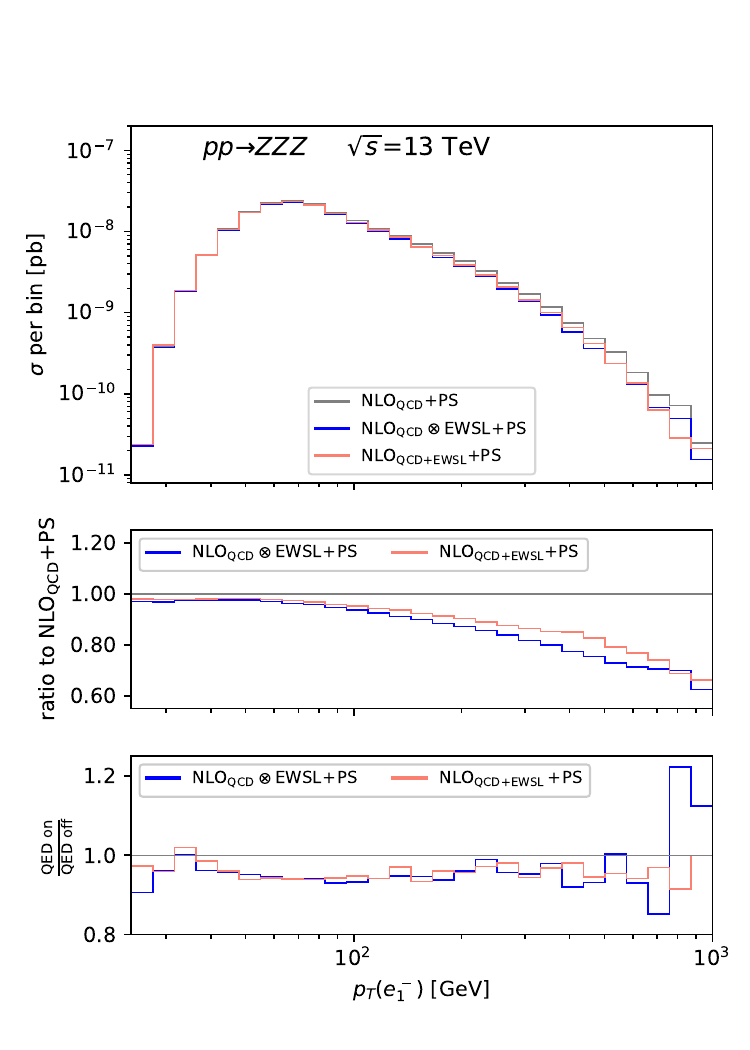}
\includegraphics[width= 0.39\linewidth]{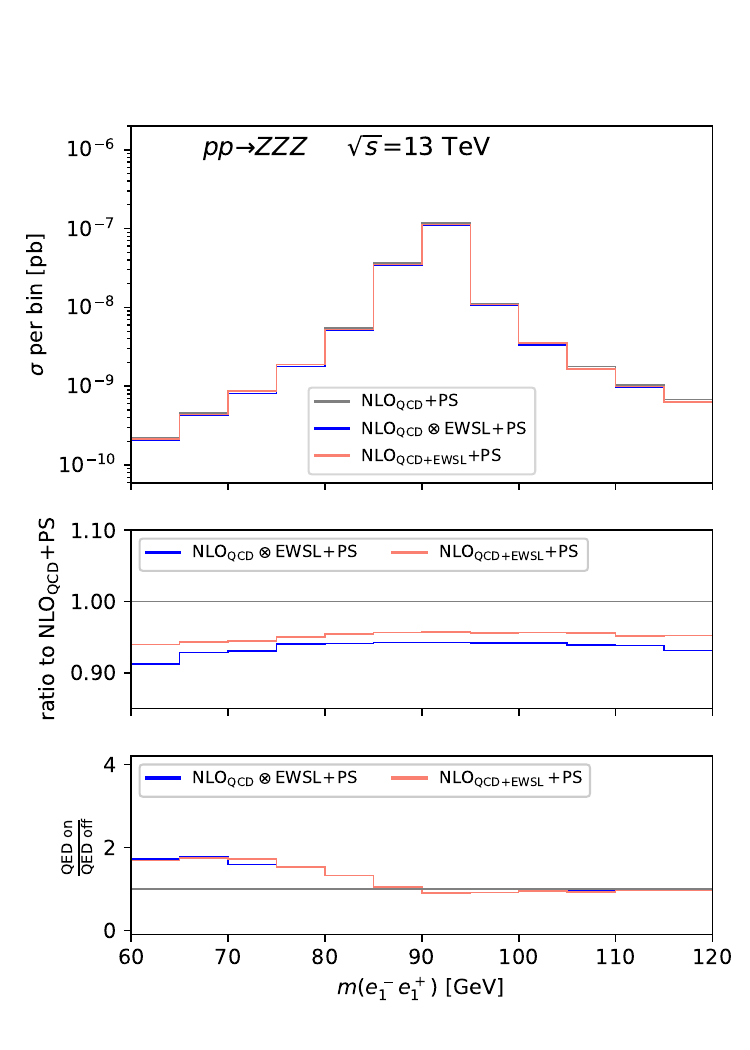}\\
\includegraphics[width= 0.39\linewidth]{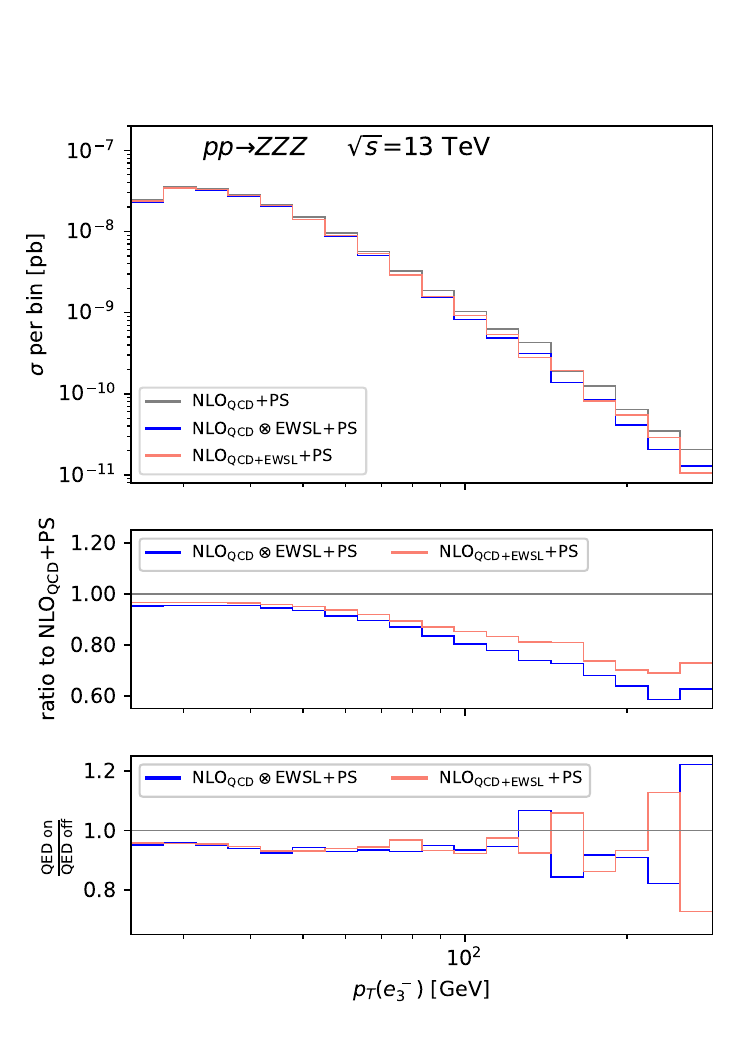}
\includegraphics[width= 0.39\linewidth]{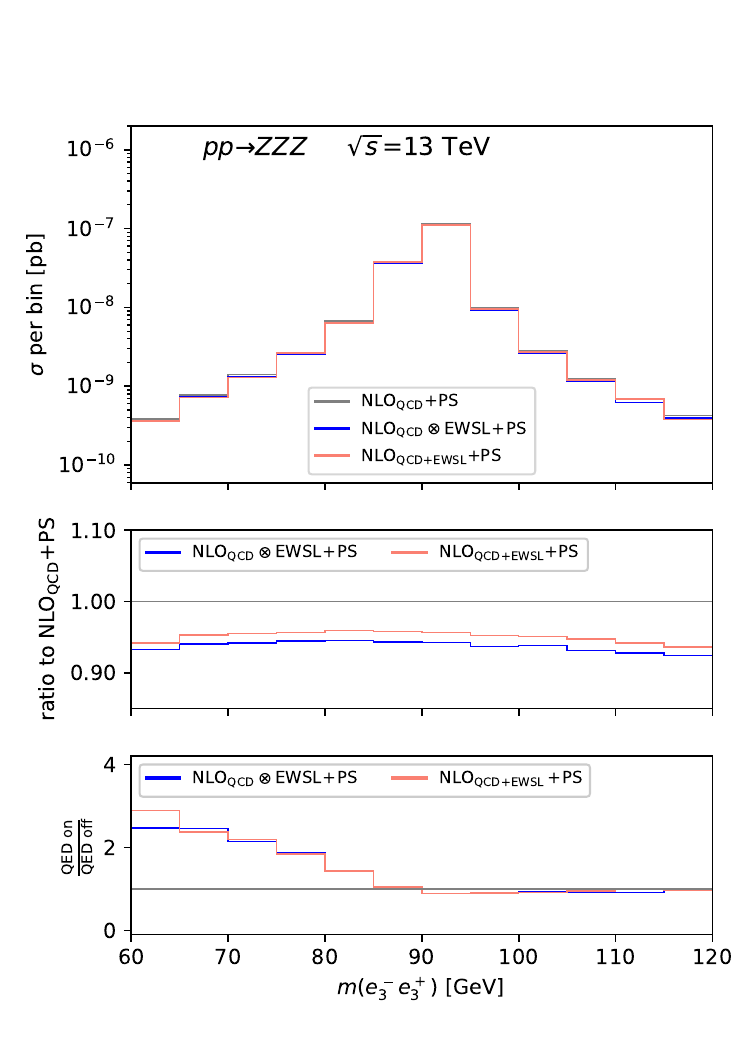}
\end{center}
\caption{Differential distributions for the transverse momentum (left) of the hardest (top) or softest (bottom) negatively-charged electron jet, and its invariant
    mass with the corresponding positively-charged one (right).  \label{fig:ZZZ-ee}}
\end{figure}

The usage of {\sc \small MadSpin} allows for the correct reconstruction of the tree-level spin correlations, which would be lost if the decay were performed via a general PS simulator in which spin correlations are not preserved. However, the correlation of the helicity-dependent EWSL with the helicities and angular distributions of the decay products within this framework is not correctly addressed. Therefore, the impact of EWSL on observables that are sensitive to spin correlations cannot be correctly taken into account. However, we stress that this limitation is not due to the reweighting procedure for the inclusion of the EWSL
presented in this work, but rather to {\sc \small MadSpin} itself, and it affects also the case of the $\NLOQCDPS$ predictions.

The layout of the plots in Fig.~\ref{fig:ZZZ-ee} is different w.r.t. those in Sec.~\ref{sec:ZZZstable}. In particular, the main panel is the same as in the plots of Figs.~\ref{fig:ttH-ptH}--\ref{fig:ZZZ-ptj1}, while the insets are different. In the first inset we show both the $\bestpred$ (blue) and $\NLOQCDEWPS$ (red) predictions normalised to the $\NLOQCDPS$ one. This is similar to Figs.~\ref{fig:ttH-ptH}--\ref{fig:ZZZ-ptj1}, but without uncertainty bands.  In the second inset instead we show for both predictions, with the same colour convention of the first inset, the ratio of the PS and $\PSnoQED$ case. We have decided to omit from the plots the dependence on $c_{\clH\TO\clS}$, since also in this case no visible effects have been observed. 

As a general comment, the observables displayed in Fig.~\ref{fig:ZZZ-ee} show that both EWSL effects (first insets) and the QED effects (second insets) cannot be neglected. Also, their relative importance strongly depends on the considered observable.
In the transverse-momentum distributions, the growth of the EW corrections in the Sudakov approximation is manifest, reaching $-20\%$
w.r.t. $\NLOQCD$ for $p_T(e^-_1)\simeq 500$ GeV and already for $p_T(e^-_3)\simeq 100$ GeV. Around these values, the benefits
of the $\bestpred$ prediction over the $\NLOQCDEWPS$ one start to be visible, with the former displaying negative corrections about 5--10\% larger than
the former. For these observables, QED effects in the PS lead to $\sim -10\%$ corrections, which are mostly related to a reduction of probability in passing the selection cuts specified in the previous bullet points. Indeed, the photon
radiation reduces the leptonic-jet energy, leading also to a very mild shape distortion. We stress that, such effects strongly depends on the recombination details and that their relative impact does not depend on the employed approximation for the EWSL 
($\bestpred$ or $\NLOQCDEWPS$), since the two classes of EW corrections factorise.

Turning to the invariant-mass distributions, the situation is somehow the opposite.  On the one hand, as expected, the enhancement due 
to QED corrections in the region $m(e^+_1e^-_1) < \MZ$, and especially $m(e^+_3e^-_3) < \MZ$, is sizeable. QED corrections are important not only in the first bins of the invariant-mass plots of Fig.~\ref{fig:ZZZ-ee}, where they easily exceed +100\% effects,\footnote{For a correct description of this process in the phase-space region  $|m(e^+_ie^-_i)- \MZ|\gg\Gamma_{Z}$ a full simulation for the complete process $pp\TO e^+e^-e^+e^-e^+e^-$ would be necessary, or at least  the contributions from $\gamma\TO e^+e^-$ splittings and their interferences with $Z\TO e^+e^-$ off-shell decays should be taken into account. This is the reason why, in the first bins of the invariant-mass plots of Fig.~\ref{fig:ZZZ-ee}, QED effects appear even larger than what has been documented in, {\it e.g.}, Refs.~\cite{CarloniCalame:2007cd, Dittmaier:2009cr, Pagani:2020mov}.}
but also in the bins around $m(e^+_1e^-_1) = \MZ$ and $m(e^+_3e^-_3) = \MZ$. These bins are the most relevant for the correct simulation of the signal region of $ZZZ$ production.  On the other hand, the 
impact of EWSL is flat and amounts to just $-5\%$.\footnote{Since the invariant mass of the system is much larger than $M_{W}$ already at the threshold for the on-shell production, $s\geq(3M_{Z})^{2}$, and the EW Casimir of the $Z$ boson is particularly large, it is not surprising to observe non-vanishing EWSL also for $Z$ bosons that are almost at rest. The value is quite flat because we calculate the EWSL for the on-shell production only. However, the variation of a few GeV for  $m(e^+_1e^-_1)$ or  $m(e^+_3e^-_3)$ has a negligible impact on the value of the invariants built with the momenta of the $Z$ bosons, so we do not expect very large effects if one also takes into account the $Z$-boson off-shellness in the EWSL evaluation.} However, should we have asked for {\it e.g.}~cuts on the transverse-momenta of the $e^+_i,e^-_i$ pairs, the EWSL effect would have been larger, as expected. For these invariant-mass distributions, no significant  difference 
is visible among the two predictions $\bestpred$ or $\NLOQCDEWPS$.
All in all, these plots in Fig.~\ref{fig:ZZZ-ee} show that both effects due to QED radiation and to the EWSL have to be included, not only for achieving percent precision as shown, {\it e.g.}, in Ref.~\cite{Gutschow:2020cug}, but also because they can in general give effects of order 10\% or more. We re-stress here that using the $\rm SDK_{\rm weak}$ scheme for the evaluation of EWSL, the QED effects can be incorporated without any problem of double-counting.

\section{Conclusion and Outlook}
\label{sec:conclusion}

In this work we have presented the automation in {\aNLO} of combined NLO+PS accuracy in QCD ($\NLOQCDPS$), Electroweak Sudakov Logarithms (EWSL) corrections, and QED final-state radiation (FSR) for event generation of SM processes at hadron colliders. Our strategy consists in the reweighting of ${\NLOQCDPS}$ events for taking into account the EWSL contribution. FSR effects of QED are simulated directly via the PS.

 We do not only reweight the LO contribution from the hard process, but also the QCD one-loop virtual contribution as well as the contribution from the first QCD real emission, the latter taking into account the different kinematic and external states for the evaluation of the EWSL. Moreover, since we have adopted the so-called $\rm SDK_{weak}$ scheme \cite{Pagani:2021vyk} for the evaluation of EWSL, FSR or in general QED effects can be included in the PS simulation avoiding their double-counting. We have denoted the accuracy of our simulation as $\bestpred$ and motivated via theoretical arguments and numerical results its superiority to an approach where only the LO is reweighted with EWSL. In particular, we have shown and discussed results for physical distributions from $pp\TO t \bar t H$ production and $pp\TO ZZZ$ production. Concerning the latter, we have also considered the case with  $Z\TO e^+e^-$ decays and stressed how neither EWSL nor QED FSR effects can be neglected.

In this paper we have focussed on the technical implementation of the $\bestpred$ accuracy in {\aNLO}, and also its automation and validation. The approach we have adopted is actually completely general and could be in principle extended to other tools that use different matching schemes for $\NLOQCDPS$ simulations. Indeed, since the approach is based on the reweighting, {\it i.e.}~a step happening after the event generation, it does not rely on the strategy for the event generation itself. Moreover, since the evaluation of EWSL involves only tree-level matrix elements and compact analytical formulas, an advantage of the reweighting via the Sudakov approximation is the speed and especially the numerical stability of the results. 

A natural follow up of our work is the extension of this technology to the matching and merging of $\NLOQCDPS$ predictions with different jet multiplicities, namely, in the {\aNLO} framework, the {\sc \small FxFx} formalism \cite{Frederix:2012ps}, similarly to what has been done in Ref.~\cite{Bothmann:2021led}.
Also, an improvement of {\sc \small MadSpin} in order to account for the information of correlation of the helicity-dependent EWSL with the helicities and angular distributions of the decay products would be beneficial for observables that are sensitive to spin correlations. 

We have left phenomenological studies and comparisons with exact NLO EW accuracy predictions to dedicated works. Since EWSL are an approximation of the exact NLO EW corrections,  there can be non-negligible effects at high energy that cannot be captured, such as photon-initiated contributions. These comparisons are therefore crucial before performing any phenomenological study.
 However, we remind the reader that a dedicated comparison of the EWSL automation in {\aNLO} and exact NLO EW corrections at fixed-order has already been performed in Ref.~\cite{Pagani:2021vyk}. Moreover, with our approach, the reweighting factors for the $\NLOQCDPS$ events can be augmented with further contributions on top of the EWSL. In other words, not only the percent-level mismatch with exact NLO EW corrections can be further decreased, but also higher-order EW effects or even BSM contributions can be taken into account.

\section*{Acknowledgements}

 We are  grateful to  the developers of {\sc MadGraph5\_aMC@NLO} for the long-standing collaboration and for discussions.  We are in particular indebted to Rikkert Frederix for discussions and clarifications of different aspects of this work. D.P.~acknowledges support from the DESY computing infrastructures provided through the DESY theory group.
T.V.~is supported by the Swedish Research Council under contract numbers 201605996 and 202004423. 
M.Z.~is supported by the ``Programma per Giovani Ricercatori Rita Levi Montalcini'' granted by the Italian Ministero dell'Universit\`a e della Ricerca (MUR).

\appendix

\section{The {\denpoz} algorithm in {\sc MadGraph5\_aMC@NLO} \label{sec:EWSL}}

In this Appendix we briefly remind the main features of the {\denpoz} algorithm as implemented in {\aNLO}, which is based on the revisitation in Ref.~\cite{Pagani:2021vyk} where many more details can be found.

\subsection{Amplitude level}
 \label{sec:EWSLamp}

The {\denpoz} algorithm \cite{Denner:2000jv,Denner:2001gw} allows for the calculation of one-loop EW double-logarithmic (DL) and single-logarithmic (SL) corrections, denoted also collectively as leading approximation (LA), for any individual helicity configuration that does not give
a mass-suppressed amplitude in the high-energy limit, and for a generic SM partonic processes with on-shell external legs.
First of all, the algorithm strictly relies on the assumption that all invariants are much larger than the gauge boson masses.
Specifically, with $k$ and $l$ being two generic external particles with momenta $p_k$ and $p_l$, 
\beq \label{eq:Sudaklim} 
r_{kl}\equiv(p_k+p_l)^2 \simeq 2p_kp_l \gg \MW^2 \simeq \MH^2,\Mt^2,\MW^2,\MZ^2.
\eeq
The {\denpoz} algorithm has been formulated for amplitudes with $n$ arbitrary external particles, where all momenta
$p_k$  are assumed as incoming.  
Processes are denoted as
\beq \label{eq:process}
\varphi_{i_1}(p_1)\dots \varphi_{i_n}(p_n)\rightarrow 0\,,
\eeq
where the (anti-)particles  $\varphi_{i_k}$ are the
components of the various multiplets
$\varphi$ of the SM. Moreover, contributions from longitudinal gauge-bosons are always evaluated via the Gold\-stone-boson equivalence theorem. If the Born matrix element for the process in \eqref{eq:process} is written as
\begin{equation}
\M_0^{i_1 \ldots i_n}(p_1,\ldots, p_n)\,, \label{eq:M0}
\end{equation}
in LA  the $\ord(\alpha)$ corrections to $\M_0$,  $\delta \M$, can be written of the form\footnote{The only logarithms that cannot be directly written in this form are those associated to the parameter renormalisation, the terms denoted as  ``PR'' in the Denner-Pozzorini notation. They are discussed in detail in Ref.~\cite{Denner:2000jv} and their implementation in {\aNLO} in Ref.~\cite{Pagani:2021vyk}. For the present discussion it is important to know only that they can also be calculated via tree-level amplitudes multiplied by logarithms of invariants and proper coefficients.}   
\beq \label{LAfactorization} 
\delta \M^{i_1 \ldots i_n}(p_1,\ldots, p_n)= 
\M_0^{i'_1 \ldots i'_n}(p_1, \ldots, p_n)\delta_{i'_1i_1 \ldots i'_ni_n}.
\end{equation}
Equation \refeq{LAfactorization} is the main reason why the EWSL for physical cross sections can be computed in a much faster and stable way than the exact NLO EW corrections. Indeed Eq.~\refeq{LAfactorization} means that in LA the one-loop EW corrections can be written in terms of only tree-level amplitudes ($\M_0^{i'_1 \ldots i'_n}$), which on the other hand involve different processes than the original one in \eqref{eq:process}, as can be seen in the indices. On top of that, the quantities $\delta_{i'_1i_1 \ldots i'_ni_n}$ depend only on two kinds of ingredients. First, on logarithms of the form
\beq
\LrM\equiv\frac{\alpha}{4\pi}\log^2{\frac{|r_{kl}|}{M^2}} \qquad{\rm and}\qquad 
\lrMwithabs\equiv\frac{\alpha}{4\pi}\log{\frac{|r_{kl}|}{M^2}}\,, \label{eq:generallogs}
\eeq
where $r_{kl}$ denotes a generic kinematic invariant and $M$ any of the masses of the SM heavy particles ($\MW, \MH,$  $\Mt$ and $\MZ$) or in the case of the photon the IR-regularisation scale $Q$, using the same notation of Ref.~\cite{Pagani:2021vyk}. Second, on the couplings of each external field  $\varphi_{i_k}$  to the gauge bosons $\GB_a$ and another field $\varphi_{i_k'}$,  $I^a_{i_ki'_k}(k)$, or associated quantities such as electroweak Casimir operators $\cew$, which involve the entire $\rm SU(2)\times U(1)$ group.
The exact expressions, as implemented in {\aNLO}, can be found in Ref.~\cite{Pagani:2021vyk} and are based on the results of Refs.~\cite{Denner:2000jv,Denner:2001gw}. 

As mentioned before, an important limitation of the {\denpoz} algorithm is that, for a given process, at least one helicity configuration must not be mass suppressed, {\it i.e.},  the amplitude should scale as $s ^{\frac{4-n}{2}}$  for a process with $n$ external legs. Most of the SM processes satisfy this assumption, but important exceptions are possible, such as the Higgs production via vector-boson fusion. Another important limitation is given by condition \eqref{eq:Sudaklim}. Processes including resonating unstable particles cannot be treated in this approximation. Rather, the process without decays should be first considered when applying the {\denpoz} algorithm. Only afterwards the decays should be taken into account.

In Ref.~\cite{Pagani:2021vyk} it has been shown that not only the logarithms of the form
\beq \label{eq:dslogs}
\LsW \qquad{\rm and}\qquad 
\lsW\,,
\eeq
but also those of the form
\beq\label{eq:dslogsrkl}
L(r_{k l},r_{k' l'}) \qquad{\rm and} \qquad l(r_{k l},r_{k' l'})\,,  
\eeq 
can be relevant when  $r_{k l}\gg r_{k' l'}\gg \MW^2$, where $r_{k l}$ and $r_{k' l'}$ is a generic pair of the many possible invariants that one can build with two external momenta. It is important to note that the condition $r_{k l}/ r_{k' l'} = 1$
can never be satisfied at the same time for all possible pairs of $r_{k l}$ and $ r_{k' l'}$ invariants.

The logarithms in Eq.~\eqref{eq:dslogs} are those yielding the formal LA as presented in Refs.~\cite{Denner:2000jv,Denner:2001gw}, while those in Eq.~\eqref{eq:dslogsrkl} have been reintroduced in the {\denpoz} algorithm in the revisitation in Ref.~\cite{Pagani:2021vyk} and are accounted for in the algorithm by the term 
\beq
\Delta^{s\TO r_{kl}}(r_{kl},M^2)\equiv \Lrs+ 2\lWM\lrs -2 i  \pi \Theta(r_{kl}) \lrsalpha\,, \label{eq:deltasr}
\eeq
where $\Theta$ is the Heaviside step function and it multiplies an imaginary term whose origin has been discussed in Ref.~\cite{Pagani:2021vyk}. In this paper, for all results, we have always understood the inclusion of terms in Eq.~\eqref{eq:deltasr}.

Before moving to the case of squared matrix elements and cross sections it is important to note that the terms $\delta_{i'_1i_1 \ldots i'_ni_n}$ in Eq.~\eqref{LAfactorization} involve also logarithms of the form $\log(r_{k l}/Q^2) $ or in the original formulation of Denner and Pozzorini, {\it e.g.}, $\log(\MW^2/\lambda^2)$, where $\lambda$ is the fictitious photon mass used as IR-regulator. Thus the quantity $\delta \M$ is IR-divergent and therefore non-physical.

\subsection{Cross-section level}
What has been discussed up to this point in this Appendix concerns the approximation of an amplitude. The case of squared matrix elements and cross sections is different and discussed in the following. We focus first on the case of the squared matrix elements and then on the case of physical cross sections.

Using the notation in Eqs.~\eqref{eq:blobs_LO_general} and \eqref{eq:blobs_NLO_general}, it is easy to understand that the term
 \beq
 \ord(\Sigma^{}_{\NLO_{i}})=\ord(\Sigma^{}_{\LO_{i}})\times \as = \ord(\Sigma^{}_{\LO_{i-1}})\times \alpha\,, \label{eq:orders}
 \eeq
where $\ord$ denotes the perturbative order. In particular this shows that NLO EW corrections ($\NLO_2$) receives both corrections from ``EW loops'' on top of $\LO_1\equiv\LOQCD$ and ``QCD loops'' on top of $\LO_2$. Thus, in LA, the contribution from one-loop corrections to the quantity $\Sigma^{}_{\NLO_{2}}$, denoted as $\Sigma^{\rm virt}_{\NLO_{2}}$ can be written of the form
\beq
(\Sigma^{\rm virt}_{\NLO_{2}})\Big|_{\rm LA}=\Sigma^{}_{\LO_{1}} \deltaEW + \Sigma^{}_{\LO_{2}} \deltaQCD . \label{eq:QCDEWcomb}
\eeq
 
The quantity $\deltaEW$ is calculated via the {\denpoz} as summarised in Sec.~\ref{sec:EWSLamp} and in particular Eqs.~\eqref{eq:M0} and \eqref{LAfactorization}. With $\M_0$ being the amplitude that once squared leads to $\Sigma^{}_{\LO_{1}}$,
\beq
\deltaEW\equiv \frac{2\Re(\M_0\delta\M^*)}{|\M_0|^2}\,. \label{eq:deltaEW}
\eeq

Strictly speaking, Eq.~\eqref{eq:QCDEWcomb} is valid only under two simple assumptions:  $Q^2=s$ and $\Delta^{s\TO r_{kl}}(r_{kl},M^2)=0$. Otherwise, also the information on the colour-linked matrix elements would be necessary. A simple expression for $\deltaQCD$ under the two aforementioned assumptions has been provided Ref.~\cite{Pagani:2021vyk} and is reported later, in Eq.~\eqref{eq:dQCDfinal}. However, as we will discuss in the following and as has already been mentioned in Sec.~\ref{sec:dEWSLtech}, $\deltaQCD$ is not so relevant as $\deltaEW$ for the physical observables and processes considered in this work. This is ultimately the reason why the previous two assumptions are  irrelevant in view of the EWSL  approximation in the context of physical cross sections, which we are going to describe in the following.

Similarly to the case of $\delta\M$, the quantity $\deltaEW$ is IR-divergent and therefore non-physical. This is not a surprise, being an approximation of virtual EW corrections, which involve contributions from massless photons. Using the same notation as in Ref.~\cite{Pagani:2021vyk}, this is the scheme denoted as SDK, meaning the {\denpoz} algorithm for the calculation of the amplitudes as in Eq.~\eqref{LAfactorization},\footnote{In Ref.~\cite{Pagani:2021vyk} not only the terms in Eq.~\eqref{eq:deltasr} but also an additional imaginary term missing in the original formulation of the {\denpoz} algorithm has been introduced. Moreover, IR divergencies are regularised via Dimensional Regularisation. We understand these two features in the text when referring to the {\denpoz} algorithm.} and in the case of squared matrix elements contributing to the virtual component  of NLO EW corrections as in Eq.~\eqref{eq:QCDEWcomb}. The SDK scheme is therefore a very good approximation at high energies of loop amplitude and virtual contributions, but it is not directly suitable in the case of phenomenological predictions.

In order to obtain predictions in LA  that can be used for physical cross sections,  the approach that has been employed often in literature is what has been denoted in Ref.~\cite{Pagani:2021vyk} as $\rm SDK_{0}$. We stress here again that the $\rm SDK_{0}$ is an approach mostly driven by simplicity. Indeed, it bypasses the problem of IR finiteness by simply removing some QED logarithms involving $\MW$ and the IR scale, but those logarithms arise from the conventions used in Refs.~\cite{Denner:2000jv,Denner:2001gw} and not from physical argument.

In Ref.~\cite{Pagani:2021vyk}, the $\rm SDK_{\rm weak}$ scheme has been precisely designed in order to solve this problem. The main idea behind it is that in sufficiently inclusive observables the Sudakov logarithms of QED and IR origin in the virtual contributions cancel against their real counterpart. In fact, the $\rm SDK_{\rm weak}$ scheme consists  in a purely weak version of the SDK approach where almost all contributions of QED IR origin are removed\footnote{See Ref.~\cite{Pagani:2021vyk} for more details and for the modifications to the {\denpoz} algorithm for switching from the SDK to the $\rm SDK_{\rm weak}$ scheme.}, while those of QED and UV origin are retained.  In Ref.~\cite{Pagani:2021vyk}, it has been clearly shown how the $\rm SDK_{\rm weak}$ is superior to the $\rm SDK_{0}$ in catching the EWSL component of NLO EW corrections when any electrically charged object is clustered with quasi-collinear photons. 

The cancellation between real and virtual contributions takes place also for the case of QCD on top of the $\LO_2$, and this is the reason why in first approximation one can neglect  the contribution from $\deltaQCD$ in Eq.~\eqref{eq:QCDEWcomb} for a vast class of processes.
In particular the formula for $\deltaQCD$ is the following\footnote{As can be seen by comparison with Ref~\cite{Pagani:2021vyk}, there was a typo therein, but the formula in Eq.~\eqref{eq:dQCDfinal} was already correctly implemented in the code for producing the results.}: 
\begin{equation}
\deltaQCD\equiv  2 \left[ n_t \,\Ltop + \left(n-1\right)\Las{\mu_R^2}-n_{g}\, \Las{s}  + \frac{1}{ \Sigma^{}_{\LO_{2}}} \frac{\de \Sigma^{}_{\LO_{2}}}{\de \Mt} \dmtQCD\right] \, .
\label{eq:dQCDfinal}
\end{equation}
with 
\begin{eqnarray}
\Ltop& \equiv &\frac{C_F}{2} \frac{\as}{4 \pi}  \left( \log^2 \frac{s}{\Mt^2} +   \log \frac{s}{\Mt^2}  \right)\, , \label{eq:Lt} \\
 \Las{\mu^2} &\equiv& \frac{1}{3}~ \frac{\as}{4 \pi} \log \frac{\mu^2}{\Mt^2}\, , \\
 \dmtQCD&\equiv& -3 C_F \frac{\as}{4 \pi}    \log \frac{s}{\Mt^2}\, , \label{eq:dmtQCD}
\end{eqnarray}
where $n$ is defined by the perturbative order of $\LOQCD\equiv\LO_1$ in the convention $\ord(\LOQCD)=\as^n\alpha^m$, therefore $\ord(\LO_2)=\as^{n-1}\alpha^m$,  and $n_t$ and $n_g$ are the number of top quarks 
and gluons in the external legs, respectively. 

Similarly to the case of QED, unless there are boosted tops and the radiation collinear to them is not clustered together with the tops, the terms proportional to $\Ltop$ can be neglected in the approximation of physical observables, as it was already understood in the $\rm SDK_{\rm weak}$ scheme as defined in Ref.~\cite{Pagani:2021vyk}.

\addcontentsline{toc}{section}{References}
\bibliographystyle{utphys.bst}
\bibliography{biblionew}

\providecommand{\href}[2]{#2}\begingroup\raggedright\begin{thebibliography}{100}

\bibitem{Aad:2012tfa}
{\bf ATLAS} Collaboration, G.~Aad {\em et al.}, {\em {Observation of a new
  particle in the search for the Standard Model Higgs boson with the ATLAS
  detector at the LHC}}.
  \href{http://dx.doi.org/10.1016/j.physletb.2012.08.020}{Phys. Lett. B {\bf
  716} (2012)  1--29}, \href{http://arxiv.org/abs/1207.7214}{{\tt
  arXiv:1207.7214 [hep-ex]}}.

\bibitem{Chatrchyan:2012ufa}
{\bf CMS} Collaboration, S.~Chatrchyan {\em et al.}, {\em {Observation of a New
  Boson at a Mass of 125 GeV with the CMS Experiment at the LHC}}.
  \href{http://dx.doi.org/10.1016/j.physletb.2012.08.021}{Phys. Lett. B {\bf
  716} (2012)  30--61}, \href{http://arxiv.org/abs/1207.7235}{{\tt
  arXiv:1207.7235 [hep-ex]}}.

\bibitem{Aad:2019mbh}
{\bf ATLAS} Collaboration, G.~Aad {\em et al.}, {\em {Combined measurements of
  Higgs boson production and decay using up to $80$ fb$^{-1}$ of proton-proton
  collision data at $\sqrt{s}=$ 13 TeV collected with the ATLAS experiment}}.
  \href{http://dx.doi.org/10.1103/PhysRevD.101.012002}{Phys. Rev. D {\bf 101}
  (2020) no.~1, 012002}, \href{http://arxiv.org/abs/1909.02845}{{\tt
  arXiv:1909.02845 [hep-ex]}}.

\bibitem{Azzi:2019yne}
P.~Azzi {\em et al.}, {\em {Report from Working Group 1}: {Standard Model
  Physics at the HL-LHC and HE-LHC}}.
  \href{http://dx.doi.org/10.23731/CYRM-2019-007.1}{CERN Yellow Rep. Monogr.
  {\bf 7} (2019)  1--220}, \href{http://arxiv.org/abs/1902.04070}{{\tt
  arXiv:1902.04070 [hep-ph]}}.

\bibitem{Cepeda:2019klc}
M.~Cepeda {\em et al.}, {\em {Report from Working Group 2}: {Higgs Physics at
  the HL-LHC and HE-LHC}}.
  \href{http://dx.doi.org/10.23731/CYRM-2019-007.221}{CERN Yellow Rep. Monogr.
  {\bf 7} (2019)  221--584}, \href{http://arxiv.org/abs/1902.00134}{{\tt
  arXiv:1902.00134 [hep-ph]}}.

\bibitem{CidVidal:2018eel}
X.~Cid~Vidal {\em et al.}, {\em {Report from Working Group 3}: {Beyond the
  Standard Model physics at the HL-LHC and HE-LHC}}.
  \href{http://dx.doi.org/10.23731/CYRM-2019-007.585}{CERN Yellow Rep. Monogr.
  {\bf 7} (2019)  585--865}, \href{http://arxiv.org/abs/1812.07831}{{\tt
  arXiv:1812.07831 [hep-ph]}}.

\bibitem{Cerri:2018ypt}
A.~Cerri {\em et al.}, {\em {Report from Working Group 4}: {Opportunities in
  Flavour Physics at the HL-LHC and HE-LHC}}.
  \href{http://dx.doi.org/10.23731/CYRM-2019-007.867}{CERN Yellow Rep. Monogr.
  {\bf 7} (2019)  867--1158}, \href{http://arxiv.org/abs/1812.07638}{{\tt
  arXiv:1812.07638 [hep-ph]}}.

\bibitem{Citron:2018lsq}
Z.~Citron {\em et al.}, {\em {Report from Working Group 5}: {Future physics
  opportunities for high-density QCD at the LHC with heavy-ion and proton
  beams}}. \href{http://dx.doi.org/10.23731/CYRM-2019-007.1159}{CERN Yellow
  Rep. Monogr. {\bf 7} (2019)  1159--1410},
  \href{http://arxiv.org/abs/1812.06772}{{\tt arXiv:1812.06772 [hep-ph]}}.

\bibitem{Chapon:2020heu}
E.~Chapon {\em et al.}, {\em {Prospects for quarkonium studies at the
  high-luminosity LHC}}.
  \href{http://dx.doi.org/10.1016/j.ppnp.2021.103906}{Prog. Part. Nucl. Phys.
  {\bf 122} (2022)  103906}, \href{http://arxiv.org/abs/2012.14161}{{\tt
  arXiv:2012.14161 [hep-ph]}}.

\bibitem{Alwall:2014hca}
J.~Alwall, R.~Frederix, S.~Frixione, V.~Hirschi, F.~Maltoni, O.~Mattelaer,
  H.~S. Shao, T.~Stelzer, P.~Torrielli, and M.~Zaro, {\em {The automated
  computation of tree-level and next-to-leading order differential cross
  sections, and their matching to parton shower simulations}}.
  \href{http://dx.doi.org/10.1007/JHEP07(2014)079}{JHEP {\bf 07} (2014)  079},
  \href{http://arxiv.org/abs/1405.0301}{{\tt arXiv:1405.0301 [hep-ph]}}.

\bibitem{Kallweit:2014xda}
S.~Kallweit, J.~M. Lindert, P.~Maierh\"ofer, S.~Pozzorini, and M.~Sch\"onherr,
  {\em {NLO electroweak automation and precise predictions for W+multijet
  production at the LHC}}.
  \href{http://dx.doi.org/10.1007/JHEP04(2015)012}{JHEP {\bf 04} (2015)  012},
  \href{http://arxiv.org/abs/1412.5157}{{\tt arXiv:1412.5157 [hep-ph]}}.

\bibitem{Frixione:2015zaa}
S.~Frixione, V.~Hirschi, D.~Pagani, H.~S. Shao, and M.~Zaro, {\em {Electroweak
  and QCD corrections to top-pair hadroproduction in association with heavy
  bosons}}. \href{http://dx.doi.org/10.1007/JHEP06(2015)184}{JHEP {\bf 06}
  (2015)  184}, \href{http://arxiv.org/abs/1504.03446}{{\tt arXiv:1504.03446
  [hep-ph]}}.

\bibitem{Chiesa:2015mya}
M.~Chiesa, N.~Greiner, and F.~Tramontano, {\em {Automation of electroweak
  corrections for LHC processes}}.
  \href{http://dx.doi.org/10.1088/0954-3899/43/1/013002}{J. Phys. G {\bf 43}
  (2016) no.~1, 013002}, \href{http://arxiv.org/abs/1507.08579}{{\tt
  arXiv:1507.08579 [hep-ph]}}.

\bibitem{Biedermann:2017yoi}
B.~Biedermann, S.~Br\"auer, A.~Denner, M.~Pellen, S.~Schumann, and J.~M.
  Thompson, {\em {Automation of NLO QCD and EW corrections with Sherpa and
  Recola}}. \href{http://dx.doi.org/10.1140/epjc/s10052-017-5054-8}{Eur. Phys.
  J. C {\bf 77} (2017)  492}, \href{http://arxiv.org/abs/1704.05783}{{\tt
  arXiv:1704.05783 [hep-ph]}}.

\bibitem{Chiesa:2017gqx}
M.~Chiesa, N.~Greiner, M.~Sch\"onherr, and F.~Tramontano, {\em {Electroweak
  corrections to diphoton plus jets}}.
  \href{http://dx.doi.org/10.1007/JHEP10(2017)181}{JHEP {\bf 10} (2017)  181},
  \href{http://arxiv.org/abs/1706.09022}{{\tt arXiv:1706.09022 [hep-ph]}}.

\bibitem{Frederix:2018nkq}
R.~Frederix, S.~Frixione, V.~Hirschi, D.~Pagani, H.~S. Shao, and M.~Zaro, {\em
  {The automation of next-to-leading order electroweak calculations}}.
  \href{http://dx.doi.org/10.1007/JHEP11(2021)085}{JHEP {\bf 07} (2018)  185},
  \href{http://arxiv.org/abs/1804.10017}{{\tt arXiv:1804.10017 [hep-ph]}}.
  [Erratum: JHEP 11, 085 (2021)].

\bibitem{Pagani:2021iwa}
D.~Pagani, H.-S. Shao, I.~Tsinikos, and M.~Zaro, {\em {Automated EW corrections
  with isolated photons: t$ \overline{t} $\ensuremath{\gamma}, t$ \overline{t}
  $\ensuremath{\gamma}\ensuremath{\gamma} and t\ensuremath{\gamma}j as case
  studies}}. \href{http://dx.doi.org/10.1007/JHEP09(2021)155}{JHEP {\bf 09}
  (2021)  155}, \href{http://arxiv.org/abs/2106.02059}{{\tt arXiv:2106.02059
  [hep-ph]}}.

\bibitem{Hirschi:2011pa}
V.~Hirschi, R.~Frederix, S.~Frixione, M.~V. Garzelli, F.~Maltoni, and
  R.~Pittau, {\em {Automation of one-loop QCD corrections}}.
  \href{http://dx.doi.org/10.1007/JHEP05(2011)044}{JHEP {\bf 05} (2011)  044},
  \href{http://arxiv.org/abs/1103.0621}{{\tt arXiv:1103.0621 [hep-ph]}}.

\bibitem{Cullen:2011ac}
{\bf GoSam} Collaboration, G.~Cullen, N.~Greiner, G.~Heinrich, G.~Luisoni,
  P.~Mastrolia, G.~Ossola, T.~Reiter, and F.~Tramontano, {\em {Automated
  One-Loop Calculations with GoSam}}.
  \href{http://dx.doi.org/10.1140/epjc/s10052-012-1889-1}{Eur. Phys. J. C {\bf
  72} (2012)  1889}, \href{http://arxiv.org/abs/1111.2034}{{\tt arXiv:1111.2034
  [hep-ph]}}.

\bibitem{Cascioli:2011va}
F.~Cascioli, P.~Maierhofer, and S.~Pozzorini, {\em {Scattering Amplitudes with
  Open Loops}}. \href{http://dx.doi.org/10.1103/PhysRevLett.108.111601}{Phys.
  Rev. Lett. {\bf 108} (2012)  111601},
  \href{http://arxiv.org/abs/1111.5206}{{\tt arXiv:1111.5206 [hep-ph]}}.

\bibitem{Actis:2012qn}
S.~Actis, A.~Denner, L.~Hofer, A.~Scharf, and S.~Uccirati, {\em {Recursive
  generation of one-loop amplitudes in the Standard Model}}.
  \href{http://dx.doi.org/10.1007/JHEP04(2013)037}{JHEP {\bf 04} (2013)  037},
  \href{http://arxiv.org/abs/1211.6316}{{\tt arXiv:1211.6316 [hep-ph]}}.

\bibitem{Actis:2016mpe}
S.~Actis, A.~Denner, L.~Hofer, J.-N. Lang, A.~Scharf, and S.~Uccirati, {\em
  {RECOLA: REcursive Computation of One-Loop Amplitudes}}.
  \href{http://dx.doi.org/10.1016/j.cpc.2017.01.004}{Comput. Phys. Commun. {\bf
  214} (2017)  140--173}, \href{http://arxiv.org/abs/1605.01090}{{\tt
  arXiv:1605.01090 [hep-ph]}}.

\bibitem{Denner:2017wsf}
A.~Denner, J.-N. Lang, and S.~Uccirati, {\em {Recola2: REcursive Computation of
  One-Loop Amplitudes 2}}.
  \href{http://dx.doi.org/10.1016/j.cpc.2017.11.013}{Comput. Phys. Commun. {\bf
  224} (2018)  346--361}, \href{http://arxiv.org/abs/1711.07388}{{\tt
  arXiv:1711.07388 [hep-ph]}}.

\bibitem{Buccioni:2019sur}
{\bf OpenLoops 2} Collaboration, F.~Buccioni, J.-N. Lang, J.~M. Lindert,
  P.~Maierh\"ofer, S.~Pozzorini, H.~Zhang, and M.~F. Zoller, {\em {OpenLoops
  2}}. \href{http://dx.doi.org/10.1140/epjc/s10052-019-7306-2}{Eur. Phys. J. C
  {\bf 79} (2019) no.~10, 866}, \href{http://arxiv.org/abs/1907.13071}{{\tt
  arXiv:1907.13071 [hep-ph]}}.

\bibitem{Frixione:2002ik}
S.~Frixione and B.~R. Webber, {\em {Matching NLO QCD computations and parton
  shower simulations}}.
  \href{http://dx.doi.org/10.1088/1126-6708/2002/06/029}{JHEP {\bf 06} (2002)
  029}, \href{http://arxiv.org/abs/hep-ph/0204244}{{\tt arXiv:hep-ph/0204244}}.

\bibitem{Frixione:2007nw}
S.~Frixione, P.~Nason, and G.~Ridolfi, {\em {A Positive-weight
  next-to-leading-order Monte Carlo for heavy flavour hadroproduction}}.
  \href{http://dx.doi.org/10.1088/1126-6708/2007/09/126}{JHEP {\bf 09} (2007)
  126}, \href{http://arxiv.org/abs/0707.3088}{{\tt arXiv:0707.3088 [hep-ph]}}.

\bibitem{Frixione:2007vw}
S.~Frixione, P.~Nason, and C.~Oleari, {\em {Matching NLO QCD computations with
  Parton Shower simulations: the POWHEG method}}.
  \href{http://dx.doi.org/10.1088/1126-6708/2007/11/070}{JHEP {\bf 11} (2007)
  070}, \href{http://arxiv.org/abs/0709.2092}{{\tt arXiv:0709.2092 [hep-ph]}}.

\bibitem{Hamilton:2012rf}
K.~Hamilton, P.~Nason, C.~Oleari, and G.~Zanderighi, {\em {Merging H/W/Z + 0
  and 1 jet at NLO with no merging scale: a path to parton shower + NNLO
  matching}}. \href{http://dx.doi.org/10.1007/JHEP05(2013)082}{JHEP {\bf 05}
  (2013)  082}, \href{http://arxiv.org/abs/1212.4504}{{\tt arXiv:1212.4504
  [hep-ph]}}.

\bibitem{Alioli:2013hqa}
S.~Alioli, C.~W. Bauer, C.~Berggren, F.~J. Tackmann, J.~R. Walsh, and
  S.~Zuberi, {\em {Matching Fully Differential NNLO Calculations and Parton
  Showers}}. \href{http://dx.doi.org/10.1007/JHEP06(2014)089}{JHEP {\bf 06}
  (2014)  089}, \href{http://arxiv.org/abs/1311.0286}{{\tt arXiv:1311.0286
  [hep-ph]}}.

\bibitem{Hoche:2014uhw}
S.~H\"oche, Y.~Li, and S.~Prestel, {\em {Drell-Yan lepton pair production at
  NNLO QCD with parton showers}}.
  \href{http://dx.doi.org/10.1103/PhysRevD.91.074015}{Phys. Rev. D {\bf 91}
  (2015) no.~7, 074015}, \href{http://arxiv.org/abs/1405.3607}{{\tt
  arXiv:1405.3607 [hep-ph]}}.

\bibitem{Monni:2019whf}
P.~F. Monni, P.~Nason, E.~Re, M.~Wiesemann, and G.~Zanderighi, {\em
  {MiNNLO$_{PS}$: a new method to match NNLO QCD to parton showers}}.
  \href{http://dx.doi.org/10.1007/JHEP05(2020)143}{JHEP {\bf 05} (2020)  143},
  \href{http://arxiv.org/abs/1908.06987}{{\tt arXiv:1908.06987 [hep-ph]}}.
  [Erratum: JHEP 02, 031 (2022)].

\bibitem{Bertone:2022hig}
V.~Bertone and S.~Prestel, {\em {Combining N3LO QCD calculations and parton
  showers for hadronic collision events}}.
  \href{http://arxiv.org/abs/2202.01082}{{\tt arXiv:2202.01082 [hep-ph]}}.

\bibitem{Barze:2012tt}
L.~Barze, G.~Montagna, P.~Nason, O.~Nicrosini, and F.~Piccinini, {\em
  {Implementation of electroweak corrections in the POWHEG BOX: single W
  production}}. \href{http://dx.doi.org/10.1007/JHEP04(2012)037}{JHEP {\bf 04}
  (2012)  037}, \href{http://arxiv.org/abs/1202.0465}{{\tt arXiv:1202.0465
  [hep-ph]}}.

\bibitem{Barze:2013fru}
L.~Barze, G.~Montagna, P.~Nason, O.~Nicrosini, F.~Piccinini, and A.~Vicini,
  {\em {Neutral current Drell-Yan with combined QCD and electroweak corrections
  in the POWHEG BOX}}.
  \href{http://dx.doi.org/10.1140/epjc/s10052-013-2474-y}{Eur. Phys. J. C {\bf
  73} (2013) no.~6, 2474}, \href{http://arxiv.org/abs/1302.4606}{{\tt
  arXiv:1302.4606 [hep-ph]}}.

\bibitem{Kallweit:2015dum}
S.~Kallweit, J.~M. Lindert, P.~Maierhofer, S.~Pozzorini, and M.~Sch\"onherr,
  {\em {NLO QCD+EW predictions for V + jets including off-shell vector-boson
  decays and multijet merging}}.
  \href{http://dx.doi.org/10.1007/JHEP04(2016)021}{JHEP {\bf 04} (2016)  021},
  \href{http://arxiv.org/abs/1511.08692}{{\tt arXiv:1511.08692 [hep-ph]}}.

\bibitem{Granata:2017iod}
F.~Granata, J.~M. Lindert, C.~Oleari, and S.~Pozzorini, {\em {NLO QCD+EW
  predictions for HV and HV +jet production including parton-shower effects}}.
  \href{http://dx.doi.org/10.1007/JHEP09(2017)012}{JHEP {\bf 09} (2017)  012},
  \href{http://arxiv.org/abs/1706.03522}{{\tt arXiv:1706.03522 [hep-ph]}}.

\bibitem{Gutschow:2018tuk}
C.~G\"utschow, J.~M. Lindert, and M.~Sch\"onherr, {\em {Multi-jet merged
  top-pair production including electroweak corrections}}.
  \href{http://dx.doi.org/10.1140/epjc/s10052-018-5804-2}{Eur. Phys. J. C {\bf
  78} (2018) no.~4, 317}, \href{http://arxiv.org/abs/1803.00950}{{\tt
  arXiv:1803.00950 [hep-ph]}}.

\bibitem{Chiesa:2019ulk}
M.~Chiesa, A.~Denner, J.-N. Lang, and M.~Pellen, {\em {An event generator for
  same-sign W-boson scattering at the LHC including electroweak corrections}}.
  \href{http://dx.doi.org/10.1140/epjc/s10052-019-7290-6}{Eur. Phys. J. C {\bf
  79} (2019) no.~9, 788}, \href{http://arxiv.org/abs/1906.01863}{{\tt
  arXiv:1906.01863 [hep-ph]}}.

\bibitem{Chiesa:2020ttl}
M.~Chiesa, C.~Oleari, and E.~Re, {\em {NLO QCD+NLO EW corrections to diboson
  production matched to parton shower}}.
  \href{http://dx.doi.org/10.1140/epjc/s10052-020-8419-3}{Eur. Phys. J. C {\bf
  80} (2020) no.~9, 849}, \href{http://arxiv.org/abs/2005.12146}{{\tt
  arXiv:2005.12146 [hep-ph]}}.

\bibitem{Brauer:2020kfv}
S.~Br\"auer, A.~Denner, M.~Pellen, M.~Sch\"onherr, and S.~Schumann, {\em
  {Fixed-order and merged parton-shower predictions for WW and WWj production
  at the LHC including NLO QCD and EW corrections}}.
  \href{http://dx.doi.org/10.1007/JHEP10(2020)159}{JHEP {\bf 10} (2020)  159},
  \href{http://arxiv.org/abs/2005.12128}{{\tt arXiv:2005.12128 [hep-ph]}}.

\bibitem{Lindert:2022qdd}
J.~M. Lindert, D.~Lombardi, M.~Wiesemann, G.~Zanderighi, and S.~Zanoli, {\em
  {W$^{\pm}$Z production at NNLO QCD and NLO EW matched to parton showers with
  MiNNLO$_{PS}$}}. \href{http://dx.doi.org/10.1007/JHEP11(2022)036}{JHEP {\bf
  11} (2022)  036}, \href{http://arxiv.org/abs/2208.12660}{{\tt
  arXiv:2208.12660 [hep-ph]}}.

\bibitem{Hollik:2011ps}
W.~Hollik and D.~Pagani, {\em {The electroweak contribution to the top quark
  forward-backward asymmetry at the Tevatron}}.
  \href{http://dx.doi.org/10.1103/PhysRevD.84.093003}{Phys. Rev. D {\bf 84}
  (2011)  093003}, \href{http://arxiv.org/abs/1107.2606}{{\tt arXiv:1107.2606
  [hep-ph]}}.

\bibitem{Baglio:2013toa}
J.~Baglio, L.~D. Ninh, and M.~M. Weber, {\em {Massive gauge boson pair
  production at the LHC: a next-to-leading order story}}.
  \href{http://dx.doi.org/10.1103/PhysRevD.94.099902}{Phys. Rev. D {\bf 88}
  (2013)  113005}, \href{http://arxiv.org/abs/1307.4331}{{\tt arXiv:1307.4331
  [hep-ph]}}. [Erratum: Phys.Rev.D 94, 099902 (2016)].

\bibitem{Biedermann:2016yds}
B.~Biedermann, A.~Denner, and M.~Pellen, {\em {Large electroweak corrections to
  vector-boson scattering at the Large Hadron Collider}}.
  \href{http://dx.doi.org/10.1103/PhysRevLett.118.261801}{Phys. Rev. Lett. {\bf
  118} (2017) no.~26, 261801}, \href{http://arxiv.org/abs/1611.02951}{{\tt
  arXiv:1611.02951 [hep-ph]}}.

\bibitem{Biedermann:2017bss}
B.~Biedermann, A.~Denner, and M.~Pellen, {\em {Complete NLO corrections to
  W$^{+}$W$^{+}$ scattering and its irreducible background at the LHC}}.
  \href{http://dx.doi.org/10.1007/JHEP10(2017)124}{JHEP {\bf 10} (2017)  124},
  \href{http://arxiv.org/abs/1708.00268}{{\tt arXiv:1708.00268 [hep-ph]}}.

\bibitem{Frederix:2017wme}
R.~Frederix, D.~Pagani, and M.~Zaro, {\em {Large NLO corrections in
  $t\bar{t}W^{\pm}$ and $t\bar{t}t\bar{t}$ hadroproduction from supposedly
  subleading EW contributions}}.
  \href{http://dx.doi.org/10.1007/JHEP02(2018)031}{JHEP {\bf 02} (2018)  031},
  \href{http://arxiv.org/abs/1711.02116}{{\tt arXiv:1711.02116 [hep-ph]}}.

\bibitem{Denner:2019tmn}
A.~Denner, S.~Dittmaier, P.~Maierh\"ofer, M.~Pellen, and C.~Schwan, {\em {QCD
  and electroweak corrections to WZ scattering at the LHC}}.
  \href{http://dx.doi.org/10.1007/JHEP06(2019)067}{JHEP {\bf 06} (2019)  067},
  \href{http://arxiv.org/abs/1904.00882}{{\tt arXiv:1904.00882 [hep-ph]}}.

\bibitem{Denner:2020zit}
A.~Denner, R.~Franken, M.~Pellen, and T.~Schmidt, {\em {NLO QCD and EW
  corrections to vector-boson scattering into ZZ at the LHC}}.
  \href{http://dx.doi.org/10.1007/JHEP11(2020)110}{JHEP {\bf 11} (2020)  110},
  \href{http://arxiv.org/abs/2009.00411}{{\tt arXiv:2009.00411 [hep-ph]}}.

\bibitem{Sjostrand:2007gs}
T.~Sjostrand, S.~Mrenna, and P.~Z. Skands, {\em {A Brief Introduction to PYTHIA
  8.1}}. \href{http://dx.doi.org/10.1016/j.cpc.2008.01.036}{Comput. Phys.
  Commun. {\bf 178} (2008)  852--867},
  \href{http://arxiv.org/abs/0710.3820}{{\tt arXiv:0710.3820 [hep-ph]}}.

\bibitem{Sjostrand:2014zea}
T.~Sj\"ostrand, S.~Ask, J.~R. Christiansen, R.~Corke, N.~Desai, P.~Ilten,
  S.~Mrenna, S.~Prestel, C.~O. Rasmussen, and P.~Z. Skands, {\em {An
  introduction to PYTHIA 8.2}}.
  \href{http://dx.doi.org/10.1016/j.cpc.2015.01.024}{Comput. Phys. Commun. {\bf
  191} (2015)  159--177}, \href{http://arxiv.org/abs/1410.3012}{{\tt
  arXiv:1410.3012 [hep-ph]}}.

\bibitem{Bierlich:2022pfr}
C.~Bierlich {\em et al.}, {\em {A comprehensive guide to the physics and usage
  of PYTHIA 8.3}}. \href{http://arxiv.org/abs/2203.11601}{{\tt arXiv:2203.11601
  [hep-ph]}}.

\bibitem{Bellm:2015jjp}
J.~Bellm {\em et al.}, {\em {Herwig 7.0/Herwig++ 3.0 release note}}.
  \href{http://dx.doi.org/10.1140/epjc/s10052-016-4018-8}{Eur. Phys. J. C {\bf
  76} (2016) no.~4, 196}, \href{http://arxiv.org/abs/1512.01178}{{\tt
  arXiv:1512.01178 [hep-ph]}}.

\bibitem{Sherpa:2019gpd}
{\bf Sherpa} Collaboration, E.~Bothmann {\em et al.}, {\em {Event Generation
  with Sherpa 2.2}}.
  \href{http://dx.doi.org/10.21468/SciPostPhys.7.3.034}{SciPost Phys. {\bf 7}
  (2019) no.~3, 034}, \href{http://arxiv.org/abs/1905.09127}{{\tt
  arXiv:1905.09127 [hep-ph]}}.

\bibitem{Schonherr:2008av}
M.~Schonherr and F.~Krauss, {\em {Soft Photon Radiation in Particle Decays in
  SHERPA}}. \href{http://dx.doi.org/10.1088/1126-6708/2008/12/018}{JHEP {\bf
  12} (2008)  018}, \href{http://arxiv.org/abs/0810.5071}{{\tt arXiv:0810.5071
  [hep-ph]}}.

\bibitem{Hoeche:2009xc}
S.~Hoeche, S.~Schumann, and F.~Siegert, {\em {Hard photon production and
  matrix-element parton-shower merging}}.
  \href{http://dx.doi.org/10.1103/PhysRevD.81.034026}{Phys. Rev. D {\bf 81}
  (2010)  034026}, \href{http://arxiv.org/abs/0912.3501}{{\tt arXiv:0912.3501
  [hep-ph]}}.

\bibitem{Sudakov:1954sw}
V.~V. Sudakov, {\em {Vertex parts at very high-energies in quantum
  electrodynamics}}. Sov. Phys. JETP {\bf 3} (1956)  65--71.

\bibitem{Denner:2000jv}
A.~Denner and S.~Pozzorini, {\em {One loop leading logarithms in electroweak
  radiative corrections. 1. Results}}.
  \href{http://dx.doi.org/10.1007/s100520100551}{Eur. Phys. J. C {\bf 18}
  (2001)  461--480}, \href{http://arxiv.org/abs/hep-ph/0010201}{{\tt
  arXiv:hep-ph/0010201}}.

\bibitem{Denner:2001gw}
A.~Denner and S.~Pozzorini, {\em {One loop leading logarithms in electroweak
  radiative corrections. 2. Factorization of collinear singularities}}.
  \href{http://dx.doi.org/10.1007/s100520100721}{Eur. Phys. J. C {\bf 21}
  (2001)  63--79}, \href{http://arxiv.org/abs/hep-ph/0104127}{{\tt
  arXiv:hep-ph/0104127}}.

\bibitem{Denner:2003wi}
A.~Denner, M.~Melles, and S.~Pozzorini, {\em {Two loop electroweak angular
  dependent logarithms at high-energies}}.
  \href{http://dx.doi.org/10.1016/S0550-3213(03)00307-9}{Nucl. Phys. B {\bf
  662} (2003)  299--333}, \href{http://arxiv.org/abs/hep-ph/0301241}{{\tt
  arXiv:hep-ph/0301241}}.

\bibitem{Denner:2004iz}
A.~Denner and S.~Pozzorini, {\em {An Algorithm for the high-energy expansion of
  multi-loop diagrams to next-to-leading logarithmic accuracy}}.
  \href{http://dx.doi.org/10.1016/j.nuclphysb.2005.03.036}{Nucl. Phys. B {\bf
  717} (2005)  48--85}, \href{http://arxiv.org/abs/hep-ph/0408068}{{\tt
  arXiv:hep-ph/0408068}}.

\bibitem{Denner:2006jr}
A.~Denner, B.~Jantzen, and S.~Pozzorini, {\em {Two-loop electroweak
  next-to-leading logarithmic corrections to massless fermionic processes}}.
  \href{http://dx.doi.org/10.1016/j.nuclphysb.2006.10.014}{Nucl. Phys. B {\bf
  761} (2007)  1--62}, \href{http://arxiv.org/abs/hep-ph/0608326}{{\tt
  arXiv:hep-ph/0608326}}.

\bibitem{Denner:2008yn}
A.~Denner, B.~Jantzen, and S.~Pozzorini, {\em {Two-loop electroweak
  next-to-leading logarithms for processes involving heavy quarks}}.
  \href{http://dx.doi.org/10.1088/1126-6708/2008/11/062}{JHEP {\bf 11} (2008)
  062}, \href{http://arxiv.org/abs/0809.0800}{{\tt arXiv:0809.0800 [hep-ph]}}.

\bibitem{Bothmann:2020sxm}
E.~Bothmann and D.~Napoletano, {\em {Automated evaluation of electroweak
  Sudakov logarithms in Sherpa}}.
  \href{http://dx.doi.org/10.1140/epjc/s10052-020-08596-2}{Eur. Phys. J. C {\bf
  80} (2020) no.~11, 1024}, \href{http://arxiv.org/abs/2006.14635}{{\tt
  arXiv:2006.14635 [hep-ph]}}.

\bibitem{Pagani:2021vyk}
D.~Pagani and M.~Zaro, {\em {One-loop electroweak Sudakov logarithms: a
  revisitation and automation}}.
  \href{http://dx.doi.org/10.1007/JHEP02(2022)161}{JHEP {\bf 02} (2022)  161},
  \href{http://arxiv.org/abs/2110.03714}{{\tt arXiv:2110.03714 [hep-ph]}}.

\bibitem{Hoeche:2012yf}
S.~Hoeche, F.~Krauss, M.~Schonherr, and F.~Siegert, {\em {QCD matrix elements +
  parton showers: The NLO case}}.
  \href{http://dx.doi.org/10.1007/JHEP04(2013)027}{JHEP {\bf 04} (2013)  027},
  \href{http://arxiv.org/abs/1207.5030}{{\tt arXiv:1207.5030 [hep-ph]}}.

\bibitem{Bothmann:2021led}
E.~Bothmann, D.~Napoletano, M.~Sch\"onherr, S.~Schumann, and S.~L. Villani,
  {\em {Higher-order EW corrections in ZZ and ZZj production at the LHC}}.
  \href{http://dx.doi.org/10.1007/JHEP06(2022)064}{JHEP {\bf 06} (2022)  064},
  \href{http://arxiv.org/abs/2111.13453}{{\tt arXiv:2111.13453 [hep-ph]}}.

\bibitem{Mattelaer:2016gcx}
O.~Mattelaer, {\em {On the maximal use of Monte Carlo samples: re-weighting
  events at NLO accuracy}}.
  \href{http://dx.doi.org/10.1140/epjc/s10052-016-4533-7}{Eur. Phys. J. C {\bf
  76} (2016) no.~12, 674}, \href{http://arxiv.org/abs/1607.00763}{{\tt
  arXiv:1607.00763 [hep-ph]}}.

\bibitem{Artoisenet:2012st}
P.~Artoisenet, R.~Frederix, O.~Mattelaer, and R.~Rietkerk, {\em {Automatic
  spin-entangled decays of heavy resonances in Monte Carlo simulations}}.
  \href{http://dx.doi.org/10.1007/JHEP03(2013)015}{JHEP {\bf 03} (2013)  015},
  \href{http://arxiv.org/abs/1212.3460}{{\tt arXiv:1212.3460 [hep-ph]}}.

\bibitem{Lindert:2023fcu}
J.~M. Lindert and L.~Mai, {\em {Logarithmic EW corrections at one-loop}}.
  \href{http://arxiv.org/abs/2312.07927}{{\tt arXiv:2312.07927 [hep-ph]}}.

\bibitem{Frixione:2014qaa}
S.~Frixione, V.~Hirschi, D.~Pagani, H.~S. Shao, and M.~Zaro, {\em {Weak
  corrections to Higgs hadroproduction in association with a top-quark pair}}.
  \href{http://dx.doi.org/10.1007/JHEP09(2014)065}{JHEP {\bf 09} (2014)  065},
  \href{http://arxiv.org/abs/1407.0823}{{\tt arXiv:1407.0823 [hep-ph]}}.

\bibitem{Pagani:2016caq}
D.~Pagani, I.~Tsinikos, and M.~Zaro, {\em {The impact of the photon PDF and
  electroweak corrections on $t \bar{t}$ distributions}}.
  \href{http://dx.doi.org/10.1140/epjc/s10052-016-4318-z}{Eur. Phys. J. C {\bf
  76} (2016) no.~9, 479}, \href{http://arxiv.org/abs/1606.01915}{{\tt
  arXiv:1606.01915 [hep-ph]}}.

\bibitem{Frederix:2016ost}
R.~Frederix, S.~Frixione, V.~Hirschi, D.~Pagani, H.-S. Shao, and M.~Zaro, {\em
  {The complete NLO corrections to dijet hadroproduction}}.
  \href{http://dx.doi.org/10.1007/JHEP04(2017)076}{JHEP {\bf 04} (2017)  076},
  \href{http://arxiv.org/abs/1612.06548}{{\tt arXiv:1612.06548 [hep-ph]}}.

\bibitem{Czakon:2017wor}
M.~Czakon, D.~Heymes, A.~Mitov, D.~Pagani, I.~Tsinikos, and M.~Zaro, {\em
  {Top-pair production at the LHC through NNLO QCD and NLO EW}}.
  \href{http://dx.doi.org/10.1007/JHEP10(2017)186}{JHEP {\bf 10} (2017)  186},
  \href{http://arxiv.org/abs/1705.04105}{{\tt arXiv:1705.04105 [hep-ph]}}.

\bibitem{Broggio:2019ewu}
A.~Broggio, A.~Ferroglia, R.~Frederix, D.~Pagani, B.~D. Pecjak, and
  I.~Tsinikos, {\em {Top-quark pair hadroproduction in association with a heavy
  boson at NLO+NNLL including EW corrections}}.
  \href{http://dx.doi.org/10.1007/JHEP08(2019)039}{JHEP {\bf 08} (2019)  039},
  \href{http://arxiv.org/abs/1907.04343}{{\tt arXiv:1907.04343 [hep-ph]}}.

\bibitem{Frederix:2019ubd}
R.~Frederix, D.~Pagani, and I.~Tsinikos, {\em {Precise predictions for
  single-top production: the impact of EW corrections and QCD shower on the
  $t$-channel signature}}.
  \href{http://dx.doi.org/10.1007/JHEP09(2019)122}{JHEP {\bf 09} (2019)  122},
  \href{http://arxiv.org/abs/1907.12586}{{\tt arXiv:1907.12586 [hep-ph]}}.

\bibitem{Pagani:2020rsg}
D.~Pagani, H.-S. Shao, and M.~Zaro, {\em {RIP $ Hb\overline{b} $: how other
  Higgs production modes conspire to kill a rare signal at the LHC}}.
  \href{http://dx.doi.org/10.1007/JHEP11(2020)036}{JHEP {\bf 11} (2020)  036},
  \href{http://arxiv.org/abs/2005.10277}{{\tt arXiv:2005.10277 [hep-ph]}}.

\bibitem{Pagani:2020mov}
D.~Pagani, I.~Tsinikos, and E.~Vryonidou, {\em {NLO QCD+EW predictions for
  $tHj$ and $tZj$ production at the LHC}}.
  \href{http://dx.doi.org/10.1007/JHEP08(2020)082}{JHEP {\bf 08} (2020)  082},
  \href{http://arxiv.org/abs/2006.10086}{{\tt arXiv:2006.10086 [hep-ph]}}.

\bibitem{Christiansen:2014kba}
J.~R. Christiansen and T.~Sj\"ostrand, {\em {Weak Gauge Boson Radiation in
  Parton Showers}}. \href{http://dx.doi.org/10.1007/JHEP04(2014)115}{JHEP {\bf
  04} (2014)  115}, \href{http://arxiv.org/abs/1401.5238}{{\tt arXiv:1401.5238
  [hep-ph]}}.

\bibitem{Kleiss:2020rcg}
R.~Kleiss and R.~Verheyen, {\em {Collinear electroweak radiation in antenna
  parton showers}}.
  \href{http://dx.doi.org/10.1140/epjc/s10052-020-08510-w}{Eur. Phys. J. C {\bf
  80} (2020) no.~10, 980}, \href{http://arxiv.org/abs/2002.09248}{{\tt
  arXiv:2002.09248 [hep-ph]}}.

\bibitem{Brooks:2021kji}
H.~Brooks, P.~Skands, and R.~Verheyen, {\em {Interleaved resonance decays and
  electroweak radiation in the Vincia parton shower}}.
  \href{http://dx.doi.org/10.21468/SciPostPhys.12.3.101}{SciPost Phys. {\bf 12}
  (2022) no.~3, 101}, \href{http://arxiv.org/abs/2108.10786}{{\tt
  arXiv:2108.10786 [hep-ph]}}.

\bibitem{Masouminia:2021kne}
M.~R. Masouminia and P.~Richardson, {\em {Implementation of angularly ordered
  electroweak parton shower in Herwig 7}}.
  \href{http://dx.doi.org/10.1007/JHEP04(2022)112}{JHEP {\bf 04} (2022)  112},
  \href{http://arxiv.org/abs/2108.10817}{{\tt arXiv:2108.10817 [hep-ph]}}.

\bibitem{Mangano:2016jyj}
M.~L. Mangano {\em et al.}, {\em {Physics at a 100 TeV pp Collider: Standard
  Model Processes}}. \href{http://arxiv.org/abs/1607.01831}{{\tt
  arXiv:1607.01831 [hep-ph]}}.

\bibitem{Bauer:2018arx}
C.~W. Bauer and B.~R. Webber, {\em {Polarization Effects in Standard Model
  Parton Distributions at Very High Energies}}.
  \href{http://dx.doi.org/10.1007/JHEP03(2019)013}{JHEP {\bf 03} (2019)  013},
  \href{http://arxiv.org/abs/1808.08831}{{\tt arXiv:1808.08831 [hep-ph]}}.

\bibitem{Fornal:2018znf}
B.~Fornal, A.~V. Manohar, and W.~J. Waalewijn, {\em {Electroweak Gauge Boson
  Parton Distribution Functions}}.
  \href{http://dx.doi.org/10.1007/JHEP05(2018)106}{JHEP {\bf 05} (2018)  106},
  \href{http://arxiv.org/abs/1803.06347}{{\tt arXiv:1803.06347 [hep-ph]}}.

\bibitem{Han:2020uid}
T.~Han, Y.~Ma, and K.~Xie, {\em {High energy leptonic collisions and
  electroweak parton distribution functions}}.
  \href{http://dx.doi.org/10.1103/PhysRevD.103.L031301}{Phys. Rev. D {\bf 103}
  (2021) no.~3, L031301}, \href{http://arxiv.org/abs/2007.14300}{{\tt
  arXiv:2007.14300 [hep-ph]}}.

\bibitem{Han:2021kes}
T.~Han, Y.~Ma, and K.~Xie, {\em {Quark and gluon contents of a lepton at high
  energies}}. \href{http://dx.doi.org/10.1007/JHEP02(2022)154}{JHEP {\bf 02}
  (2022)  154}, \href{http://arxiv.org/abs/2103.09844}{{\tt arXiv:2103.09844
  [hep-ph]}}.

\bibitem{Ruiz:2021tdt}
R.~Ruiz, A.~Costantini, F.~Maltoni, and O.~Mattelaer, {\em {The Effective
  Vector Boson Approximation in high-energy muon collisions}}.
  \href{http://dx.doi.org/10.1007/JHEP06(2022)114}{JHEP {\bf 06} (2022)  114},
  \href{http://arxiv.org/abs/2111.02442}{{\tt arXiv:2111.02442 [hep-ph]}}.

\bibitem{Garosi:2023bvq}
F.~Garosi, D.~Marzocca, and S.~Trifinopoulos, {\em {LePDF: Standard Model PDFs
  for High-Energy Lepton Colliders}}.
  \href{http://arxiv.org/abs/2303.16964}{{\tt arXiv:2303.16964 [hep-ph]}}.

\bibitem{Frixione:2021yim}
S.~Frixione and B.~R. Webber, {\em {The role of colour flows in matrix element
  computations and Monte Carlo simulations}}.
  \href{http://dx.doi.org/10.1007/JHEP11(2021)045}{JHEP {\bf 11} (2021)  045},
  \href{http://arxiv.org/abs/2106.13471}{{\tt arXiv:2106.13471 [hep-ph]}}.

\bibitem{Kallweit:2017khh}
S.~Kallweit, J.~M. Lindert, S.~Pozzorini, and M.~Sch\"onherr, {\em {NLO QCD+EW
  predictions for $2\ell2\nu$ diboson signatures at the LHC}}.
  \href{http://dx.doi.org/10.1007/JHEP11(2017)120}{JHEP {\bf 11} (2017)  120},
  \href{http://arxiv.org/abs/1705.00598}{{\tt arXiv:1705.00598 [hep-ph]}}.

\bibitem{Maltoni:2014eza}
F.~Maltoni, E.~Vryonidou, and M.~Zaro, {\em {Top-quark mass effects in double
  and triple Higgs production in gluon-gluon fusion at NLO}}.
  \href{http://dx.doi.org/10.1007/JHEP11(2014)079}{JHEP {\bf 11} (2014)  079},
  \href{http://arxiv.org/abs/1408.6542}{{\tt arXiv:1408.6542 [hep-ph]}}.

\bibitem{Frixione:1995ms}
S.~Frixione, Z.~Kunszt, and A.~Signer, {\em {Three jet cross-sections to
  next-to-leading order}}.
  \href{http://dx.doi.org/10.1016/0550-3213(96)00110-1}{Nucl. Phys. B {\bf 467}
  (1996)  399--442}, \href{http://arxiv.org/abs/hep-ph/9512328}{{\tt
  arXiv:hep-ph/9512328}}.

\bibitem{Catani:1996vz}
S.~Catani and M.~H. Seymour, {\em {A General algorithm for calculating jet
  cross-sections in NLO QCD}}.
  \href{http://dx.doi.org/10.1016/S0550-3213(96)00589-5}{Nucl. Phys. B {\bf
  485} (1997)  291--419}, \href{http://arxiv.org/abs/hep-ph/9605323}{{\tt
  arXiv:hep-ph/9605323}}. [Erratum: Nucl.Phys.B 510, 503--504 (1998)].

\bibitem{Torrielli:2010aw}
P.~Torrielli and S.~Frixione, {\em {Matching NLO QCD computations with PYTHIA
  using MC@NLO}}. \href{http://dx.doi.org/10.1007/JHEP04(2010)110}{JHEP {\bf
  04} (2010)  110}, \href{http://arxiv.org/abs/1002.4293}{{\tt arXiv:1002.4293
  [hep-ph]}}.

\bibitem{Frixione:2010ra}
S.~Frixione, F.~Stoeckli, P.~Torrielli, and B.~R. Webber, {\em {NLO QCD
  corrections in Herwig++ with MC@NLO}}.
  \href{http://dx.doi.org/10.1007/JHEP01(2011)053}{JHEP {\bf 01} (2011)  053},
  \href{http://arxiv.org/abs/1010.0568}{{\tt arXiv:1010.0568 [hep-ph]}}.

\bibitem{Bahr:2008pv}
M.~Bahr {\em et al.}, {\em {Herwig++ Physics and Manual}}.
  \href{http://dx.doi.org/10.1140/epjc/s10052-008-0798-9}{Eur. Phys. J. C {\bf
  58} (2008)  639--707}, \href{http://arxiv.org/abs/0803.0883}{{\tt
  arXiv:0803.0883 [hep-ph]}}.

\bibitem{Bellm:2013hwb}
J.~Bellm {\em et al.}, {\em {Herwig++ 2.7 Release Note}}.
  \href{http://arxiv.org/abs/1310.6877}{{\tt arXiv:1310.6877 [hep-ph]}}.

\bibitem{Corcella:2000bw}
G.~Corcella, I.~G. Knowles, G.~Marchesini, S.~Moretti, K.~Odagiri,
  P.~Richardson, M.~H. Seymour, and B.~R. Webber, {\em {HERWIG 6: An Event
  generator for hadron emission reactions with interfering gluons (including
  supersymmetric processes)}}.
  \href{http://dx.doi.org/10.1088/1126-6708/2001/01/010}{JHEP {\bf 01} (2001)
  010}, \href{http://arxiv.org/abs/hep-ph/0011363}{{\tt arXiv:hep-ph/0011363}}.

\bibitem{Corcella:2002jc}
G.~Corcella, I.~G. Knowles, G.~Marchesini, S.~Moretti, K.~Odagiri,
  P.~Richardson, M.~H. Seymour, and B.~R. Webber, {\em {HERWIG 6.5 release
  note}}. \href{http://arxiv.org/abs/hep-ph/0210213}{{\tt
  arXiv:hep-ph/0210213}}.

\bibitem{Sjostrand:2006za}
T.~Sjostrand, S.~Mrenna, and P.~Z. Skands, {\em {PYTHIA 6.4 Physics and
  Manual}}. \href{http://dx.doi.org/10.1088/1126-6708/2006/05/026}{JHEP {\bf
  05} (2006)  026}, \href{http://arxiv.org/abs/hep-ph/0603175}{{\tt
  arXiv:hep-ph/0603175}}.

\bibitem{Bagnaschi:2018dnh}
E.~Bagnaschi, F.~Maltoni, A.~Vicini, and M.~Zaro, {\em {Lepton-pair production
  in association with a $ b\overline{b} $ pair and the determination of the $W$
  boson mass}}. \href{http://dx.doi.org/10.1007/JHEP07(2018)101}{JHEP {\bf 07}
  (2018)  101}, \href{http://arxiv.org/abs/1803.04336}{{\tt arXiv:1803.04336
  [hep-ph]}}.

\bibitem{Frederix:2009yq}
R.~Frederix, S.~Frixione, F.~Maltoni, and T.~Stelzer, {\em {Automation of
  next-to-leading order computations in QCD: The FKS subtraction}}.
  \href{http://dx.doi.org/10.1088/1126-6708/2009/10/003}{JHEP {\bf 10} (2009)
  003}, \href{http://arxiv.org/abs/0908.4272}{{\tt arXiv:0908.4272 [hep-ph]}}.

\bibitem{Frixione:2019fxg}
S.~Frixione, B.~Fuks, V.~Hirschi, K.~Mawatari, H.-S. Shao, P.~A. Sunder, and
  M.~Zaro, {\em {Automated simulations beyond the Standard Model:
  supersymmetry}}. \href{http://dx.doi.org/10.1007/JHEP12(2019)008}{JHEP {\bf
  12} (2019)  008}, \href{http://arxiv.org/abs/1907.04898}{{\tt
  arXiv:1907.04898 [hep-ph]}}.

\bibitem{Beenakker:1996ch}
W.~Beenakker, R.~Hopker, M.~Spira, and P.~M. Zerwas, {\em {Squark and gluino
  production at hadron colliders}}.
  \href{http://dx.doi.org/10.1016/S0550-3213(97)80027-2}{Nucl. Phys. B {\bf
  492} (1997)  51--103}, \href{http://arxiv.org/abs/hep-ph/9610490}{{\tt
  arXiv:hep-ph/9610490}}.

\bibitem{Frixione:2008yi}
S.~Frixione, E.~Laenen, P.~Motylinski, B.~R. Webber, and C.~D. White, {\em
  {Single-top hadroproduction in association with a W boson}}.
  \href{http://dx.doi.org/10.1088/1126-6708/2008/07/029}{JHEP {\bf 07} (2008)
  029}, \href{http://arxiv.org/abs/0805.3067}{{\tt arXiv:0805.3067 [hep-ph]}}.

\bibitem{Hollik:2012rc}
W.~Hollik, J.~M. Lindert, and D.~Pagani, {\em {NLO corrections to squark-squark
  production and decay at the LHC}}.
  \href{http://dx.doi.org/10.1007/JHEP03(2013)139}{JHEP {\bf 03} (2013)  139},
  \href{http://arxiv.org/abs/1207.1071}{{\tt arXiv:1207.1071 [hep-ph]}}.

\bibitem{Demartin:2016axk}
F.~Demartin, B.~Maier, F.~Maltoni, K.~Mawatari, and M.~Zaro, {\em {tWH
  associated production at the LHC}}.
  \href{http://dx.doi.org/10.1140/epjc/s10052-017-4601-7}{Eur. Phys. J. C {\bf
  77} (2017) no.~1, 34}, \href{http://arxiv.org/abs/1607.05862}{{\tt
  arXiv:1607.05862 [hep-ph]}}.

\bibitem{NNPDF:2021njg}
{\bf NNPDF} Collaboration, R.~D. Ball {\em et al.}, {\em {The path to proton
  structure at 1\% accuracy}}.
  \href{http://dx.doi.org/10.1140/epjc/s10052-022-10328-7}{Eur. Phys. J. C {\bf
  82} (2022) no.~5, 428}, \href{http://arxiv.org/abs/2109.02653}{{\tt
  arXiv:2109.02653 [hep-ph]}}.

\bibitem{Cacciari:2008gp}
M.~Cacciari, G.~P. Salam, and G.~Soyez, {\em {The anti-$k_t$ jet clustering
  algorithm}}. \href{http://dx.doi.org/10.1088/1126-6708/2008/04/063}{JHEP {\bf
  04} (2008)  063}, \href{http://arxiv.org/abs/0802.1189}{{\tt arXiv:0802.1189
  [hep-ph]}}.

\bibitem{Cacciari:2011ma}
M.~Cacciari, G.~P. Salam, and G.~Soyez, {\em {FastJet User Manual}}.
  \href{http://dx.doi.org/10.1140/epjc/s10052-012-1896-2}{Eur. Phys. J. C {\bf
  72} (2012)  1896}, \href{http://arxiv.org/abs/1111.6097}{{\tt arXiv:1111.6097
  [hep-ph]}}.

\bibitem{Ng:1983jm}
J.~N. Ng and P.~Zakarauskas, {\em {A {QCD} Parton Calculation of Conjoined
  Production of Higgs Bosons and Heavy Flavors in $p \bar{p}$ Collision}}.
  \href{http://dx.doi.org/10.1103/PhysRevD.29.876}{Phys. Rev. D {\bf 29} (1984)
   876}.

\bibitem{Kunszt:1984ri}
Z.~Kunszt, {\em {Associated Production of Heavy Higgs Boson with Top Quarks}}.
  \href{http://dx.doi.org/10.1016/0550-3213(84)90553-4}{Nucl. Phys. B {\bf 247}
  (1984)  339--359}.

\bibitem{Beenakker:2001rj}
W.~Beenakker, S.~Dittmaier, M.~Kramer, B.~Plumper, M.~Spira, and P.~M. Zerwas,
  {\em {Higgs radiation off top quarks at the Tevatron and the LHC}}.
  \href{http://dx.doi.org/10.1103/PhysRevLett.87.201805}{Phys. Rev. Lett. {\bf
  87} (2001)  201805}, \href{http://arxiv.org/abs/hep-ph/0107081}{{\tt
  arXiv:hep-ph/0107081}}.

\bibitem{Beenakker:2002nc}
W.~Beenakker, S.~Dittmaier, M.~Kramer, B.~Plumper, M.~Spira, and P.~M. Zerwas,
  {\em {NLO QCD corrections to t anti-t H production in hadron collisions}}.
  \href{http://dx.doi.org/10.1016/S0550-3213(03)00044-0}{Nucl. Phys. B {\bf
  653} (2003)  151--203}, \href{http://arxiv.org/abs/hep-ph/0211352}{{\tt
  arXiv:hep-ph/0211352}}.

\bibitem{Reina:2001sf}
L.~Reina and S.~Dawson, {\em {Next-to-leading order results for t anti-t h
  production at the Tevatron}}.
  \href{http://dx.doi.org/10.1103/PhysRevLett.87.201804}{Phys. Rev. Lett. {\bf
  87} (2001)  201804}, \href{http://arxiv.org/abs/hep-ph/0107101}{{\tt
  arXiv:hep-ph/0107101}}.

\bibitem{Reina:2001bc}
L.~Reina, S.~Dawson, and D.~Wackeroth, {\em {QCD corrections to associated t
  anti-t h production at the Tevatron}}.
  \href{http://dx.doi.org/10.1103/PhysRevD.65.053017}{Phys. Rev. D {\bf 65}
  (2002)  053017}, \href{http://arxiv.org/abs/hep-ph/0109066}{{\tt
  arXiv:hep-ph/0109066}}.

\bibitem{Dawson:2002tg}
S.~Dawson, L.~H. Orr, L.~Reina, and D.~Wackeroth, {\em {Associated top quark
  Higgs boson production at the LHC}}.
  \href{http://dx.doi.org/10.1103/PhysRevD.67.071503}{Phys. Rev. D {\bf 67}
  (2003)  071503}, \href{http://arxiv.org/abs/hep-ph/0211438}{{\tt
  arXiv:hep-ph/0211438}}.

\bibitem{Dawson:2003zu}
S.~Dawson, C.~Jackson, L.~H. Orr, L.~Reina, and D.~Wackeroth, {\em {Associated
  Higgs production with top quarks at the large hadron collider: NLO QCD
  corrections}}. \href{http://dx.doi.org/10.1103/PhysRevD.68.034022}{Phys. Rev.
  D {\bf 68} (2003)  034022}, \href{http://arxiv.org/abs/hep-ph/0305087}{{\tt
  arXiv:hep-ph/0305087}}.

\bibitem{Yu:2014cka}
Y.~Zhang, W.-G. Ma, R.-Y. Zhang, C.~Chen, and L.~Guo, {\em {QCD NLO and EW NLO
  corrections to $t\bar{t}H$ production with top quark decays at hadron
  collider}}. \href{http://dx.doi.org/10.1016/j.physletb.2014.09.022}{Phys.
  Lett. B {\bf 738} (2014)  1--5}, \href{http://arxiv.org/abs/1407.1110}{{\tt
  arXiv:1407.1110 [hep-ph]}}.

\bibitem{Kulesza:2015vda}
A.~Kulesza, L.~Motyka, T.~Stebel, and V.~Theeuwes, {\em {Soft gluon resummation
  for associated $t \bar{t} H$ production at the LHC}}.
  \href{http://dx.doi.org/10.1007/JHEP03(2016)065}{JHEP {\bf 03} (2016)  065},
  \href{http://arxiv.org/abs/1509.02780}{{\tt arXiv:1509.02780 [hep-ph]}}.

\bibitem{Broggio:2015lya}
A.~Broggio, A.~Ferroglia, B.~D. Pecjak, A.~Signer, and L.~L. Yang, {\em
  {Associated production of a top pair and a Higgs boson beyond NLO}}.
  \href{http://dx.doi.org/10.1007/JHEP03(2016)124}{JHEP {\bf 03} (2016)  124},
  \href{http://arxiv.org/abs/1510.01914}{{\tt arXiv:1510.01914 [hep-ph]}}.

\bibitem{Broggio:2016lfj}
A.~Broggio, A.~Ferroglia, B.~D. Pecjak, and L.~L. Yang, {\em {NNLL resummation
  for the associated production of a top pair and a Higgs boson at the LHC}}.
  \href{http://dx.doi.org/10.1007/JHEP02(2017)126}{JHEP {\bf 02} (2017)  126},
  \href{http://arxiv.org/abs/1611.00049}{{\tt arXiv:1611.00049 [hep-ph]}}.

\bibitem{Kulesza:2017ukk}
A.~Kulesza, L.~Motyka, T.~Stebel, and V.~Theeuwes, {\em {Associated $t \bar{t}
  H$ production at the LHC: Theoretical predictions at NLO+NNLL accuracy}}.
  \href{http://dx.doi.org/10.1103/PhysRevD.97.114007}{Phys. Rev. D {\bf 97}
  (2018) no.~11, 114007}, \href{http://arxiv.org/abs/1704.03363}{{\tt
  arXiv:1704.03363 [hep-ph]}}.

\bibitem{Ju:2019lwp}
W.-L. Ju and L.~L. Yang, {\em {Resummation of soft and Coulomb corrections for
  $ t\overline{t}h $ production at the LHC}}.
  \href{http://dx.doi.org/10.1007/JHEP06(2019)050}{JHEP {\bf 06} (2019)  050},
  \href{http://arxiv.org/abs/1904.08744}{{\tt arXiv:1904.08744 [hep-ph]}}.

\bibitem{Kulesza:2020nfh}
A.~Kulesza, L.~Motyka, D.~Schwartl\"ander, T.~Stebel, and V.~Theeuwes, {\em
  {Associated top quark pair production with a heavy boson: differential cross
  sections at NLO+NNLL accuracy}}.
  \href{http://dx.doi.org/10.1140/epjc/s10052-020-7987-6}{Eur. Phys. J. C {\bf
  80} (2020) no.~5, 428}, \href{http://arxiv.org/abs/2001.03031}{{\tt
  arXiv:2001.03031 [hep-ph]}}.

\bibitem{Denner:2015yca}
A.~Denner and R.~Feger, {\em {NLO QCD corrections to off-shell top-antitop
  production with leptonic decays in association with a Higgs boson at the
  LHC}}. \href{http://dx.doi.org/10.1007/JHEP11(2015)209}{JHEP {\bf 11} (2015)
  209}, \href{http://arxiv.org/abs/1506.07448}{{\tt arXiv:1506.07448
  [hep-ph]}}.

\bibitem{Denner:2016wet}
A.~Denner, J.-N. Lang, M.~Pellen, and S.~Uccirati, {\em {Higgs production in
  association with off-shell top-antitop pairs at NLO EW and QCD at the LHC}}.
  \href{http://dx.doi.org/10.1007/JHEP02(2017)053}{JHEP {\bf 02} (2017)  053},
  \href{http://arxiv.org/abs/1612.07138}{{\tt arXiv:1612.07138 [hep-ph]}}.

\bibitem{Catani:2022mfv}
S.~Catani, S.~Devoto, M.~Grazzini, S.~Kallweit, J.~Mazzitelli, and C.~Savoini,
  {\em {Higgs Boson Production in Association with a Top-Antitop Quark Pair in
  Next-to-Next-to-Leading Order QCD}}.
  \href{http://dx.doi.org/10.1103/PhysRevLett.130.111902}{Phys. Rev. Lett. {\bf
  130} (2023) no.~11, 111902}, \href{http://arxiv.org/abs/2210.07846}{{\tt
  arXiv:2210.07846 [hep-ph]}}.

\bibitem{Lazopoulos:2007ix}
A.~Lazopoulos, K.~Melnikov, and F.~Petriello, {\em {QCD corrections to
  tri-boson production}}.
  \href{http://dx.doi.org/10.1103/PhysRevD.76.014001}{Phys. Rev. D {\bf 76}
  (2007)  014001}, \href{http://arxiv.org/abs/hep-ph/0703273}{{\tt
  arXiv:hep-ph/0703273}}.

\bibitem{Binoth:2008kt}
T.~Binoth, G.~Ossola, C.~G. Papadopoulos, and R.~Pittau, {\em {NLO QCD
  corrections to tri-boson production}}.
  \href{http://dx.doi.org/10.1088/1126-6708/2008/06/082}{JHEP {\bf 06} (2008)
  082}, \href{http://arxiv.org/abs/0804.0350}{{\tt arXiv:0804.0350 [hep-ph]}}.

\bibitem{Wang:2016fvj}
H.~Wang, R.-Y. Zhang, W.-G. Ma, L.~Guo, X.-Z. Li, and S.-M. Wang, {\em {NLO QCD
  + EW corrections to ZZZ production with subsequent leptonic decays at the
  LHC}}. \href{http://dx.doi.org/10.1088/0954-3899/43/11/115001}{J. Phys. G
  {\bf 43} (2016) no.~11, 115001}, \href{http://arxiv.org/abs/1610.05876}{{\tt
  arXiv:1610.05876 [hep-ph]}}.

\bibitem{Degrassi:2016wml}
G.~Degrassi, P.~P. Giardino, F.~Maltoni, and D.~Pagani, {\em {Probing the Higgs
  self coupling via single Higgs production at the LHC}}.
  \href{http://dx.doi.org/10.1007/JHEP12(2016)080}{JHEP {\bf 12} (2016)  080},
  \href{http://arxiv.org/abs/1607.04251}{{\tt arXiv:1607.04251 [hep-ph]}}.

\bibitem{Czakon:2019bcq}
M.~L. Czakon, C.~G\"utschow, J.~M. Lindert, A.~Mitov, D.~Pagani, A.~S.
  Papanastasiou, M.~Sch\"onherr, I.~Tsinikos, and M.~Zaro, ``{NNLO versus NLO
  multi-jet merging for top-pair production including electroweak
  corrections},'' in {\em {11th International Workshop on Top Quark Physics}}.
\newblock 1, 2019.
\newblock \href{http://arxiv.org/abs/1901.04442}{{\tt arXiv:1901.04442
  [hep-ph]}}.

\bibitem{Bierweiler:2013dja}
A.~Bierweiler, T.~Kasprzik, and J.~H. K\"uhn, {\em {Vector-boson pair
  production at the LHC to $\mathcal{O}(\alpha^3)$ accuracy}}.
  \href{http://dx.doi.org/10.1007/JHEP12(2013)071}{JHEP {\bf 12} (2013)  071},
  \href{http://arxiv.org/abs/1305.5402}{{\tt arXiv:1305.5402 [hep-ph]}}.

\bibitem{Bredt:2022dmm}
P.~M. Bredt, W.~Kilian, J.~Reuter, and P.~Stienemeier, {\em {NLO electroweak
  corrections to multi-boson processes at a muon collider}}.
  \href{http://dx.doi.org/10.1007/JHEP12(2022)138}{JHEP {\bf 12} (2022)  138},
  \href{http://arxiv.org/abs/2208.09438}{{\tt arXiv:2208.09438 [hep-ph]}}.

\bibitem{Frixione:1992pj}
S.~Frixione, P.~Nason, and G.~Ridolfi, {\em {Strong corrections to W Z
  production at hadron colliders}}.
  \href{http://dx.doi.org/10.1016/0550-3213(92)90668-2}{Nucl. Phys. B {\bf 383}
  (1992)  3--44}.

\bibitem{Frixione:1993yp}
S.~Frixione, {\em {A Next-to-leading order calculation of the cross-section for
  the production of W+ W- pairs in hadronic collisions}}.
  \href{http://dx.doi.org/10.1016/0550-3213(93)90435-R}{Nucl. Phys. B {\bf 410}
  (1993)  280--324}.

\bibitem{Rubin:2010xp}
M.~Rubin, G.~P. Salam, and S.~Sapeta, {\em {Giant QCD K-factors beyond NLO}}.
  \href{http://dx.doi.org/10.1007/JHEP09(2010)084}{JHEP {\bf 09} (2010)  084},
  \href{http://arxiv.org/abs/1006.2144}{{\tt arXiv:1006.2144 [hep-ph]}}.

\bibitem{Denner:2019vbn}
A.~Denner and S.~Dittmaier, {\em {Electroweak Radiative Corrections for
  Collider Physics}}.
  \href{http://dx.doi.org/10.1016/j.physrep.2020.04.001}{Phys. Rept. {\bf 864}
  (2020)  1--163}, \href{http://arxiv.org/abs/1912.06823}{{\tt arXiv:1912.06823
  [hep-ph]}}.

\bibitem{Dokshitzer:1997in}
Y.~L. Dokshitzer, G.~D. Leder, S.~Moretti, and B.~R. Webber, {\em {Better jet
  clustering algorithms}}.
  \href{http://dx.doi.org/10.1088/1126-6708/1997/08/001}{JHEP {\bf 08} (1997)
  001}, \href{http://arxiv.org/abs/hep-ph/9707323}{{\tt arXiv:hep-ph/9707323}}.

\bibitem{Wobisch:1998wt}
M.~Wobisch and T.~Wengler, ``{Hadronization corrections to jet cross-sections
  in deep inelastic scattering},'' in {\em {Workshop on Monte Carlo Generators
  for HERA Physics (Plenary Starting Meeting)}}, pp.~270--279.
\newblock 4, 1998.
\newblock \href{http://arxiv.org/abs/hep-ph/9907280}{{\tt
  arXiv:hep-ph/9907280}}.

\bibitem{CarloniCalame:2007cd}
C.~M. Carloni~Calame, G.~Montagna, O.~Nicrosini, and A.~Vicini, {\em {Precision
  electroweak calculation of the production of a high transverse-momentum
  lepton pair at hadron colliders}}.
  \href{http://dx.doi.org/10.1088/1126-6708/2007/10/109}{JHEP {\bf 10} (2007)
  109}, \href{http://arxiv.org/abs/0710.1722}{{\tt arXiv:0710.1722 [hep-ph]}}.

\bibitem{Dittmaier:2009cr}
S.~Dittmaier and M.~Huber, {\em {Radiative corrections to the neutral-current
  Drell-Yan process in the Standard Model and its minimal supersymmetric
  extension}}. \href{http://dx.doi.org/10.1007/JHEP01(2010)060}{JHEP {\bf 01}
  (2010)  060}, \href{http://arxiv.org/abs/0911.2329}{{\tt arXiv:0911.2329
  [hep-ph]}}.

\bibitem{Gutschow:2020cug}
C.~G\"utschow and M.~Sch\"onherr, {\em {Four lepton production and the accuracy
  of QED FSR}}. \href{http://dx.doi.org/10.1140/epjc/s10052-020-08816-9}{Eur.
  Phys. J. C {\bf 81} (2021) no.~1, 48},
  \href{http://arxiv.org/abs/2007.15360}{{\tt arXiv:2007.15360 [hep-ph]}}.

\bibitem{Frederix:2012ps}
R.~Frederix and S.~Frixione, {\em {Merging meets matching in MC@NLO}}.
  \href{http://dx.doi.org/10.1007/JHEP12(2012)061}{JHEP {\bf 12} (2012)  061},
  \href{http://arxiv.org/abs/1209.6215}{{\tt arXiv:1209.6215 [hep-ph]}}.

\end{thebibliography}\endgroup

\end{document}